\documentclass[aps,prd,reprint,nofootinbib,superscriptaddress, floatfix]{revtex4-2}
\usepackage{graphicx}
\usepackage{float}
\usepackage{amsmath}
\usepackage{tabularx}
\usepackage{booktabs}
\usepackage[colorlinks=true,allcolors=blue]{hyperref}
\usepackage{cleveref}
\usepackage{placeins}
\usepackage{microtype}
\begin{document}

\title{Measurement of Atmospheric Neutrino Mixing with Improved IceCube DeepCore Calibration and Data Processing}
\date{\today}

\affiliation{III. Physikalisches Institut, RWTH Aachen University, D-52056 Aachen, Germany}
\affiliation{Department of Physics, University of Adelaide, Adelaide, 5005, Australia}
\affiliation{Dept. of Physics and Astronomy, University of Alaska Anchorage, 3211 Providence Dr., Anchorage, AK 99508, USA}
\affiliation{Dept. of Physics, University of Texas at Arlington, 502 Yates St., Science Hall Rm 108, Box 19059, Arlington, TX 76019, USA}
\affiliation{CTSPS, Clark-Atlanta University, Atlanta, GA 30314, USA}
\affiliation{School of Physics and Center for Relativistic Astrophysics, Georgia Institute of Technology, Atlanta, GA 30332, USA}
\affiliation{Dept. of Physics, Southern University, Baton Rouge, LA 70813, USA}
\affiliation{Dept. of Physics, University of California, Berkeley, CA 94720, USA}
\affiliation{Lawrence Berkeley National Laboratory, Berkeley, CA 94720, USA}
\affiliation{Institut f{\"u}r Physik, Humboldt-Universit{\"a}t zu Berlin, D-12489 Berlin, Germany}
\affiliation{Fakult{\"a}t f{\"u}r Physik {\&} Astronomie, Ruhr-Universit{\"a}t Bochum, D-44780 Bochum, Germany}
\affiliation{Universit{\'e} Libre de Bruxelles, Science Faculty CP230, B-1050 Brussels, Belgium}
\affiliation{Vrije Universiteit Brussel (VUB), Dienst ELEM, B-1050 Brussels, Belgium}
\affiliation{Department of Physics and Laboratory for Particle Physics and Cosmology, Harvard University, Cambridge, MA 02138, USA}
\affiliation{Dept. of Physics, Massachusetts Institute of Technology, Cambridge, MA 02139, USA}
\affiliation{Dept. of Physics and The International Center for Hadron Astrophysics, Chiba University, Chiba 263-8522, Japan}
\affiliation{Department of Physics, Loyola University Chicago, Chicago, IL 60660, USA}
\affiliation{Dept. of Physics and Astronomy, University of Canterbury, Private Bag 4800, Christchurch, New Zealand}
\affiliation{Dept. of Physics, University of Maryland, College Park, MD 20742, USA}
\affiliation{Dept. of Astronomy, Ohio State University, Columbus, OH 43210, USA}
\affiliation{Dept. of Physics and Center for Cosmology and Astro-Particle Physics, Ohio State University, Columbus, OH 43210, USA}
\affiliation{Niels Bohr Institute, University of Copenhagen, DK-2100 Copenhagen, Denmark}
\affiliation{Dept. of Physics, TU Dortmund University, D-44221 Dortmund, Germany}
\affiliation{Dept. of Physics and Astronomy, Michigan State University, East Lansing, MI 48824, USA}
\affiliation{Dept. of Physics, University of Alberta, Edmonton, Alberta, Canada T6G 2E1}
\affiliation{Erlangen Centre for Astroparticle Physics, Friedrich-Alexander-Universit{\"a}t Erlangen-N{\"u}rnberg, D-91058 Erlangen, Germany}
\affiliation{Physik-department, Technische Universit{\"a}t M{\"u}nchen, D-85748 Garching, Germany}
\affiliation{D{\'e}partement de physique nucl{\'e}aire et corpusculaire, Universit{\'e} de Gen{\`e}ve, CH-1211 Gen{\`e}ve, Switzerland}
\affiliation{Dept. of Physics and Astronomy, University of Gent, B-9000 Gent, Belgium}
\affiliation{Dept. of Physics and Astronomy, University of California, Irvine, CA 92697, USA}
\affiliation{Karlsruhe Institute of Technology, Institute for Astroparticle Physics, D-76021 Karlsruhe, Germany }
\affiliation{Karlsruhe Institute of Technology, Institute of Experimental Particle Physics, D-76021 Karlsruhe, Germany }
\affiliation{Dept. of Physics, Engineering Physics, and Astronomy, Queen's University, Kingston, ON K7L 3N6, Canada}
\affiliation{Department of Physics {\&} Astronomy, University of Nevada, Las Vegas, NV, 89154, USA}
\affiliation{Nevada Center for Astrophysics, University of Nevada, Las Vegas, NV 89154, USA}
\affiliation{Dept. of Physics and Astronomy, University of Kansas, Lawrence, KS 66045, USA}
\affiliation{Department of Physics and Astronomy, UCLA, Los Angeles, CA 90095, USA}
\affiliation{Centre for Cosmology, Particle Physics and Phenomenology - CP3, Universit{\'e} catholique de Louvain, Louvain-la-Neuve, Belgium}
\affiliation{Department of Physics, Mercer University, Macon, GA 31207-0001, USA}
\affiliation{Dept. of Astronomy, University of Wisconsin{\textendash}Madison, Madison, WI 53706, USA}
\affiliation{Dept. of Physics and Wisconsin IceCube Particle Astrophysics Center, University of Wisconsin{\textendash}Madison, Madison, WI 53706, USA}
\affiliation{Institute of Physics, University of Mainz, Staudinger Weg 7, D-55099 Mainz, Germany}
\affiliation{Department of Physics, Marquette University, Milwaukee, WI, 53201, USA}
\affiliation{Institut f{\"u}r Kernphysik, Westf{\"a}lische Wilhelms-Universit{\"a}t M{\"u}nster, D-48149 M{\"u}nster, Germany}
\affiliation{Bartol Research Institute and Dept. of Physics and Astronomy, University of Delaware, Newark, DE 19716, USA}
\affiliation{Dept. of Physics, Yale University, New Haven, CT 06520, USA}
\affiliation{Columbia Astrophysics and Nevis Laboratories, Columbia University, New York, NY 10027, USA}
\affiliation{Dept. of Physics, University of Oxford, Parks Road, Oxford OX1 3PU, UK}
\affiliation{Dipartimento di Fisica e Astronomia Galileo Galilei, Universit{\`a} Degli Studi di Padova, 35122 Padova PD, Italy}
\affiliation{Dept. of Physics, Drexel University, 3141 Chestnut Street, Philadelphia, PA 19104, USA}
\affiliation{Physics Department, South Dakota School of Mines and Technology, Rapid City, SD 57701, USA}
\affiliation{Dept. of Physics, University of Wisconsin, River Falls, WI 54022, USA}
\affiliation{Dept. of Physics and Astronomy, University of Rochester, Rochester, NY 14627, USA}
\affiliation{Department of Physics and Astronomy, University of Utah, Salt Lake City, UT 84112, USA}
\affiliation{Oskar Klein Centre and Dept. of Physics, Stockholm University, SE-10691 Stockholm, Sweden}
\affiliation{Dept. of Physics and Astronomy, Stony Brook University, Stony Brook, NY 11794-3800, USA}
\affiliation{Dept. of Physics, Sungkyunkwan University, Suwon 16419, Korea}
\affiliation{Institute of Physics, Academia Sinica, Taipei, 11529, Taiwan}
\affiliation{Dept. of Physics and Astronomy, University of Alabama, Tuscaloosa, AL 35487, USA}
\affiliation{Dept. of Astronomy and Astrophysics, Pennsylvania State University, University Park, PA 16802, USA}
\affiliation{Dept. of Physics, Pennsylvania State University, University Park, PA 16802, USA}
\affiliation{Dept. of Physics and Astronomy, Uppsala University, Box 516, S-75120 Uppsala, Sweden}
\affiliation{Dept. of Physics, University of Wuppertal, D-42119 Wuppertal, Germany}
\affiliation{Deutsches Elektronen-Synchrotron DESY, Platanenallee 6, 15738 Zeuthen, Germany }

\author{R. Abbasi}
\affiliation{Department of Physics, Loyola University Chicago, Chicago, IL 60660, USA}
\author{M. Ackermann}
\affiliation{Deutsches Elektronen-Synchrotron DESY, Platanenallee 6, 15738 Zeuthen, Germany }
\author{J. Adams}
\affiliation{Dept. of Physics and Astronomy, University of Canterbury, Private Bag 4800, Christchurch, New Zealand}
\author{S. K. Agarwalla}
\thanks{also at Institute of Physics, Sachivalaya Marg, Sainik School Post, Bhubaneswar 751005, India}
\affiliation{Dept. of Physics and Wisconsin IceCube Particle Astrophysics Center, University of Wisconsin{\textendash}Madison, Madison, WI 53706, USA}
\author{J. A. Aguilar}
\affiliation{Universit{\'e} Libre de Bruxelles, Science Faculty CP230, B-1050 Brussels, Belgium}
\author{M. Ahlers}
\affiliation{Niels Bohr Institute, University of Copenhagen, DK-2100 Copenhagen, Denmark}
\author{J.M. Alameddine}
\affiliation{Dept. of Physics, TU Dortmund University, D-44221 Dortmund, Germany}
\author{N. M. Amin}
\affiliation{Bartol Research Institute and Dept. of Physics and Astronomy, University of Delaware, Newark, DE 19716, USA}
\author{K. Andeen}
\affiliation{Department of Physics, Marquette University, Milwaukee, WI, 53201, USA}
\author{G. Anton}
\affiliation{Erlangen Centre for Astroparticle Physics, Friedrich-Alexander-Universit{\"a}t Erlangen-N{\"u}rnberg, D-91058 Erlangen, Germany}
% ...
\date{\today}

\collaboration{IceCube Collaboration}
\noaffiliation

\begin{abstract}
We describe a new data sample of IceCube DeepCore and report on the latest measurement of atmospheric neutrino oscillations obtained with data recorded between 2011-2019. The sample includes significant improvements in data calibration, detector simulation, and data processing, and the analysis benefits from a sophisticated treatment of systematic uncertainties, with significantly greater level of detail since our last study. By measuring the relative fluxes of neutrino flavors as a function of their reconstructed energies and arrival directions we constrain the atmospheric neutrino mixing parameters to be $\sin^2\theta_{23} = 0.51\pm 0.05$ and $\Delta m^2_{32} = 2.41\pm0.07\times 10^{-3}\mathrm{eV}^2$, assuming a normal mass ordering. The errors include both statistical and systematic uncertainties. The resulting 40\% reduction in the error of both parameters with respect to our previous result makes this the most precise measurement of oscillation parameters using atmospheric neutrinos. Our results are also compatible and complementary to those obtained using neutrino beams from accelerators, which are obtained at lower neutrino energies and are subject to different sources of uncertainties.
\end{abstract}

\maketitle

\renewcommand*{\thefootnote}{\fnsymbol{footnote}}
\setcounter{footnote}{3}
\footnotetext{email: \href{mailto:analysis@icecube.wisc.edu}{analysis@icecube.wisc.edu} (corresponding author)}
\renewcommand*{\thefootnote}{\arabic{footnote}}
\setcounter{footnote}{0}

\section{Introduction~\label{sec:introduction}}
The fact that neutrinos are massive particles has been demonstrated by a wide array of experiments in the last few decades~\cite{Super-Kamiokande:1998kpq, SNO:2002tuh, KamLAND:2002uet, GALLEX:1994fys, SAGE:1994ctc, DayaBay:2012fng}. 
The evidence to support this comes exclusively from measurements of flavor transformations, explained by the formalism of neutrino mixing. In that formalism, neutrinos are produced and observed as flavor eigenstates, but they propagate as mass eigenstates. These states are related by the PMNS matrix~\cite{Pontecorvo:1957cp,Maki:1962mu}, a unitary 3$\times$3 matrix that, for flavor transition purposes, is entirely defined by 3 angles $\theta_{ij}$ ($i,j \in [1,2,3]$) and a complex phase $\delta$ as  
\begin{align} \label{eq:full_matrix}
\scriptsize{
\mathbf{U} =
 \begin{pmatrix}
  1 & 0 & 0 \\
  0 & c_{23} & s_{23} \\
  0 & -s_{23} & c_{23}
 \end{pmatrix}
 \begin{pmatrix}
  c_{13} & 0 & e^{-i\delta}s_{13} \\
  0 & 1 & 0 \\
  -e^{i\delta}s_{13} & 0 & c_{13}
 \end{pmatrix}
 \begin{pmatrix}
  c_{12} & s_{12} & 0 \\
  -s_{12} & c_{12} & 0 \\
  0 & 0 & 1
 \end{pmatrix}
 ,
}
\end{align}
where $s$ and $c$ stand for sine and cosine functions and the subscripts denote each of the three angles $\theta_{ij}$.

While the PMNS matrix summarizes the mixing of the states, the mass eigenstates interfere during propagation, so flavor transformations can depend periodically on the square of mass differences $\Delta m^2$, resulting in the phenomenon of neutrino oscillations. To first order, atmospheric neutrino oscillations are well approximated by the simple case of transitions from $\mu$ to $\tau$ flavor, for which the probability takes the form

\begin{equation}
    P_{\mu \rightarrow \tau} \simeq \sin^2(2\theta)\sin^2\left(\Delta m^2 \frac{L}{4E}\right),
    \label{eq:palpha}
\end{equation}
where $L$ denotes the travel distance between source and detection and $E$ stands for the neutrino energy.

The theory of neutrino oscillations explains the possible phenomena produced by mixing of these states, but does not predict the values of either the neutrino masses, their differences, or the value of the  elements of the PMNS matrix. These values must therefore be determined by experiment.

According to global analyses of all available data, all but the imaginary phase are now known with a precision better than 5\%~\cite{Capozzi:2021fjo, deSalas:2020pgw, Esteban:2020cvm}. Unlike the case for the CKM matrix in the quark sector, where an analogous mixing occurs~\cite{Cabibbo:1963yz, Kobayashi:1973fv}, the PMNS matrix introduces a very large mixing between neutrino states, close to maximal for one of them. Precisely measuring its elements thus remains one of the most important goals in neutrino physics. Significant reduction in the errors will provide constraints to theories explaining the matrix structure, and more generally the structure of the fermions in the Standard Model (see~\cite{Feruglio:2015jfa} and references therein), and will make it possible to better constrain the origins of anomalies observed in some oscillation experiments~\cite{LSND:1996ubh,MiniBooNE:2018esg, Kaether:2010ag, SAGE:2009eeu}.

\begin{figure}[!t]
    \centering
    \includegraphics[width=\linewidth]{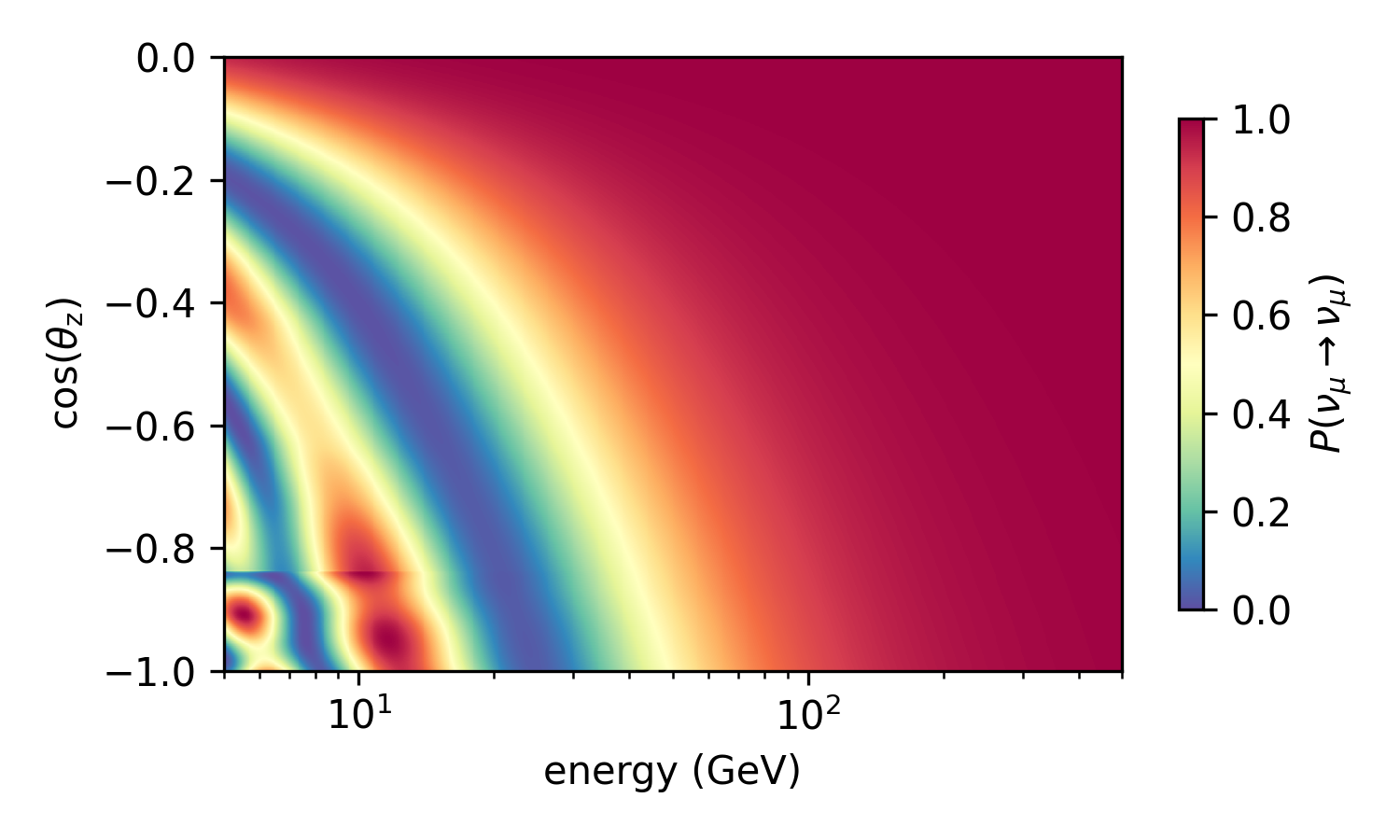}
    \caption{Muon neutrino survival probability as function of energy and $\cos(\theta_{\mathrm{zenith}})$, the latter of which is a proxy for the distance travelled, $L$. The oscillation probabaility is calculated within a three-neutrino framework where oscillation parameters used here are taken from a recent global fit of experimental data assuming the normal mass ordering~\cite{Esteban_2019,*nufit4.0}. We also account the Earth's matter profile~\cite{DZIEWONSKI1981297}, which impacts the oscillations~\cite{Akhmedov:1998ui,Wolfenstein:1977ue,Mikheyev:1985zog} most prominently at the core-mantle boundary as seen below 15 GeV at $\cos\theta_{z}\approx-0.8$.}
    \label{fig:numu_survival_prob}
\end{figure}

In this work we present a new data sample of atmospheric neutrinos collected by the DeepCore subarray of the IceCube Neutrino Observatory. Atmospheric neutrinos are naturally produced by cosmic rays, arrive at the detector from all directions, and their energy spectrum ranges from MeV to hundreds of TeV, making them ideal probes of an effect that changes as a function of $L/E$. The dominant component of the flux are muon neutrinos and antineutrinos, which are subject to a strong periodic modulation due to oscillations below energies of $\sim$100\,GeV, which affects the survival probability of $\nu_\mu$ as function of energy and incoming direction, shown in Fig.~\ref{fig:numu_survival_prob}. Thanks to the large difference in magnitude between the two independent square mass differences, atmospheric neutrino oscillations are, to a good approximation, defined by the value of the mixing angle $\theta_{23}$ and the mass splitting $\Delta m^2_{32}$, as shown in Eq.~\ref{eq:palpha}. The measurement of these two parameters are the main result shown here, where we follow a prescription that accounts for the full PMNS matrix and therefore three-flavour oscillations including for matter effects.

The new sample introduces numerous improvements over previous DeepCore results~\cite{Aartsen_2015, Aartsen_2018, IceCube:2019dqi}. We use an updated response of the optical modules calibrated individually using in-situ data~\cite{IceCube:2020nwx}, a more accurate description of the glacial ice in which the detector is located, improved reconstructions~\cite{lowen_reco}, an event selection with higher background rejection efficiency, new methods for estimating the impact of systematic uncertainties associated with the detector response and more detailed descriptions of theoretical uncertainties on neutrino fluxes and cross sections. In addition, the new sample includes 8 years of data collected from 2011-2019, which more than doubles the livetime used in previously published analyses~\cite{Aartsen_2018, IceCube:2019dqi}.

The new DeepCore event selection aims to serve future analyses within IceCube in the few GeV$-$1~TeV range. Its goal is to reduce the dominant background, atmospheric muons, to a point where they are observed at roughly the same rate as neutrinos. At this point, specific analyses can devise targeted strategies for background rejection and reconstructions that enhance their signal. In this paper we report the common DeepCore event selection, and also present the first results obtained with the sub-sample of highest-quality events that can be reconstructed with simple methods, using it to measure the atmospheric oscillation parameters $\sin^2\theta_{23}$ and $\Delta m^2_{32}$.

In Sec.~\ref{sec:detector} we begin with a description of the IceCube DeepCore detector, with a focus on calibration improvements. The new common DeepCore event selection is outlined in Sec.~\ref{sec:sample}, followed by the introduction of the ``Golden Event Sample" in Sec.~\ref{sec:golden_sample}. The details of how the data are analyzed are discussed in Sec.~\ref{sec:analysis}. An in-depth discussion of the systematic uncertainties that affect this measurement is given in Sec.~\ref{sec:systematics}, highlighting our treatment of detector-related effects as well as a new method to assess uncertainties on the neutrino flux, which we effectively fit as part of our results. Sections~\ref{sec:results} and~\ref{sec:conclusion} present the results obtained and explore their relevance and future improvements, respectively.

\section{The IceCube DeepCore Detector~\label{sec:detector}}
The IceCube Neutrino Observatory~\cite{Aartsen:2016nxy} is an ice Cherenkov telescope located at the geographic South Pole. It consists of 5,160 Digital Optical Modules (DOMs) deployed in 86 boreholes that were drilled with a high-pressure hot water drill~\cite{EHWD}. In each borehole DOMs are connected to a central cable, referred to as a \textit{string}, and cover depths between 1.45~km and 2.45~km. The instrumented volume of glacial ice contained within all 86 strings is approximately~1~km$^{3}$. 

The glacial ice serves as a detection medium for high-energy neutrino interactions, which produce a shower of relativistic particles. As they travel through the ice, the electrically charged particles in the shower can emit Cherenkov photons, which will propagate through the ice until they are either absorbed or detected by a DOM.

\subsection{Detector layout}

\begin{figure}[!bt] 
    \centering
    \includegraphics[width=\linewidth]{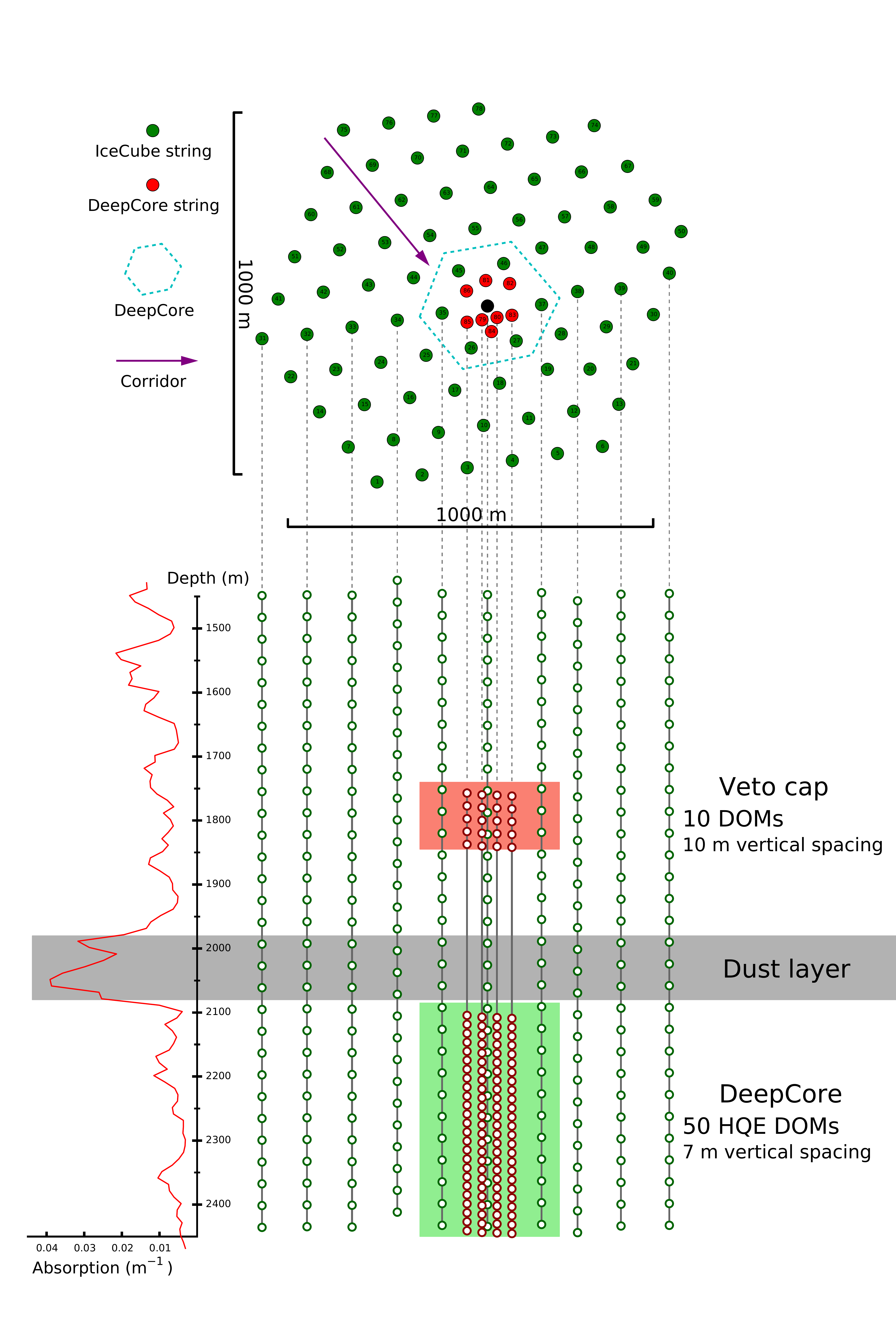}
    \caption{Top and side projections of IceCube. DeepCore DOMs are depicted with red circles while IceCube DOMs are shown as green circles. String 36 is depicted as a black circle in the top projection for reference. The green region represents the DeepCore analysis region. The absorption length for Cherenkov light vs. depth is shown at the bottom left of the figure. An example \emph{corridor} created by the hexagonal geometry is illustrated with a purple arrow (see Sec.~\ref{sec:level5})).
    \label{fig:detector}}
\end{figure}

The DOMs are arranged on a nearly-hexagonal array, as shown in the top of Fig.~\ref{fig:detector}, with a spacing of 125~m horizontally and 17~m vertically, throughout most of the detector. This configuration is optimized to detect astrophysical neutrinos with energies above $\sim$100~GeV. The bottom-center part of the array, referred to as DeepCore~\cite{IceCube:2011ucd}, has a reduced horizontal spacing of 42-72~m and a vertical spacing of 7~m. 

Each DOM consists of one 10-inch Hamamatsu R7081-02 photomultiplier tube (PMT)~\cite{Abbasi_2010} enclosed within a glass, pressure sphere. The photocathode occupies the bottom half of each sphere and is coupled to the glass with an optical gel that improves photon acceptance and provides mechanical support during transport and deployment. The top half of each sphere contains calibration devices and electronics for module control and communication~\cite{IceCube:2008qbc}. 

Most of the DeepCore DOMs are equipped with high quantum efficiency (HQE) PMTs and reside in the clearest ice between 2.1~km and 2.45~km below the surface. The increased DOM density, DOM sensitivity, and optical transparency of the ice allow the DeepCore sub-array to trigger on neutrino interactions down to a few GeV. This enables IceCube to perform measurements of atmospheric neutrino oscillations that occur in the 10-50~GeV region as described in Section~\ref{sec:analysis}.

The full 86-string detector configuration has been operating since 2011. Previous analyses of DeepCore data have included only the first three years of data. Here we analyse data collected from 2011 through 2019, more than doubling the livetime. Due to the increased statistical precision of the sample, care has been taken to improve the detector calibration and ensure proper assessment of relevant systematic uncertainties in the extraction of oscillation parameters from these data. 

\subsection{Data acquisition}
\label{sec:DAQ}
DOMs record data when the PMT signal voltage passes a threshold equivalent to 0.25 photoelectrons (PEs)~\cite{icecube:daq2014}. At that point they can contribute towards the various triggers in operation. These triggers are based on temporal and/or spatial coincidences that could arise from the Cherenkov photons emitted by a charged particle in the detector. Most triggers rely on the concept of \textit{Hard Local Coincidence (HLC)}, a flag given to a DOM when it registers light within 1~$\mu$s of its neighbor or next-to-nearest neighbor on the same string~\cite{icecube:daq2014}. 

If a trigger is formed, the PMT signals are processed by a group of two types of analog to digital converters (ADCs), operating at a sampling rate of either 300 megasamples per second (MSPS) for DOMs in HLC condition or 40~MSPS for the rest. The digitized signals are transferred to the surface computers in the IceCube laboratory for feature extraction. The signals are unfolded into \textit{pulses} by fitting the known response of the PMTs to single photons with a free amplitude at time intervals of 0.833~ns for the high-resolution waveforms and 6.25~ns for the low-resolution ones. Waveform bins with ADC values smaller than 2 are treated as noise. From this point on, the data are represented by a set of pulses over time, referred to as a \textit{pulse series}, unless otherwise noted.

The detector operates continuously, collecting data in runs that last up to 8-hours, during which the configuration does not change. Calibration procedures and diagnostics are performed for every run, and the results are monitored by an automated system and vetted by hand. This information is used while selecting data for specific studies to decide if runs should be included. After excluding bad runs between 2011-2019, this analysis is left with approximately 7.5 years of livetime.

\subsection{Detector calibration}\label{sec:detector_calibration}

Several improvements have been made in the calibration of individual DOM responses and characterization of ice properties in recent years. Here we review the changes that most significantly impact the processing and interpretation of low-energy data. 

\subsubsection{Single-photon DOM response calibration}

\begin{figure}[!b] 
    \centering
    \includegraphics[width=\linewidth]{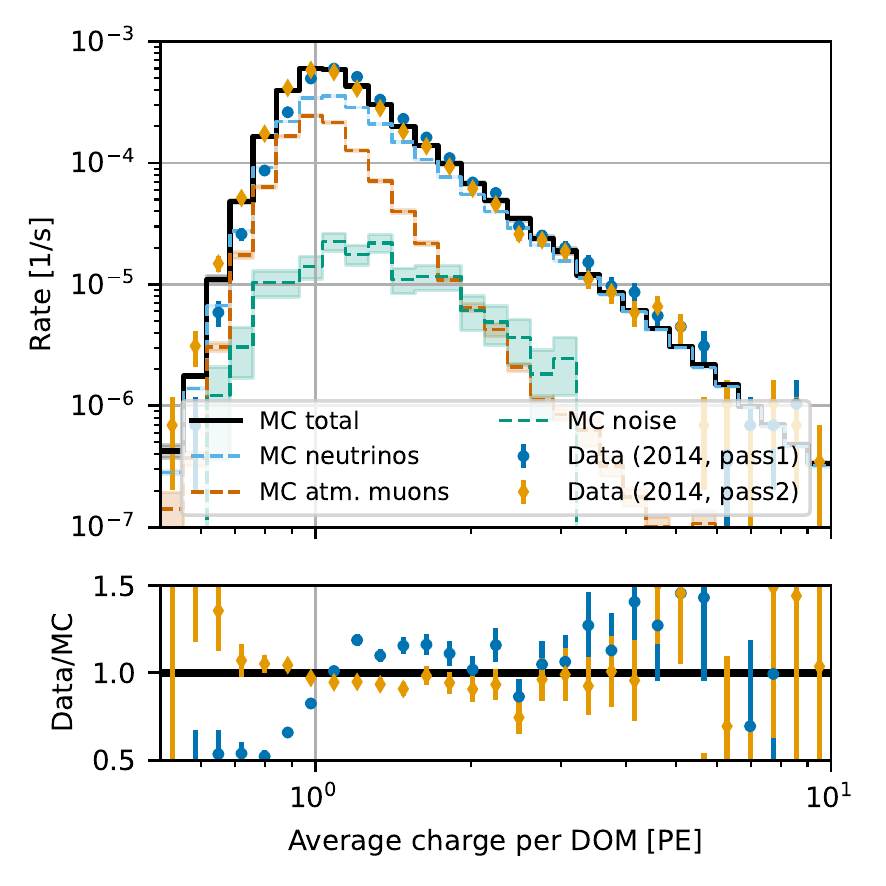}
    \caption{Distribution of the average charge observed over all triggered DOMs for events at Level~5 of the selection (see Sec.~\ref{sec:sample}), compared to expectation from simulation of neutrinos, muons and detector noise. Data are shown for the same time period in 2014, before (Pass 1) and after the SPE calibration described in the text (Pass 2).
    \label{fig:Qtot_NChan}}
\end{figure}

As the operating conditions of DOMs deployed in the deep ice are very stable, it is sufficient to only re-calibrate IceCube DOMs once per year. The application of new calibrations to the data often coincides with a new software release, and marks the beginning of a new \textit{season} of data taking.  

The per-DOM re-calibration primarily establishes the operating voltage for each PMT. It is chosen, such that the Gaussian mean of the single-photoelectron (SPE) charge distribution, which we use to define 1\,PE, corresponds to a gain of 10$^7$. This gain calibration is performed on the DOMs directly, using waveform integration for charge determination instead of pulse unfolding.  The SPE charge distributions observed in analysis-level data, after pulse unfolding, feature Gaussian means of $\sim$1.04\,PE on average instead of the simulated 1\,PE. 
At TeV energies the discrepancy manifests as an energy-scale uncertainty, while at GeV energies it introduces shape discrepancies in the mean observed charge per triggered DOM as shown in Fig. \ref{fig:Qtot_NChan}. To retroactively correct for this effect, all experimental data have been reprocessed such that the SPE Gaussian mean in data and simulation both peak at 1.0 as intended.

In addition to the position of the Gaussian mean, matching the overall shape of the SPE charge distribution between data and MC is important to achieve a good description of detector observables in low-energy data sets, where the majority of DOMs in an event record single photon hits. Prior to 2020, the SPE charge distributions employed in simulation were derived from lab measurements of bare PMTs that did not use the DOM hardware for data-aquisition. A recent study~\cite{IceCube:2020nwx} updated the SPE charge distributions used in simulation for the analysis presented here based on per-DOM, \textit{in-situ} data.

\subsubsection{In-situ calibration of optical detection efficiency}

The IceCube detector does not have a calibrated light source to measure the absolute optical detection efficiency of the DOMs. Instead, we use minimum ionizing muons from cosmic-ray showers as a controlled, constant source of light. We then simulate muon data sets modifying the response of the DOMs to Cherenkov light, and by comparing this to data we can calibrate the absolute optical detection efficiency. 

The events used are required to have passed the Minimum Bias Trigger~\cite{Aartsen:2016nxy} and have 8 HLC hits. By demanding that the muon reconstruction favors a muon that stops emitting Cherenkov light and decays at least 100~m above the bottom of the detector we preferentially select muons in the minimum ionizing regime, which occurs at $E_\mu \lesssim$ 700~GeV in water~\cite{Kowalski2004Search}. Only DOMs in the inner strings of IceCube are considered in the study, and they are selected only if the muon track they detect travels below them so that they are illuminated from the side or from below. Events with more than 20~DOMs outside the analysis region are rejected, as they correlate with a higher muon energy.

The study is done by comparing the average charge observed in data and simulation sets where the overall optical efficiency is varied, as a function of distance from the muon track reconstruction. The charge is computed using pulses with an arrival time $t<t_0+1~\mu$s, where $t_0$ is the expected arrival time of the light, assuming no scattering. 
The final calibration is obtained from the average charge ratio at distances between 60~m and 160~m. Multiple iterations of this study found optical efficiency values that varied by a few percent~\cite{domeff_jake, domeff_nick}, all of them within 10\% of both simulations and laboratory measurements done on bare PMTs. A 10\% uncertainty was therefore adopted as a conservative estimate of our knowledge of the optical efficiency of the DOMs for this study.

\subsubsection{Ice properties}

As photons travel through the ice they are subject to various scattering and absorption processes that determine the arrival time distribution and intensity observed by each DOM. These processes must be properly modelled for accurate simulation and reconstruction of data. This includes modeling both the bulk ice properties of undisturbed glacier and the properties of the refrozen borehole column of ice where the DOMs are located.  

As described in~\cite{AARTSEN201373}, the most important properties governing optical photon transport in the bulk ice are the average distance a photon travels until absorption, i.e. absorption length; the average distance between successive scatters, i.e. geometric scattering length; and the scattering angle distribution. Moreover, absorption and scattering lengths vary as a function of depth, reflecting the atmospheric conditions over the last $\sim$100~ky as the glacier slowly formed from compacted snow, as well as the underlying bedrock topology which introduces an undulation to layers of ice formed in the same year. Since 2013 IceCube has also established an optical anisotropic attenuation related to the glacial flow direction~\cite{Chirkin:2013lpu}, which has since been confirmed through independent measurements~\cite{tc-14-2537-2020}, and is believed to arise from the birefringent polycrystalline microstructure of the ice~\cite{tc-2022-174}. This manifests itself predominantly as an azimuthal anisotropy, where photons appear to propagate more efficiently along the flow direction than orthogonal to it, and is parameterized in the simulation for this analysis by a direction-dependent scaling of the effective scattering length~\cite{Chirkin:2013lpu}. 

Calibration of all bulk ice properties is performed \textit{in situ} using a pulsed light-emitting diode (LED) calibration system. Each DOM is equipped with 12 LEDs located in the upper part of the sphere that emit photons with a wavelength of approximately 405~nm\footnote{A subset of 16 DOMs contain LEDs with wavelengths of 505, 450, 370 and 340~nm that are used to calibrate the wavelength dependence of ice properties~\cite{Aartsen:2016nxy}.}. The LEDs are arranged in pairs spaced 60$^{\circ}$ apart in azimuth. In each pair, one LED points horizontally outward into the ice while the other points out at an elevation angle of 48$^{\circ}$. The LEDs can be configured to pulse with a duration between 6--70~ns and reach intensities of up to $1.2\cdot10^{10}$ photons per pulse~\cite{Aartsen:2016nxy}.

During dedicated calibration runs, LEDs from each DOM are pulsed and the arrival times of photons received in all other DOMs are recorded, creating a \textit{light curve} for each emitter-receiver pair of DOMs. The optical properties of the ice are then determined by iteratively simulating photon transport~\cite{Chirkin2013} for different realizations of the ice model parameters, and comparing the resulting light curves to calibration data through a log-likelihood (LLH) minimization described in~\cite{AARTSEN201373}. 

\begin{figure*}[!tbh]
    \centering
    \includegraphics[width=0.45\textwidth]{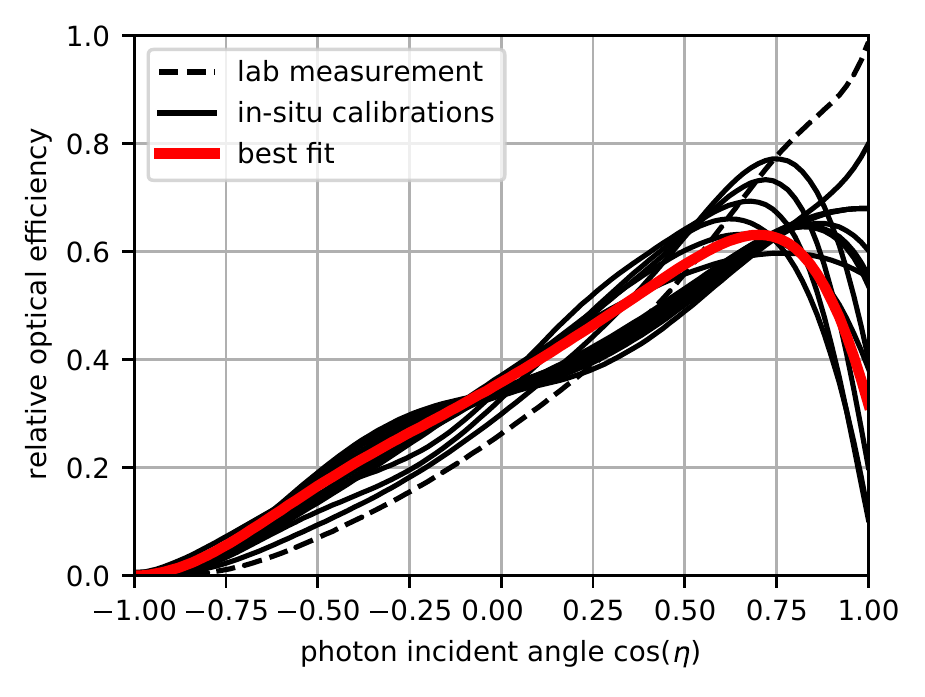}
    \includegraphics[width=0.45\textwidth]{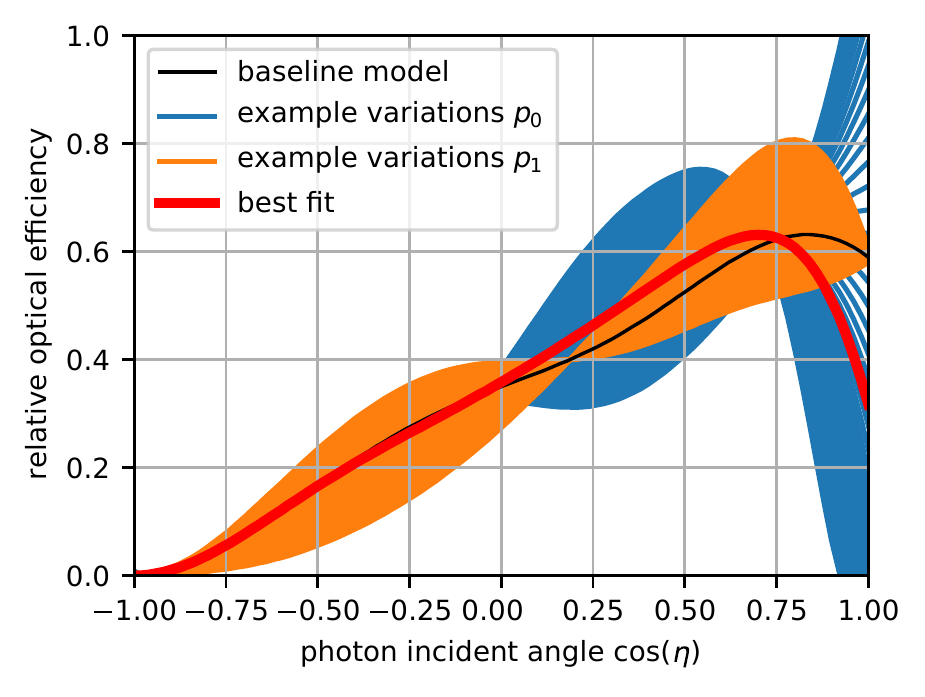}
    \caption{Relative optical efficiency of IceCube DOMs as a function of the photon incident angle $\eta$ ($\cos{\eta} = 1$ means facing the PMT). (Left) Efficiency as measured in the lab before deployment (dashed line) and various calibration curves (solid gray lines) are shown.
    (Right) Two-parameter model used in this analysis with its baseline curve and example variations of the two parameters, $p_{0}$ and $p_{1}$ within their allowed range (N.B. parameters in this figure are varied one at a time for visualization, while in the analysis both are varied simultaneously). See main text for details of the model. The wide allowed range at $\cos\eta =1$ is due to the lack of controlled sources that illuminate the DOM directly face-on.
    The best fit from the analysis discussed in Sec.~\ref{sec:analysis} is shown as the bold, red line in both panels.}
    \label{fig:holeice_calibration}
\end{figure*}

The bulk ice is finally described by a set of effective scattering and absorption coefficients that can change as a function of depth, and two parameters that govern the bedrock-induced undulation of the ice layers and the strength of the anisotropy in light attenuation. The absorption coefficient as a function of depth is shown in Fig.~\ref{fig:detector}. The uncertainties in the estimation of these parameters arise from the LED emission time and angular profile, scattering angle function, DOM optical efficiency, DOM angular acceptance, and from differences between fits using only-horizontal versus only-inclined LEDs. These sources of error are introduced as nuisance parameters in our calibration, and are individually varied within their uncertainties while the ice model is refit to LED data. In this way, we obtain a correction to the effective scattering and absorption coefficients averaged over all depths. 

Since computing ice models with perturbations on their optical properties is very time consuming, we can only produce a limited number of ice realizations, which so far suggest uncertainties of the order of 5\% in both coefficients. In this study, we use these corrections to estimate the total uncertainty on the scattering and absorption variations to be within 10\% from the best fit obtained (see Section~\ref{sec:systematics}).

The refrozen boreholes have optical properties different from the bulk ice. This alteration is believed to be due to the introduction of air bubbles and impurities during ice drilling and deployment, causing increased optical scattering.
Whether a nearby photon hits the PMT cathode in a DOM depends on these optical properties and the geometry of the module itself. For example, a down-going photon is not very likely to be detectable because it will not hit the PMT face unless it is strongly scattered. An up-going photon is more likely to hit the PMT face, but depending on the refrozen borehole ice properties may be scattered away. 

Several models~\cite{holeiceDIM} have been proposed to account for such effects by expressing how the likelihood of a photon being accepted is related to the direction in which it arrives, effectively modifying the DOM angular acceptance. The models were calibrated using a variety of methods such as the analysis of \textit{in-situ} camera images of IceCube boreholes during the re-freezing process, dim LED data for self-illumination by DOMs and bright LED data as used for the bulk ice calibration. A collection of all available calibration curves, together with the angular acceptance measured in the laboratory prior to deployment (i.e. without refrozen ice), are shown in Fig.~\ref{fig:holeice_calibration}. These acceptance curves are normalized to the same total area to decorrelate these effects from the overall optical module efficiency. The model variation is largest for normal incidence angles ($\cos\eta = 1$) where the calibration devices have reduced sensitivity. 

A parametric model was built to approximate these calibration curves shown in Fig.~\ref{fig:holeice_calibration} via a principal component analysis (PCA) \cite{doi:10.1080/14786440109462720}. The first two principal components, $p_0$ and $p_1$, are able to describe all curves with an accuracy better than 5\%. The acceptance variation resulting from the changes in the values of these two parameters is visualized in the right-hand side of Fig.~\ref{fig:holeice_calibration}. The range of values $p_0$ and $p_1$ that cover most established models are $-0.6<p_0<+0.6$ and $-0.1<p_1<+0.1$, while analyses typically allow fits to explore significantly larger regions to be conservative due to the lack of calibration data. The analysis presented herein allows these parameters to span $-2<p_0<+1$ and $-0.2<p_1<+0.2$. 

\subsection{Simulation}

The IceCube collaboration relies heavily on Monte Carlo simulation (MC) to interpret its data. These tools can be divided into those involved in the simulation of interactions of the primary cosmic-ray air-showers and the resulting neutrinos, the propagation of charged particles and photons, and the detector's response. We use well established event generators for simulating cosmic ray showers and neutrino interactions, while the propagation and detection are IceCube specific. Further details of the simulation software chain can be found in~\cite{IceCube:2019dqi}. Here we review only the main aspects while highlighting significant changes and improvements to the software since our last published neutrino oscillation measurement with 3~years of data~\cite{IceCube:2019dqi}.

\subsubsection{Event generation}
Neutrino interactions of all flavors are simulated using {\sc GENIE} version 2.12.8~\cite{Andreopoulos_2010}. Neutrino interactions are generated with true energies that follow a power-law spectrum, with an isotropic distribution around the detector. Neutrinos are forced to interact in a volume that surrounds and includes DeepCore, which is large enough to include events where the interaction takes place outside the instrumented volume but a particle could still leave pulses in the DOMs. The events are afterwards weighted to match an atmospheric neutrino flux. For this study our baseline is the model proposed by Honda \textit{et al.}~\cite{PhysRevD.92.023004}, computed for the South Pole geomagnetic and atmospheric conditions. The tables are interpolated to assign values to arbitrary directions and energies, keeping the integral per bin of the original flux. 

Atmospheric muons are produced using both a full simulation of cosmic-ray interactions in the atmosphere and the subsequent particle shower using {\sc CORSIKA}~\cite{Heck:1998vt} and parameterized tables that use output from the same code but only describe the muons that reach the vicinity of the detector, known as \textit{MuonGun} (method based on~\cite{BECHERINI20061}). The baseline simulation used in this analysis follows the composition and flux proposed by Gaisser \textit{et al.}~\cite{Gaisser:2011klf} and the Sibyll2.1 interaction model~\cite{Ahn_2009}. The parameterized method only fully simulates muons that would reach a cylinder of 180~m radius and 400~m in length, centered in DeepCore. The procedure was further optimized by biasing the injection procedure as function of energy, zenith angle and whether the muons enter the cylinder from the top or the sides, based on an initial sample of simulated muons that survived the early stages of the event selection described later. This resulted in a factor two gain in the efficiency of muon simulation, as measured by the increase in events surviving to later stages of the event selection per unit of computation time~\cite{muonopt_deholton}. Despite these efforts, it is not possible to simulate the number of atmospheric muons expected in the full data set, and we therefore implement various strategies in the event selection to address this challenge.

A dedicated simulated data set consisting exclusively of pure-noise triggers, produced by thermal electron emission of the photocathode and radioactive decays in the glass, was also generated. This additional component was found to be necessary to explain event triggers with a small number of DOMs. These events predominantly arise from random coincidences of radioactive decays in the modules' glass. More information about noise events and their impact in this study is given in Sec.~\ref{sec:sample}.

\subsubsection{Particle and photon propagation}

The treatment of charged particles in the IceCube simulation is divided in two branches. Muons are dealt with individually by {\sc PROPOSAL}~\cite{Koehne:2013gpa}, which implements a simplified model of the energy losses of charged particles. In {\sc PROPOSAL} the travel direction is kept fixed but the effects of multiple Coulomb scattering are included in the calculation of energy losses and the Cherenkov emission profile. {\sc GEANT4}~\cite{AGOSTINELLI2003250} is used to simulate tau leptons, hadrons produced in all interactions, and electrons and photons below 100 MeV. For electromagnetic showers above 100 MeV, and hadrons above 30 GeV, shower-to-shower variations are small enough to use parametrizations based on {\sc GEANT4} simulations~\cite{RADEL201253}.

Once the Cherenkov emission of muons and cascades has been calculated the photons are propagated through the ice. In order to determine which of them arrive at a DOM, and when they do so, each photon is propagated individually using the \textit{clsim} software package~\cite{clsim}. The propagation takes into account the depth-dependent optical properties of the ice as well as its anisotropy to determine the photon's path and stops once a photon is either absorbed or when it arrives at the surface of a DOM.

\subsubsection{DOM response simulation}

The expected response of the detector is simulated once photons have arrived to the surface of a DOM. Since the detector has been in a stable configuration as of 2012, we use a single snapshot of module calibration constants and noise levels to produce simulation that is representative of the eight years of data in this analysis. 

The detection of light in the DOMs is simulated using a wavelength dependent quantum efficiency, obtained in laboratory  measurements, of about 25\% for regular PMTs \cite{icecube_pmt} and 35\% for HQE PMTs \cite{deepcore} at 400 nm. The 1~PE waveform for each PMT is used to simulate the digitized signal, including pre-, late and after pulses, assuming the DOM behaves linearly \cite{icecube_pmt}.

The intrinsic noise from the PMT and the radioactivity of the glass is included in this step. The injected rate of background photoelectrons, typically 500\,Hz for IceCube DOMs and 600\,Hz for DeepCore DOMs, is obtained individually for each sensor after it has been deployed and left to stabilize~\cite{deepcore}. The noise level for all modules has remained within 2\% of their average value throughout the years used in this study.

Once the full waveform response is built for a DOM, a discriminator threshold of 0.25~PEs is applied to decide whether the DOM has recorded data. After this, the simulation is passed to the same set of algorithms that operate on the detector to decide if an event has passed any of the triggers operating at the South Pole.

\section{The DeepCore Common Data Sample~\label{sec:sample}}
Most of the events IceCube detects are atmospheric muons, which are produced alongside neutrinos in cosmic-ray air-showers. A secondary source of events, relevant for low-energy analyses, are accidental coincidences between DOMs, which are created by radioactive decays in the glass or thermal emission of photoelectrons in the PMT. These backgrounds trigger the detector at a rate $10^6$ times higher than neutrinos. An event selection has been developed to reduce these backgrounds and retain a large sample of well-reconstructed neutrinos. The resulting data sample is the common starting point for several analyses, which employ different reconstruction methods, particle identification techniques and more in order to boost sensitivity to various standard and non-standard model physics models.  

In this sample we only include runs that are at least 2-hours long, where all 86 strings of the detector are active and that have at least 5,035 ($\gtrsim97$\%) DOMs collecting data. Runs with missing information are rejected from the sample. After these data quality considerations, the event selection broadly follows the same procedure as previous IceCube oscillation analyses~\cite{Aartsen_2015, Aartsen_2018, IceCube:2019dqi}. The earliest stages of the selection aim to reduce atmospheric muons and events consisting of pure detector noise using low-level detector observables and simple reconstructions. Later stages of the selection include more sophisticated event reconstructions and algorithms to further enhance the purity of neutrinos in the final sample. 

Most of the algorithms and variables used for the event selection are defined in the appendices in~\cite{IceCube:2019dqi} or in previous publications. To avoid repetition, we only provide a brief description here, followed by the relevant reference and the section or appendix where they can be found. 

One key difference of the selection outlined here with respect to previous studies is that we avoid dependence on the charge observed by the PMTs in the earliest stages of the selection. During the recent calibration campaign~\cite{IceCube:2020nwx} it was noted that small variations in the single photoelectron response could lead to large changes in passing rates in older event selections. Since this selection was being developed in parallel with that calibration effort, we decided to decouple the event selection from these effects so that the common sample would be more robust. The observed charge in selection variables was substituted by the number of DOMs with a signal above some threshold, typically 0.1~PE. Note that, since the pulse unfolding has a free amplitude, PE values below the discriminator threshold of 0.25~PE are possible. Because most GeV-scale neutrino interactions in DeepCore only produce SPEs, this change \emph{does not} result in a significant loss of useful information. Later stages of the selection and reconstruction still make use of the improved charge calibration, as these were developed at a later time.

\subsection{On-line trigger and filter (Levels 1 and 2)}
\label{sec:trigger_filter}

The Level 1 trigger used by the DeepCore stream requires 3~HLC hits within a 2.5~$\mu$s time window. Triggered events are filtered at the IceCube Laboratory at the South Pole (Level 2), and those passing the filter are transmitted over satellite to our data repository. DeepCore has a dedicated filter that seeks to discard potential muon events from veto regions based on the speed that a hypothetical particle would need to connect clusters of light inside and outside DeepCore. The filter was updated from~\cite{IceCube:2019dqi} to first run a pulse-cleaning algorithm to reject noise pulses and operate on the result. At Level 2 we have an event rate of 15~Hz, and it is mainly dominated by coincident noise, due to the loose trigger conditions, and atmospheric muons.

The pulse series (defined in Sec.~\ref{sec:DAQ})  of events that pass the trigger and filter conditions are analyzed by an algorithm that looks for DOMs with signals that could be causally connected, starting from the DOMs that satisfied the HLC conditions, to reject hits likely caused by noise. This \textit{cleaned pulse series} is used in event selection algorithms and in the reconstruction of the event properties~\cite{lowen_reco}.

\subsection{Basic background reduction (Level 3)} \label{sec:deepcore_trigger}

The simulation of muons and pure-noise events is challenging and known to have its limitations within our software, coming from imperfect simulation of e.g. muon bundles or muons from multiple atmospheric showers in the same readout window. The goal of the next level of selection, Level 3 (L3), is therefore to use simple variables that are easy to compute to remove regions of the parameter space that are expected to be dominated by muons and pure-noise events.

\begin{figure}[!t] 
    \centering
    \includegraphics[width=\linewidth]{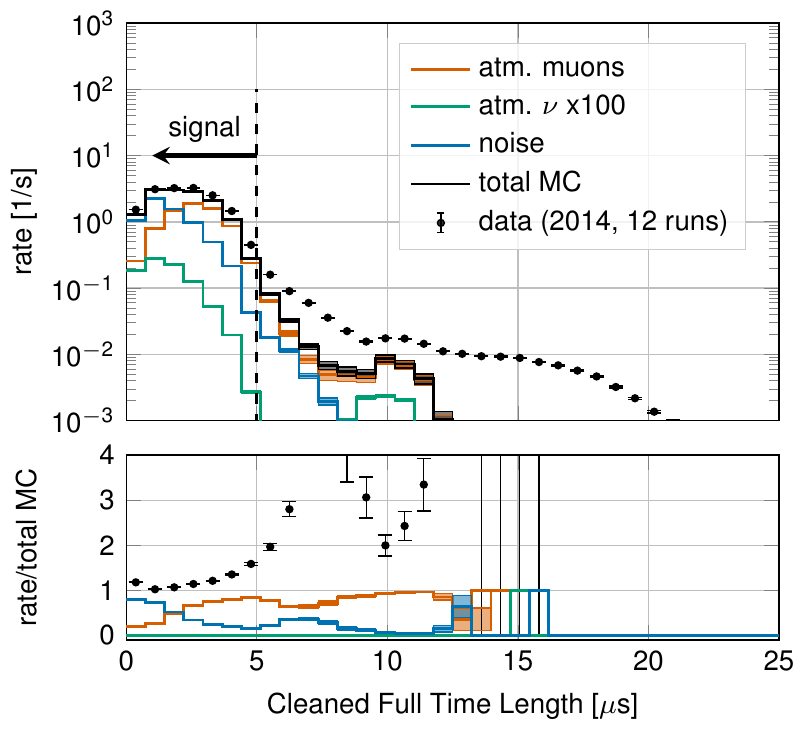}

    \caption{Distribution of the time difference between the last and first pulse observed in an event, after the pulse series has been cleaned. This variable is used at L3 to reject events that contain more than one muon in a single readout window, coming from multiple air showers. The simulation does not contain these events, and therefore only describes the data up to values of 4~$\mu$s. 
    \label{fig:l3_example}}
    
\end{figure}

The L3 algorithm was updated from~\cite{IceCube:2019dqi} to include four new variables. Two of them use the time difference between the first and last pulse observed in the raw and cleaned pulse series. A third one counts the number of hits found in the veto region by an algorithm that searches for hit clusters based on whether their arrival time can be causally connected to a cluster of HLC pulses. The last variable slides a 300~ns time window over a cleaned pulse series, saving the maximum number of DOMs found. Fig.~\ref{fig:l3_example} shows an example of the behavior of an L3 variable, where the neutrino region has reasonable agreement and the muon-dominated region cannot be modeled correctly due to coincident events missing in the simulation. After L3 the event rate is about 0.5 events per second, still dominated by muons and noise triggers, with a neutrino:noise:muon ratio of 1:7:100.

\subsection{Multivariate background rejection (Level 4)}

Two classifiers are trained and applied at Level 4 (L4) to target noise events and atmospheric muons individually. We use the LightGBM package~\cite{ke2017lightgbm} which features gradient boosting to train ensembles of Boosted Decision Trees (BDTs). Input data were split between \textit{train} and \textit{test} samples to verify the output and avoid overtraining. Each classifier was trained with a balanced set, using the same weighted number of signal and background events. Over 40 variables were tested as input, keeping those that showed good agreement between data and simulation as well as high classification power.

\begin{figure}[!t] 
    \centering
    \includegraphics[width=.94\linewidth]{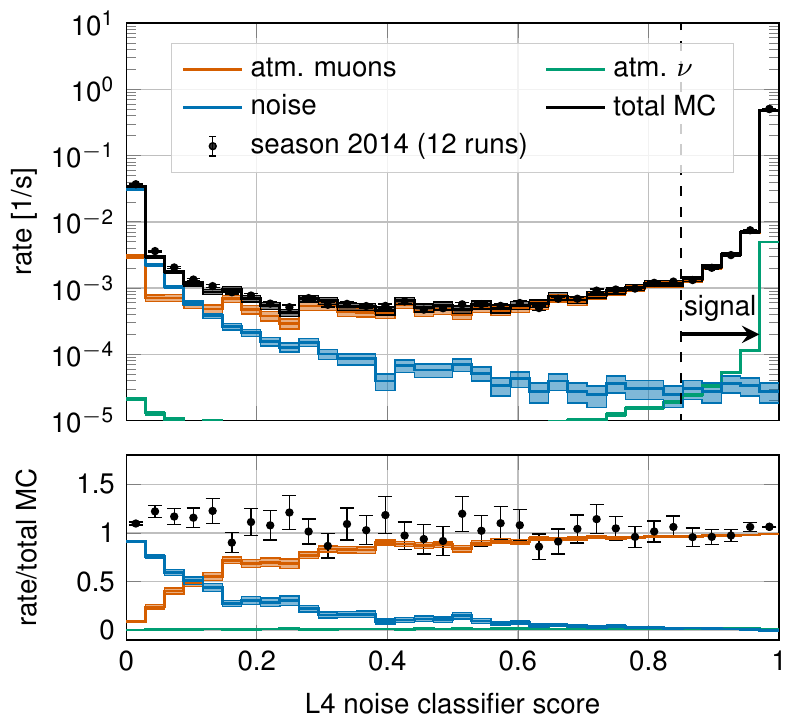}

    \caption{Distribution of the Level 4 noise classifier score, where larger values close to 1 indicate neutrino-like events and lower values reflect noise-like events.
    \label{fig:noise_bdt}}
\end{figure}

The L4 noise classifier uses five input variables with similar feature importance. They are the number of DOMs with pulses after noise cleaning, the maximum number of DOMs observed in a sliding 200~ns time window, particle speed fitted by the \textit{LineFit} algorithm~\cite{Aartsen:2013bfa}, a measure of the geometrical spread of the pulses about the first HLC position (C.4 in ~\cite{IceCube:2019dqi}), and the ratio of the event duration between the cleaned and raw pulse series. The expected rate of events as a function of the L4 noise classifier score is given in Fig.~\ref{fig:noise_bdt}, where larger scores indicate a higher probability that the event is produced by a neutrino interaction. We cut events with probability $P_{\nu}<0.85$, reducing the pure noise events by a factor 100 while keeping about 96\% of the neutrino sample.

After removing the noise, the data are expected to be 99\% atmospheric muons, so the L4 muon classifier was trained on real detector data, while simultaneously verifying its performance in simulation. This alleviates the concern of having simulation that lacks some type of muon background, such as the conicident events in Fig.~\ref{fig:l3_example}, and since at this stage neutrinos make up less than 1\% of the data, the contamination in the training sample is minimal. We use data from 2014, selecting runs throughout the year to sample the expected seasonal variations in muon rates due to changes in temperature and atmospheric conditions. The signal training sample came from our nominal GENIE simulation, including events from all flavors between 1~GeV and 10~TeV, weighted to our nominal atmospheric flux, including oscillations.

A total of 10 input variables were chosen for the L4 muon classifier. Four of these were used in L3, namely the number of pulses in a noise-cleaned series, in the veto region, above 200~m and those found by the veto algorithm. Another five had been used in previous analyses; they are the Veto Identified Causal Hits (A.3 in~\cite{IceCube:2019dqi}), radial position of the first HLC hit (A.1 in~\cite{IceCube:2019dqi}) and three properties of the positions of cleaned pulses (A.4 in~\cite{IceCube:2019dqi}): the center of gravity in $z$ (CoG-$z$), spread of pulses in $z$ ($\sigma_z$) and the total vertical span of the pulses ($z$-travel). Additionally, the time to reach 75\% of the total event charge in the cleaned pulse series was also included.

\begin{figure}[!t] 
    \centering
    \includegraphics[width=0.94\linewidth]{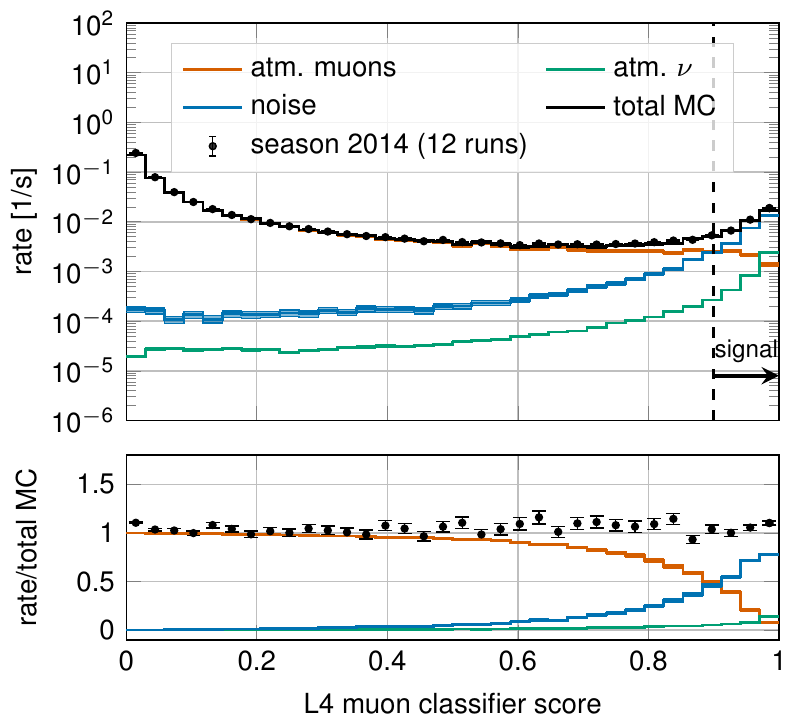}

    \caption{Distribution of the Level 4 muon classifier score for events passing the cut on the L4 noise classifier score. Larger values close to 1 indicate neutrino-like events and lower values reflect muon-like events. The classifier was trained on detector data, which is about 99\% atmospheric muons at this level, to avoid issues coming from limited statistics on the simulation.
    \label{fig:muon_bdt}}
\end{figure}

The event rate as a function of L4 muon classifier score for data and the different components of the simulation are shown in Fig.~\ref{fig:muon_bdt}. Larger scores correspond to a higher probability that event originates from a neutrino interaction. Again we note that the simulated atmospheric $\mu$ component was \textit{not} used in the training, and is only used to check agreement with the $\mu$-dominated data sample. This method achieves excellent agreement between data and simulation for most score values, with only a small region with disagreement towards zero. We keep events with $P_{\nu}>0.90$, removing 94\% of muons and retaining 87\% of all neutrinos. After L4 the event rate is about 1$\times10^{-3}$ events per second, noise events have been significantly suppressed, and muon and neutrinos have a closer ratio neutrino:muon of about 1:2.

\subsection{Advanced muon veto techniques (Level 5)}
\label{sec:level5}

The goal of Level~5 (L5) is to further suppress the muon background and reach a sample dominated by neutrinos. The muons that survived the previous cuts and remain at L5 did not leave a clear signature of entering the detector from the veto region. However, thanks to the reduced event rate, more computationally demanding algorithms can be used at this point. These muons are identified by introducing further cuts on the interaction position and by looking at directions with lower instrumentation density.

Cuts on the interaction position require the events to be within 150~m of the string at the center of the array (string~36), considering three different estimates of the vertex: the string with most collected charge, the position of the First HLC, and the vertex estimator used in L3. On the vertical direction, the first HLC and the L3 vertex estimator are required to be within $z=[-490,-220]$m.

\begin{figure}
    \centering
    \includegraphics[width=0.95\linewidth]{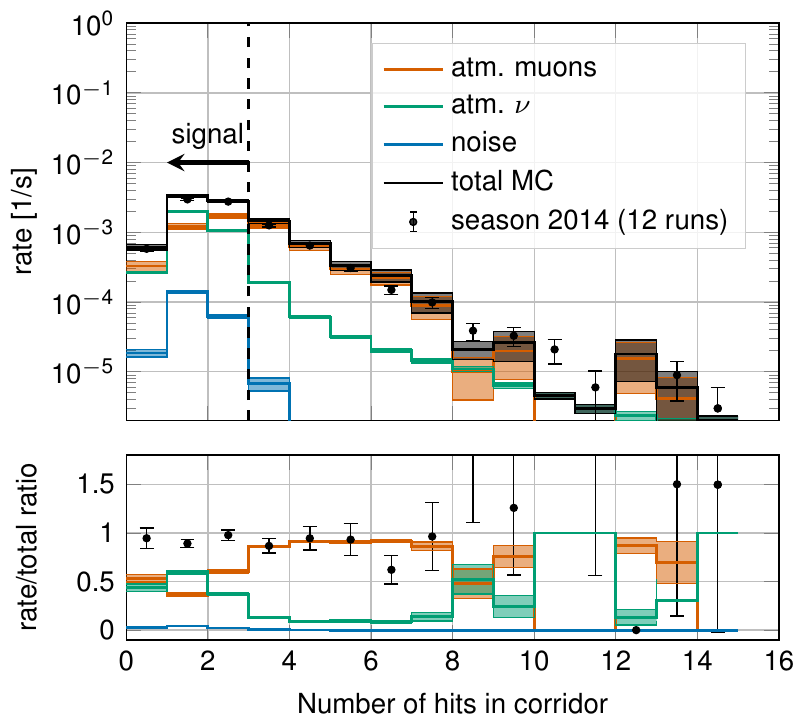}

    \vspace{0.5cm}
    
    \includegraphics[width=0.95\linewidth]{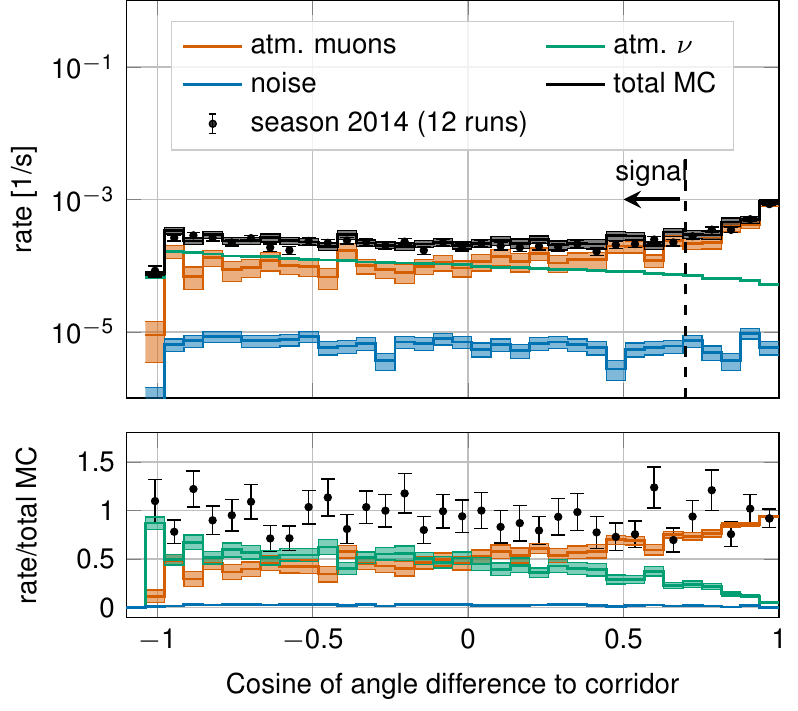}
  
  \caption{Number of DOMs found in L5 corridors (top); cosine of an angle between the brightest corridor and SPEFit11 direction (bottom).}
  \label{fig:corridor}
\end{figure}

The hexagonal configuration of IceCube in the horizontal plane gives rise to \textit{corridors} that lead to DeepCore without passing close to IceCube DOMs, and muons following these paths can evade simple veto techniques (Figure~\ref{fig:detector} shows an example of a corridor). A computationally expensive algorithm targets these positions by using a full pulse series to calculate the center of gravity of an event and select the closest DeepCore string to it. Using that string, it looks for any triggered DOMs located along any of the previously identified corridors within a cylinder of 250~m and a light arrival time from $-100$~ns to 1000~ns with respect to a hypothetical muon track. The track is varied in zenith in steps of 0.02~radians. The track hypothesis that results in the largest number of DOMs is kept, and cuts are applied on the highest number of DOMs found. 

As part of this search we also compute a track reconstruction using the \textit{SPEFit} algorithm~\cite{Ahrens_2004} with 11 iterations. This reconstructed track is compared to the corridor muon hypothesis by calculating their dot product, regardless of the number of DOMs found in the corridor. The rejection power of both variables is shown in Fig.~\ref{fig:corridor}.

\begin{figure}[!t]
    \centering
    \includegraphics[width=\linewidth]{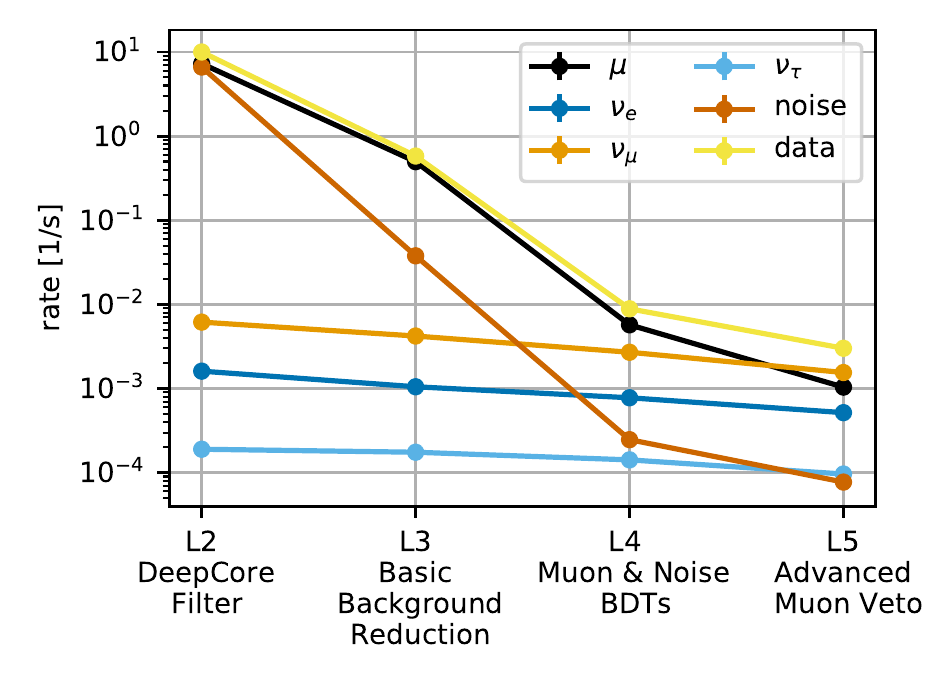}
    \caption{Summary of the rates obtained after each level of selection. Neutrinos are weighted to an atmospheric spectrum with oscillations included.}
    \label{fig:l5_summary}
\end{figure}

After L5 we have a neutrino event rate of 2~mHz, and a rate of muons of close to 1~mHz, with a negligible component of pure noise. The selection efficiency has little dependency on neutrino flavor, as shown in Fig.~\ref{fig:l5_summary}, so analyses looking for specific signatures can start from this point and apply selection criteria tailored for a given study.

\section{The DeepCore Golden Event Sample~\label{sec:golden_sample}}

As a first analysis using this new data sample, we have performed a measurement of the atmospheric mixing parameters $\theta_{23}$ and $\Delta m^{2}_{32}$. In order to validate the new calibrations and ensure good agreement between data and simulation can be achieved with the updated event selection and analysis tools, this analysis is performed on a subset of \emph{golden events} that contain mostly direct, i.e. unscattered, photons and has been optimised for high $\nu_{\mu}$ CC purity. While the fraction of golden events in the sample is small, using them allows us to apply relatively simple, fast reconstruction methods that have already been used in previous DeepCore analyses~\cite{Aartsen_2015,deepcore_sterile_2017}. The following sections describe the event reconstruction and remaining selection criteria applied to the data to obtain this golden event sample. Because this analysis uses data collected over an 8 year period, we also provide an assessment of the data stability over time. 

\subsection{Reconstruction} \label{sec:reconstruction}

The goal of the event reconstruction is to translate the observed charge as a function of time at each DOM location into an estimate of the interacting neutrino flavor, along with its energy, $E$, and its incoming direction, which correlates with the distance travelled between production and detection, $L$. The atmospheric neutrino oscillation pattern can then be observed by comparing the change in relative fluxes of neutrino species as a function of $L/E$. In this analysis, the reconstruction of $L$, $E$ and the neutrino flavor are carried out successively by separate algorithms.

\subsubsection{Scattered-photon cleaning}

We first perform a cleaning algorithm to remove DOM hits created by photons that have undergone significant scattering. This helps to reduce the impact of uncertainties related to the complex photon scattering processes that are challenging not only to calibrate (see Sec.~\ref{sec:detector_calibration}) but also to parameterise in reconstructions. By removing these scattered-photon hits, the expected arrival time for Cherenkov photons can instead be calculated geometrically given a certain event hypothesis. This is possible thanks to the dense module configuration in DeepCore, where the inter-module distance is comparable to the effective scattering length. 

As described in~\cite{lowen_reco}, hits arising from scattered photons are identified by calculating the time difference between the first 
pulse observed in each DOM along the same string. Assuming the pulses are created by photons in the same Cherenkov cone, the largest possible delay between pulses is given by $\Delta\tau_{ij} = |\Delta z_{ij}|/c_{\mathrm{ice}} +\Delta t_{delay}$, where $|\Delta z_{ij}|$ is the distance between DOMs \emph{i} and \emph{j}, $c_{\mathrm{ice}}$ is the speed of light in ice and $\Delta t_{delay}$ is a tunable parameter that governs how strict the cleaning is with respect to scattering effects. For this analysis $\Delta t_{delay}$ is set to 20~ns. The scattered-photon cleaning algorithm starts from the DOM that observes the largest total charge, as a proxy for the point of closest approach to the event vertex. Any DOM on the string with an earlier pulse than this highest-charge DOM is removed. The algorithm then moves along the string and removes DOMs where the time difference to the next DOM is larger than $\Delta\tau_{ij}$. When applying this cleaning algorithm to simulated $\nu_{\mu}$ CC interactions, 81\% of pulses are accurately classified as unscattered. In order to proceed with the event reconstruction, at least 5 hit DOMs must remain after this cleaning is complete. Only approximately one third of all events in the L5 sample fulfill this criteria. 

\subsubsection{Directional reconstruction}

For atmospheric neutrinos, the propagation distance \textit{L} can be determined via the cosine of the reconstructed zenith direction of the interacting particle. In the coordinate system of IceCube, a value of cos$(\theta_{zenith}) = 1$ corresponds to vertically \textit{down-going} events that have travelled L~$\approx$~20~km, and cos($\theta_{zenith}) = -1$ corresponds to vertically \textit{up-going} events that have travelled diametrically through the Earth with L~$\approx 1.3\times10^{4}$~km. This analysis uses the SANTA algorithm~\cite{Garza2014Measurement,lowen_reco} for directional reconstruction. This algorithm was originally developed for event reconstruction in the ANTARES water-Cherenkov neutrino telescope~\cite{Aguilar:2011zz}, and has been modified to improve performance for interactions in ice where scattering effects are stronger~\cite{lowen_reco}. 

Using only DOM hits that survive the scattered-photon cleaning, a $\chi^{2}$ minimization between the observed and expected hit times is performed for each event using both a \textit{cascade} and \textit{track} hypothesis for the light emission pattern. Cascades, produced by electromagnetic and hadronic showers, appear roughly point-like as viewed by the sparsely instrumented DeepCore array. This topology is indicative of $\nu_{e}$ and most $\nu_{\tau}$ CC interactions, as well as NC interactions of all flavors. In SANTA, cascades are modeled as isotropic bright-point functions with the time \textit{t} and position (\textit{x, y, z}) of the interaction vertex as free parameters to be fit. This fit, therefore, does not provide any directional information. Only the goodness-of-fit, $(\chi^2/\mathrm{d.o.f.})_{\mathrm{cascade}}$, is later used to help identify the neutrino flavor. The d.o.f. are calculated as the number of hit DOMs minus the four free parameters in the fit.

Tracks, or elongated light emission patterns, are produced by long-lived muons that mainly come from $\nu_{\mu}$ CC interactions or cosmic-ray air-showers, with a subdominant component from $\nu_{\tau}$ CC interactions\footnote{Branching ratio of $\tau\rightarrow\mu$ is approximately $17\%$~\cite{PhysRevD.98.030001}.}. SANTA considers an infinite line hypothesis for tracks, where all photons are emitted at the characteristic Cherenkov angle of $\sim$40.2$^{\circ}$ (with refractive index $n_{ice}\approx1.31$) relative to the direction of their parent particle. Under this hypothesis, the intersection of unscattered photons with DOMs arranged vertically on strings creates a hyperbolic pattern in hit times as a function of DOM depth. The track direction can be deduced from the fitted parameters of this hyperbola as described in~\cite{Aguilar:2011zz} and~\cite{lowen_reco}.
If unscattered photon hits are only found in DOMs along a single string, the azimuth cannot be uniquely determined. In this case the zenith angle is still reconstructed with SANTA, while the azimuth angle is determined by a simple LineFit algorithm~\cite{Aartsen:2013bfa} to DOM hits created by both scattered and unscattered photons. In this single-string fit configuration, there are 5 free parameters in the track fit, which include the interaction vertex (\textit{x, y, z, t}) as in the case of cascades plus the zenith direction. For multi-string fits the azimuth is also fit, increasing the number of parameters to 6.

\begin{figure}[t!]
	\centering
    \includegraphics[width=0.99\linewidth]{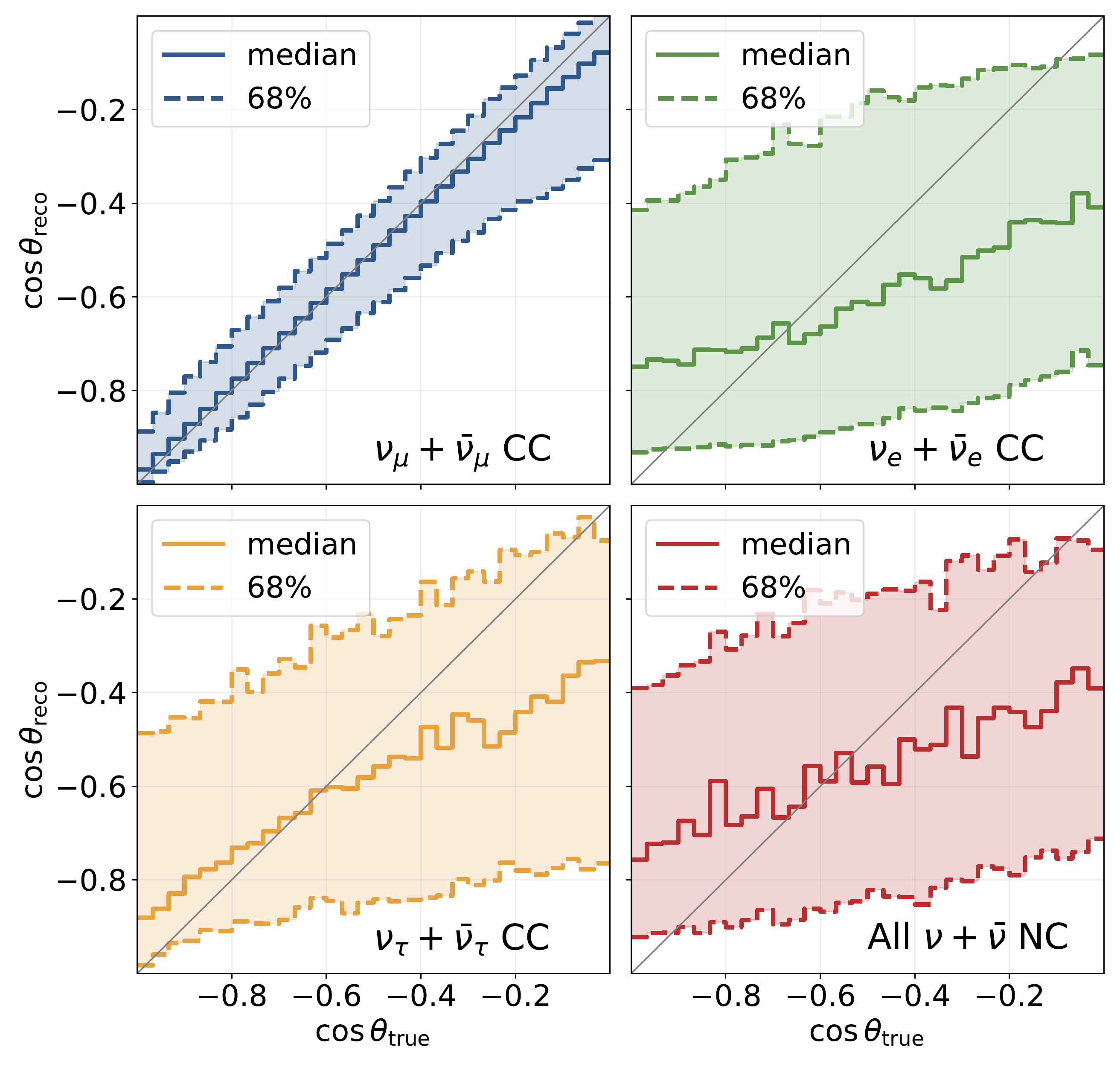}
    \caption{Cosine zenith resolutions for different classes of neutrino events at final level. The solid lines show the median resolutions and dashed lines indicate the central 68\% containment region.}
    \label{fig:resolutions_coszen}
\end{figure}

Reconstructed zenith angle resolutions under the track hypothesis are shown in Fig.~\ref{fig:resolutions_coszen} for each interaction type. The accuracy and precision are better than than the first study that used this approach~\cite{Aartsen_2015} and comparable to recent results obtained with more sophisticated algorithms~\cite{IceCube:2019dqi}. The reconstruction performs best overall for $\nu_{\mu}$ CC events as expected, since the hypothesis more accurately describes the underlying topology. Similarly, $\nu_{\tau}$ CC events perform better than $\nu_{e}$ CC and NC since some events contain a visible muon track coming from a $\tau$-decay. Resolutions improve for $\nu_{\mu,\tau}$ CC events as the muon tracks get longer, providing a longer lever arm for the reconstruction. The zenith error for $\nu_{e}$ CC and NC events also improves slightly with energy, from 25$^{\circ}$ down to 20$^{\circ}$, as the cascades become more elongated and preserve more directionality. However, generally the resolution is much worse since the track hypothesis is not a good description of the event signature.

\subsubsection{Energy reconstruction}

Once the event direction is reconstructed, it is used to determine the total energy of the neutrino in an interaction. The energy reconstruction algorithm, called LEERA~\cite{Aartsen_2015,AndriiThesis}, is also optimized for $\nu_\mu$ CC interactions. Unlike the directional reconstruction, which assumes an infinite track, the energy reconstruction uses a more realistic hypothesis that includes a hadronic shower at the interaction vertex and a finite track length for all events. In addition, all DOM hits (after noise cleaning) are considered, even those produced by scattered photons. The algorithm first reconstructs the endpoint along the track direction by comparing the Poisson log-likelihood (LLH) to not observe DOM hits given an infinite track hypothesis, versus the LLH to not observe DOM hits given some finite track length. The ratio of these so-called ``no-hit" probabilities is minimized to determine the track end-point~\cite{FiniteReco}. 

In a second step, the vertex position and the energy deposited in the hadronic shower are estimated. In this case, both the no-hit and hit Poisson probabilities are summed over all DOMs within a 200~m cylinder along the track for a given vertex hypothesis. The negative logarithm of these summed probabilities is minimized to obtain the vertex position and shower energy. 

In both steps, the expectation for the number of photon hits is obtained from pre-computed look-up tables, which are parameterized by spline functions, for short track segments and particle showers~\cite{WHITEHORN20132214}. By considering only the probability of being hit vs. not hit for a given DOM, the likelihood shows robust behaviour against PMT systematic uncertainties, such as those described in~\cite{IceCube:2020nwx}. 

\begin{figure}[t!]
    \centering
    \includegraphics[width=0.99\linewidth]{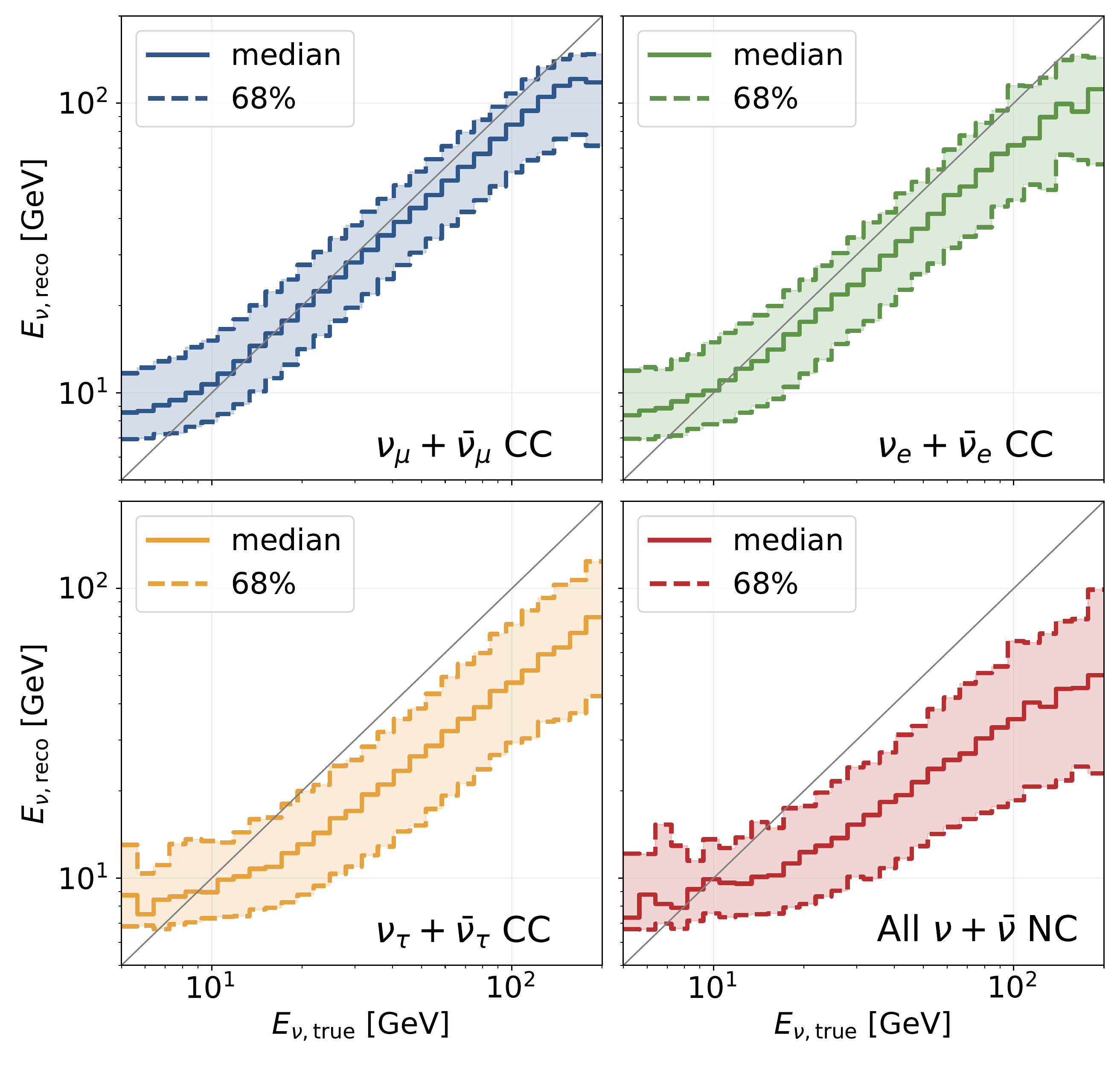}
    \caption{Energy resolutions for different classes of neutrino events at final level. The solid lines show the median resolutions and dashed lines indicate the central 68\% containment region. All events are reconstructed using a track-plus-cascade hypothesis.}
    \label{fig:resolutions_energy}
\end{figure}

\begin{table*}[t!]
\begin{tabular}{p{0.27\linewidth} p{0.18\linewidth} p{0.24\linewidth} p{0.2\linewidth}}
\toprule
\hline
Interaction                                               & Signature         & Energy bias and res. & Zenith bias and res. ($^\circ$) \\ \midrule
$\nu_{e}+N\rightarrow e+had.$                             & Cascade           & $-3^{+4}_{-4}$ &     $18^{+16}_{-16}$      \\ 
$\nu_{\mu}+N\rightarrow \mu+had.$                         & Cascade + Track  & $0^{+5}_{-4}$  &    $6^{+12}_{-6}$      \\ 
$\nu_{\tau}+N\rightarrow \tau+had.\rightarrow had.$       & Cascade           & $-7^{+5}_{-3}$ &      $-2^{+14}_{-14}$     \\
$\nu_{\tau}+N\rightarrow \tau+had.\rightarrow \mu + had.$ & Cascade + Track    & $-7^{+5}_{-3}$ &      $-2^{+14}_{-14}$     \\
$\nu_{l}+N\rightarrow \nu_{l} + had.$                     & Cascade           & $-8^{+3}_{-3}$ &     $22^{+15}_{-25}$   \\
\bottomrule
\end{tabular}\caption{Summary of neutrino interaction types in IceCube DeepCore. Each row applies to both neutrinos and antineutrinos. The observable event topology and benchmark values for reconstruction performance at 20 GeV are provided for each interaction type. The biases are shown as the mean of $X_{reco}-X_{true}$, with a range that contains 50\% of the events around the mean value of the distribution. \label{tab:reconstruction_summary}}
\end{table*}
The total event energy is determined from the sum of these two steps:
\begin{equation}
    E_{reco} = E_{\mathrm{shower}} + \frac{b}{a}(e^{b\cdot L_{\mu}}-1),
\end{equation}
where $a = 0.226$~GeV/m and $b=4.6\cdot10^{-4}$~m$^{-1}$ characterize muon energy losses in the range 10-100~GeV~\cite{GROOM2001183}, $E_{\mathrm{shower}}$ is the hadronic shower energy and $L_{\mu}$ is the reconstructed muon track length. The energy resolution is shown in Fig.~\ref{fig:resolutions_energy} for each interaction type as a function of true energy. Again, we observe the best performance for $\nu_{\mu}$ CC events, for which the reconstruction is optimized, with a median error of approximately 20\% for events above 10~GeV. For the other interaction types, the performance varies from $22-60\%$ between 10-100~GeV. It should be noted that the true energy in Fig.~\ref{fig:resolutions_energy} corresponds to the true neutrino energy, and therefore neglects the unobservable energy taken by neutrinos in NC interactions and decays, as well as the energy of neutral hadrons produced in the interaction. This causes the bias observed for all interaction types, where the reconstructed energy underestimates the true neutrino energy on average.

In addition to this track-plus-cascade fit, the hypothesis of a cascade-only event is reconstructed. The difference in LLH between both hypotheses is then used as an input for the identification of the interaction that took place.

Figures~\ref{fig:resolutions_coszen} and \ref{fig:resolutions_energy} show the correlation between reconstructed zenith angle and energy for different interaction types. Table~\ref{tab:reconstruction_summary} contains a summary of each interaction type and associated event signature in DeepCore, along with 
benchmark resolutions for both energy and zenith angle calculated at 20~GeV.

\subsection{Particle Identification (PID)\label{sec:PID}}

In previous DeepCore oscillation analyses, single reconstructed variables were used to classify events as tracks or cascades. Here instead we employ a multivariate approach to improve the PID discriminator. 
We train a Gradient Tree Boosting algorithm~\cite{friedman2001} provided by the \texttt{scikit-learn} package~\cite{scikit-learn} with simulation to identify $\nu_{\mu}$ CC events as signal (tracks), against a background (cascades) consisting of $\nu_{e}$ CC and all NC events. Events from $\nu_{\tau}$ CC interactions are not used in the training to avoid confusion from $\tau\rightarrow\mu$ decays. The simulation is split such that 50\% is used to train the classifier and 50\% is used for testing to evaluate the performance and assess the level of overtraining, i.e. robustness against fitting statistical fluctuations in the simulation. For classifier training, events are weighted to match the initial, unoscillated flux expectation taken from the Honda \textit{et al.} model~\cite{PhysRevD.92.023004}. The event weights are then scaled so that the classifier sees the same total number of signal and background events during the training. 

Several input variables and combinations were tested, and the best classifier performance was found using the following seven features:
\begin{itemize}
    \item SANTA $\chi^2\textrm{-ratio}$, defined as  $\frac{(\chi^{2}/\mathrm{d.o.f.})_{\mathrm{track}}}{(\chi^{2}/\mathrm{d.o.f.})_{\mathrm{cascade}}}$, i.e. the ratio of goodness-of-fit metrics from each fit hypothesis in the directional reconstruction. The number of d.o.f. is calculated as the number of hit DOMs in the unscattered photon pulse series minus the number of fit dimensions in the hypothesis, which is 4 for cascades and 5 (single-string events) or 6 (multi-string events) for tracks.
    \item $\Delta$LLH from energy reconstruction, defined as LLH$_\mathrm{track}-$LLH$_\mathrm{cascade}$, i.e. the best-fit LLH value from each hypothesis
    \item Reconstructed muon track length, $L_{\mu}$
    \item Radial distance of the interaction vertex from string 36\footnote{String 36 is approximately at the center of the array, and near to the densest region of DeepCore (see Fig.~\ref{fig:detector}).}, $\rho^{36}_{vertex}$
    \item Radial distance of the end-point from string 36, $\rho^{36}_{stop}$
    \item Depth of the interaction vertex, $z_{vertex}$
    \item Depth of the end point, $z_{stop}$    
\end{itemize}

SANTA $\chi^{2}$-ratio and $\Delta$LLH are found to be the most useful variables, with an importance of 55\% and 21\%, respectively, as defined by the classification algorithm~\cite{friedman2001}. The former is calculated using unscattered photon hits, while the latter is calculated using pulses also from scattered photons. Using both of these variables helps to mitigate potential biases from detector calibration uncertainties. 

Longer muon tracks are more easily distinguished from the hadronic shower at the interaction vertex, motivating the inclusion of $L_{\mu}$. The reconstructed positions of the event vertex and end-point are useful in accounting for the inhomogeneous reconstruction performance within the DeepCore volume, which results from different ice properties and module density (see Fig.~\ref{fig:detector}). The spatial coordinates and $L_{\mu}$ play a less important role with an importance of 3-8\%. 

\begin{figure}[!t]
    \centering
    \includegraphics[width=0.94\linewidth]{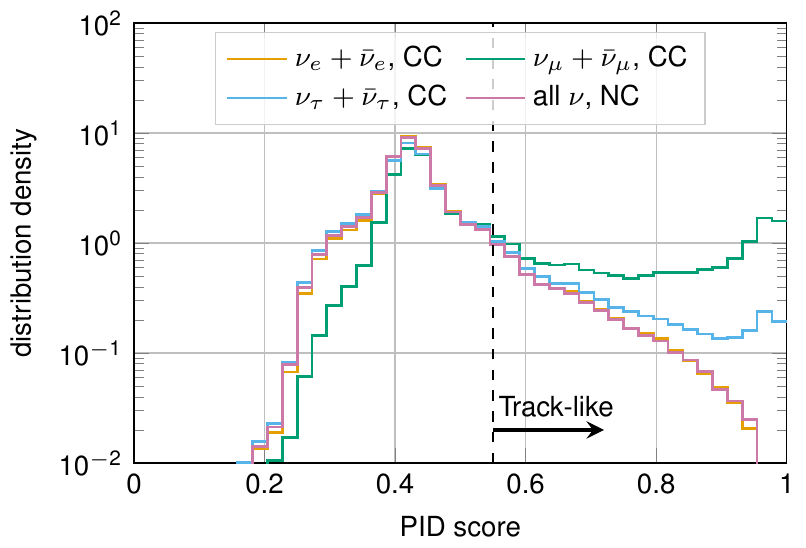}

    \caption{Output probability score distribution from the PID algorithm for the different interaction types. Distributions are normalized to better visualize shape differences. A dashed line at 0.55 indicates scores where track-like event topologies begin to diverge from the cascade-like topologies.}
    \label{fig:gbc_probability_flavors}
\end{figure}

Fig.~\ref{fig:gbc_probability_flavors} shows the normalized probability score distribution obtained from applying the classification model to all interaction types. The distribution ranges from 0 to 1, where a value of 1 indicates the most signal-like, i.e. track-like. The $\nu_{\mu}$ CC population has two distinct peaks. More easily distinguishable track-like events peak close to 1.0 as expected, while lower energy and highly inelastic\footnote{Inelasticity refers to the amount of energy transfered from the neutrino to hadrons in the interaction.} events look more similar to cascades and thus populate the second peak near 0.42, similar to NC, $\nu_{e}$ CC and most $\nu_{\tau}$ CC events. The small fraction of $\nu_{\tau}$ CC events with probabilities closer to 1.0 are primarily due to the aforementioned events that contain visible muons from $\tau$ decays.  

If we consider events with a probability score above 0.55 to be classified as track-like, then the fraction of signal events ($\nu_{\mu}$ CC) correctly identified as tracks, i.e. true positive rate, is 33.8\% while the fraction of background events incorrectly identified as tracks, i.e. false positive rate, is 23.3\% (with a track purity of 58.9\%). Using only the SANTA $\chi^{2}$-ratio as a classification metric and requiring the same true positive rate, we obtain a false positive rate of 26.7\%. This constitutes an expected improvement of 3.3\% in track purity using the multivariate classifier.

\subsection{Final level selection}

After reconstruction, final cuts are applied to further enhance the purity of $\nu_{\mu}$ CC events in the sample, and to reduce the atmospheric $\mu$ contamination to minimize the impact of background modeling uncertainties in our measurement. 

The Earth filters out atmospheric $\mu$ very efficiently, such that only down-going trajectories can reach IceCube. Moreover, the atmospheric $\nu$ oscillation signal is expected to appear below the horizon. Therefore, we keep events with $\cos\theta_{\mathrm{reco}}<0.1$, removing a significant amount of background without loss of the signal region. 

\begin{figure}[!b]
   \centering     
   \includegraphics[width=0.95\linewidth]{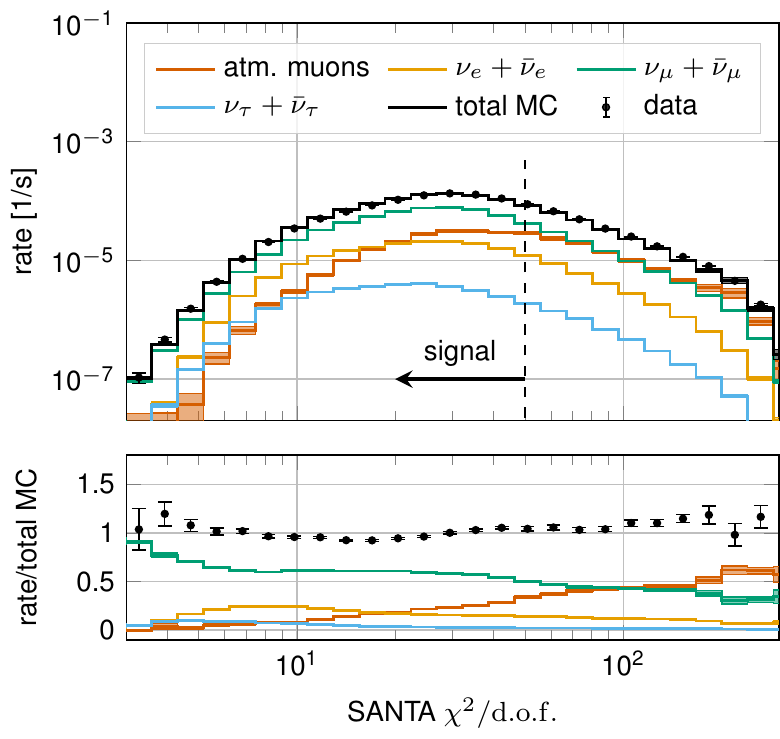}

  \caption{Reduced $\chi^2$ from the SANTA track hypothesis fit.}
  \label{fig:santaGOF}
\end{figure}

The angular reconstruction performance is found to be slightly worse for atmospheric $\mu$, as demonstrated by the SANTA $\chi^{2}$/d.o.f. under the track hypothesis shown in Fig.~\ref{fig:santaGOF}. This is due to selection bias effects. The only muons surviving to this level of the selection are those that have deposited very little light in the detector, making them generally difficult to reconstruct, and in many cases appear more cascade-like. Events with ($\chi^{2}$/d.o.f.)$_{\mathrm{track}} \geq 50$ are therefore removed. The cut on the Level 4 muon classifier described in Sec.~\ref{sec:sample} is also tightened to require scores $> 0.97$. 

\begin{table*}[tb!]
\begin{tabular}{lcccc}
\toprule
\hline
Cut       & $\nu$ rate [1/$10^6$s] \;\; & Atm. $\mu$ rate [1/$10^6$s] \;\;& $\mu$ contamination (\%) \;\;& $\nu_{\mu}$ CC purity (\%)\;\; \\ 
\midrule
Reconstruction                          & 957.1 & 313.7 & 24.7 & 49.8 \\
cos$\theta_{zenith} < 0.1$              & 751.1 & 112.1 & 13.0 & 55.7 \\
SANTA $\chi^{2}/d.o.f. < 50$            & 642.8 & 61.6  & 8.74 & 58.2 \\
L4 muon classifier score $>0.97$      & 464.7 & 13.1  & 2.75 & 59.5 \\
Coincident $\mu$ rejection              & 464.6 & 13.1  & 2.75 & 59.5 \\
$6.3$~GeV$ < E_{\mathrm{reco}} < 158.5$~GeV       & 400.9 & 13.1  & 3.16 & 60.1 \\
PID score $>0.55$                       & 101.3 & 2.10  & 2.06 & 82.1 \\
\bottomrule
\end{tabular}\caption{Selection criteria specific to the \textit{golden event sample} which aim to reduce atmospheric $\mu$ contamination and increase the purity of $\nu_{\mu}$ CC events. The cuts are applied sequentially in the order shown, starting from the top. Rates are estimated with MC simulation using the same weighting scheme described in the caption of Fig.~\ref{fig:l5_summary}.
\label{tab:golden_selection}
}
\end{table*}

\begin{figure}[!t]
    \centering  
    \includegraphics[width=\linewidth]{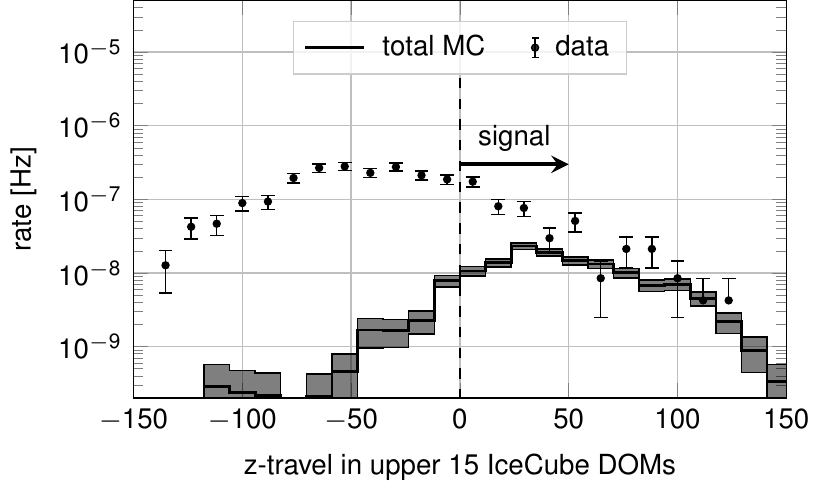}
    \caption{Proxy for direction of travel calculated using the uppermost 15 layers of IceCube DOMs. Only events with at least 4 hits in those layers are included in the histograms.
    }
  \label{fig:coincident_muons}
\end{figure}

Two additional cuts are applied to remove a small fraction of events where a neutrino and a muon interaction occur in coincidence. These events are known to occur, but the simulation used for this study does not include them, so we opt for removing them entirely. The most powerful variable to reject these events is referred to as the \emph{z-travel} direction shown in Fig.~\ref{fig:coincident_muons}. Using the shallowest 15 layers of DOMs on IceCube strings\footnote{DeepCore strings that contain the veto endcap are excluded from this calculation}, \emph{z-travel} is calculated as $\langle z \rangle - \langle z_{t(Q25)} \rangle$, where $\langle z \rangle$ is the average z-position of all hit DOMs, and $\langle z_{t(Q25)} \rangle$ is the average z-position of the earliest quantile of hit DOMs. In this way, a negative value is interpreted as a down-going event in the upper region of the detector, which is consistent with the hypothesis of a coincident atmospheric $\mu$. Such events are removed from the sample. In addition, fewer than 8 triggered DOMs in the outermost strings of the IceCube detector are allowed in each event. Together, these two cuts remove less than 1\% of simulated events from the data sample, but ensure that the data are properly described by the simulation.

Finally, we restrict the data to a region of phase space where we expect to observe the atmospheric $\nu$ oscillation signal, and where the reconstructions perform well. We only accept events in the range $0.8 < \log_{10}(E/\mathrm{GeV}) < 2.2$ ($6.3$~GeV $< E_{\mathrm{reco}} < 158.5$~GeV), which is a wide enough energy band to capture the $\nu_{\mu}$ disappearance valley and also allow for off-signal sidebands that help to constrain certain systematic uncertainties (see Sec.~\ref{sec:systematics}). Because the reconstructions are optimized on $\nu_{\mu}$ CC events, we also require events to have a PID probability score $>0.55$, removing the most cascade-like events.

The effects of the reconstruction efficiency and the final level selection on the expected atmospheric $\nu$ and $\mu$ rates are shown in Table~\ref{tab:golden_selection}. Events triggered by pure noise are negligible and therefore not shown. The atmospheric $\mu$ contamination ($\nu_{\mu}$ CC purity) is calculated as the atmospheric $\mu$ ($\nu_{\mu}$ CC) rate divided by the total event rate estimated from simulation. After all cuts, the estimated atmospheric $\mu$ contamination of the golden event sample is $\sim2\%$, and the majority ($\sim82\%$) of events are $\nu_{\mu}$ CC interactions.

\subsection{Long-term data stability}

As described in Sec.~\ref{sec:detector_calibration}, IceCube DOMs are recalibrated yearly, in spring. Along with the new calibration constants, a new software revision for Level 1 and 2 filtering is deployed for data processing. The introduction of these changes defines a new period of data taking, and the new settings are used for all the data collected from that point on. The data analysed for this paper was collected over 8 periods. Therefore, to ensure that the event selection, reconstruction and event classification behave similarly for each run period, we check the consistency between each period of data taking prior to performing the fit for oscillation parameters. 

\begin{table}[!tb]
\begin{tabular*}{0.8\linewidth}{ccc}
\toprule
\hline
Period & Rate [1/$10^6$s] & Livetime [y]\\ \midrule
2011--2012 &    96.4 $\pm$ 2.1 & 0.67 \\      
2012--2013 &    93.0 $\pm$ 1.8 & 0.87 \\      
2013--2014 &    90.0 $\pm$ 1.8 & 0.92 \\      
2014--2015 &    93.7 $\pm$ 1.8 & 0.96 \\      
2015--2016 &    95.3 $\pm$ 1.8 & 0.98 \\      
2016--2017 &    90.1 $\pm$ 1.7 & 0.96 \\      
2017--2018 &    94.1 $\pm$ 1.6 & 1.10 \\      
2018--2019 &    94.1 $\pm$ 1.7 & 0.99 \\
\bottomrule
\end{tabular*}\caption{Final rates for the golden event sample, broken down by periods of data taking, where 1$\sigma$ expresses a Poisson uncertainty given by $\sqrt N$, meant only to demonstrate agreement of rates within expected statistical fluctuations. Each period typically starts in the spring with a new run configuration defined by calibration constants and software revision, and continues through to the following spring. The transition to a new run configuration in 2018 occurred slightly later than usual, resulting in a livetime of 1.1 y. \label{tab:season_livetime_rates}}
\end{table}

The data rates for the golden event sample, i.e. at final selection level, for each period are shown in Table~\ref{tab:season_livetime_rates}, and are consistent within statistical errors. We note that these rates are systematically lower than the rates expected from simulation reported in Table~\ref{tab:golden_selection}. This discrepancy is covered by the uncertainty in the total atmospheric $\nu$ flux, which is approximately 10\% (see Sec.~\ref{section:syst_flux}). In addition, the rates given in Table~\ref{tab:golden_selection} are computed before adjusting the simulation to fit the data, described in the next section, and therefore some disagreement is expected. Nevertheless, the fit for oscillation parameters does not use any rate information and only considers the shapes of fitted distributions.

\begin{figure}[t!]
    \centering
    \includegraphics[width=.99\linewidth]{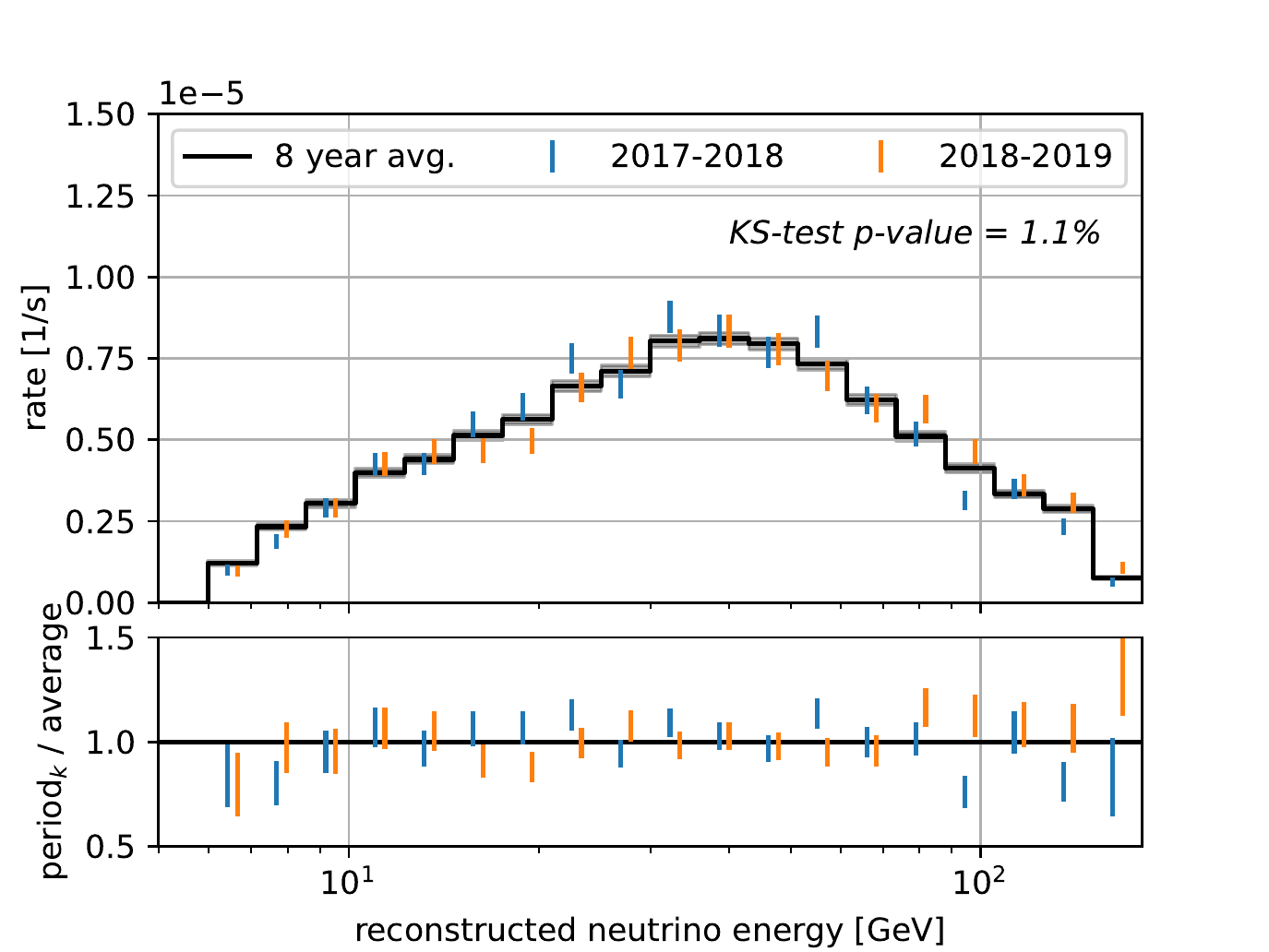}
    \caption{Observed rates as a function of reconstructed energy for the data taking periods 2017-2018 (blue) and 2018-2019 (orange). The agreement for this observable between these years, as quantified by a KS-test p-value, is 1.1\%. The black histogram shows the average rate across all 8 data taking periods for reference. See text for more details.}
    \label{fig:data_stability_reco_energy}
\end{figure}

We also investigate the agreement between data taking periods for approximately 20 distributions. The agreement between each pair of run periods is quantified by performing a two-sided Kolmogorov-Smirnov (KS) test for each distribution, which tests whether the two distributions are consistent with being drawn from the same underlying probability distribution~\cite{10010480527}. The resulting p-values are used to identify potential outliers that would require further investigation. As an example, the reconstructed energy distribution is shown in Fig.~\ref{fig:data_stability_reco_energy} for data taking seasons 2017-2018 and 2018-2019, compared to the average rate calculated using the complete eight year sample for reference. These years have a p-value of 1.1\%, which was the lowest observed for any pair of seasons. However, observing the bin-wise rates in this observable, this level of agreement appears consistent with statistical fluctuations. All other pairs of seasons have a good agreement in this observable as well, as can be seen in Appendix~\ref{appendix:data_quality}.

\begin{figure}[t!]
    \centering
    \includegraphics[width=.99\linewidth]{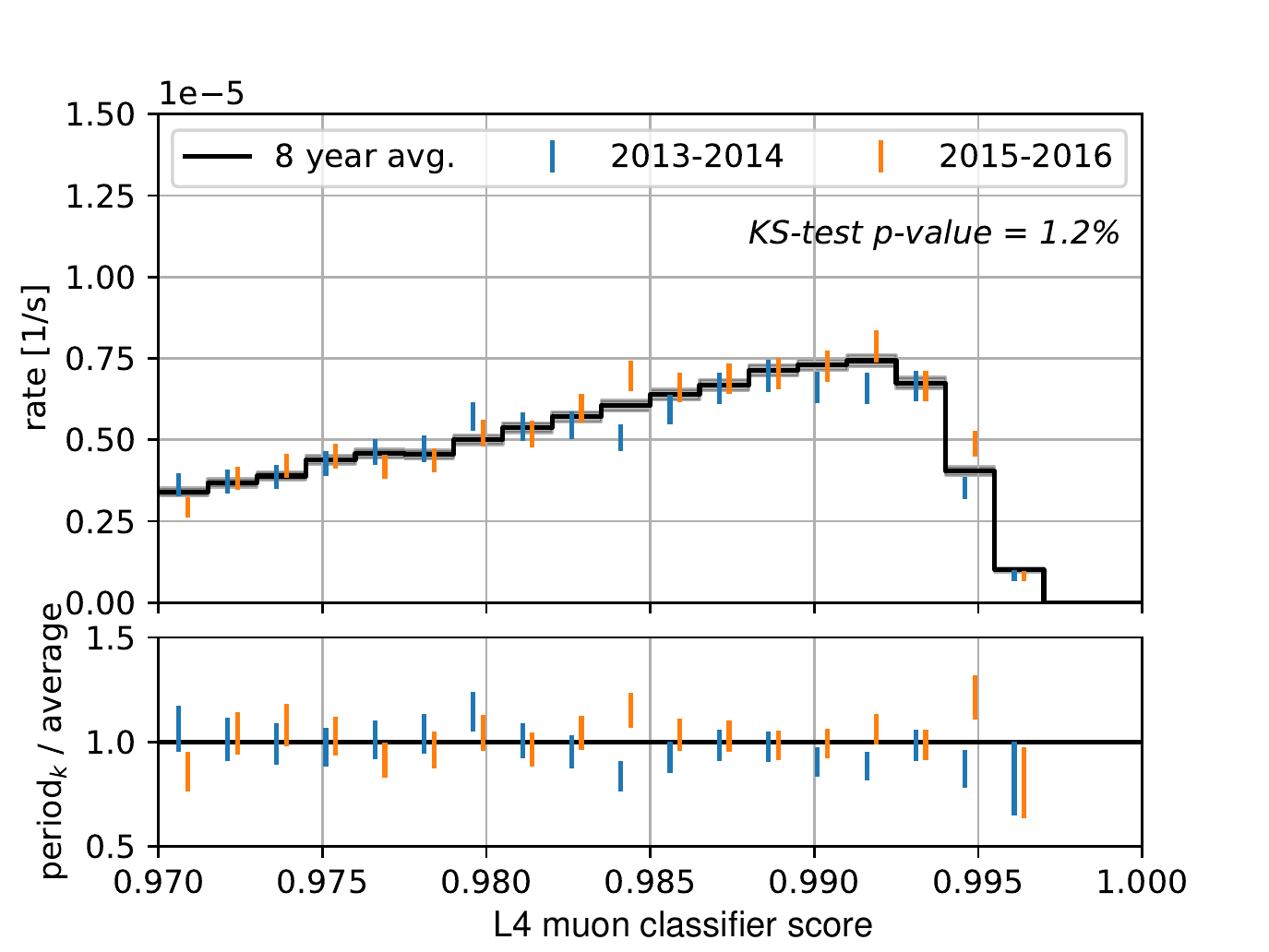}
    
    \caption{Observed rates as a function of the L4 muon classifier score for the data taking periods 2013-2014 (blue) and 2015-2016 (orange). The agreement for this observable between these years, as quantified by a KS-test p-value, is 1.2\%. The black histogram shows the average rate across all 8 data taking periods for reference. See text for more details.}
    \label{fig:data_stability_L4MuBDT}
\end{figure}

Similarly, Fig.~\ref{fig:data_stability_L4MuBDT} shows the probability score from the Level 4 muon classifier. As described in Sec.~\ref{sec:sample}), this classifier was trained using only a subset of  data collected in 2014. However, the performance appears consistent across all years, with no p-value lower than 1.2\%. To better illustrate what level of agreement this p-value represents, we again provide the 1D distribution of this L4 Muon Classifer score for the corresponding years. 

We investigated 20 distributions and did not observe any p-value below 0.1\% for any combination of data taking periods. Moreover, no periods appear systematically different from others across several observables. A selection of relevant variables are shown in Appendix~\ref{appendix:data_quality}. This gives us confidence in the data calibration and filtering processes such that data from all 8 run periods can be combined in the analysis.

\section{Analysis~\label{sec:analysis}}

The analysis is performed by comparing a template histogram of our simulation to a detector data histogram, where the template is re-weighted according to free parameters accounting for the physics and nuisance parameters. A minimizer is used to fit the parameters to best match the simulation to data. This method is often referred to as ‘forward-folded parameter estimation’, and is the same strategy used in previous measurements of the atmospheric $\nu$ oscillation parameters using DeepCore data~\cite{Aartsen_2018}. 
 
We use a modified $\chi^2$ test statistic defined as
\begin{equation}\label{eq:mod_chi2}
\chi^2_{\mathrm{mod}} = \sum_{i \in \mathrm{bins}}^{}\frac{(N^{\mathrm{exp}}_i - N^{\mathrm{obs}}_i)^2}{N^{\mathrm{exp}}_i + (\sigma^{\mathrm{sim}}_i)^2} + \sum_{j \in \mathrm{syst}}^{}\frac{(s_j - \hat{s_j})^2}{\sigma^2_{s_j}},
\end{equation}

\noindent where the expectation within a bin is calculated as the sum of the event weights $N^{\mathrm{exp}}_i = \sum_{i}^{\mathrm{evts}} w_i$. We calculate the error term due to Poisson fluctuations of the data with the expectation from simulation $N^{\mathrm{exp}}_i$. The statistical uncertainty due to the finite statistics of our simulation sets is included as $(\sigma^{\mathrm{sim}}_i)^2 = \sum_{i}^{\mathrm{evts}} w_i^2 + (\sigma^\mu_i)^2$, where the first term applies to neutrino sets and the second one applies to atmospheric $\mu$ sets, whose treatment is described in Sec.~\ref{sec:atm_mu}. The second term in Eq.~\ref{eq:mod_chi2} is a penalty term to account for prior knowledge of some systematic parameters, where $s_j$ is the nominal value of the $j$-th systematic parameter, $\hat{s_j}$ is its maximum likelihood estimator, and $\sigma^2_{s_j}$ is the prior’s Gaussian standard deviation, if applicable. In some cases the uncertainty on the prior is non-Gaussian. In these cases we do not apply any penalty, and the parameter is bounded by a range significantly larger than the estimated uncertainty.  

\begin{figure}[t!]
    \centering
    \includegraphics[width=.99\linewidth]{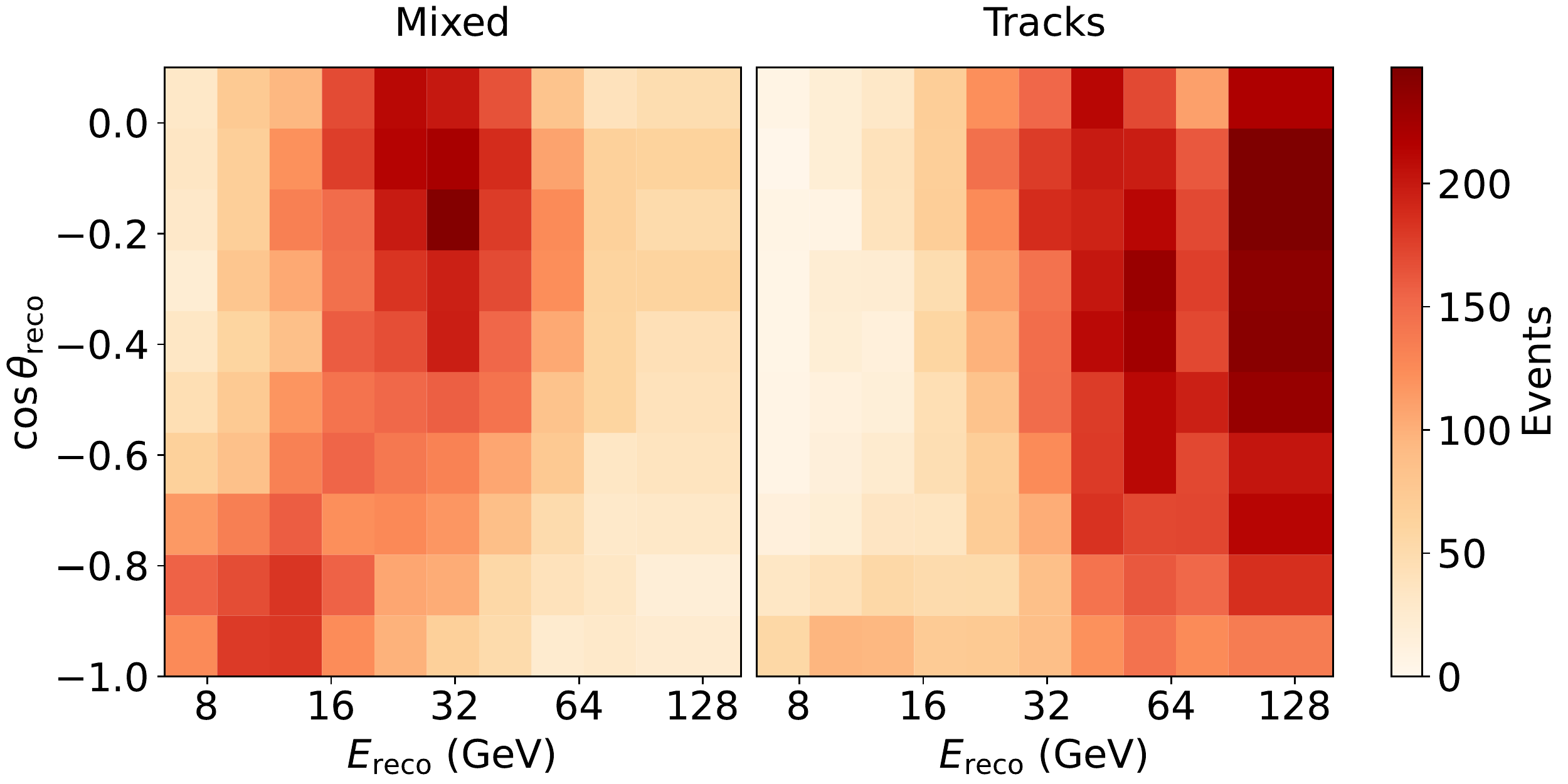}
    
    \caption{Observed number of data events in the analysis binning for the full 8 years of livetime.}
    \label{fig:data_3Dbinning}
\end{figure}

The data is binned into a three-dimensional histogram with ten reconstructed energy bins spaced logarithmically between 6.31 and 158.49 GeV, and ten reconstructed
cosine zenith bins spaced linearly between -1 and 0.1 (see Fig.~\ref{fig:data_3Dbinning}). The last energy bin is twice as wide to contain sufficient statistics. The data are also separated into two PID bins to improve the sensitivity to observe $\nu_{\mu}$ disappearance. The first PID bin, referred to as the \textit{mixed channel}, contains events with a PID classifier score between 0.55 and 0.75. This bin is comprised mostly by $\nu_{\mu}$ CC events, with approximately 30\% contamination from other neutrino flavors and interactions. The second bin, or \textit{track channel}, contains events with a PID score between 0.75 and 1.0, and therefore has a higher $\nu_{\mu}$ CC purity of 94\%. The PID definition was optimised for sensitivity to atmospheric $\nu$ mixing while aiming to keep roughly similar statistical power between both mixed and track channels.

The signal of this analysis is the disappearance of muon neutrinos coming from below the horizon. The oscillation pattern at energies below 10~GeV is not resolvable with DeepCore. Instead, the analysis is driven by the position and amplitude of the first oscillation valley between 10~GeV and 50~GeV, where the position is determined by the mass splitting $\Delta m_{32}^2$ and the amplitude by $\sin^2\theta_{23}$ with the mixing angle $\theta_{23}$. Figure~\ref{fig:data_3Dbinning} shows the observed number of real data events in the analysis binning.

\begin{figure}
    \centering
    \includegraphics[width=0.99\linewidth]{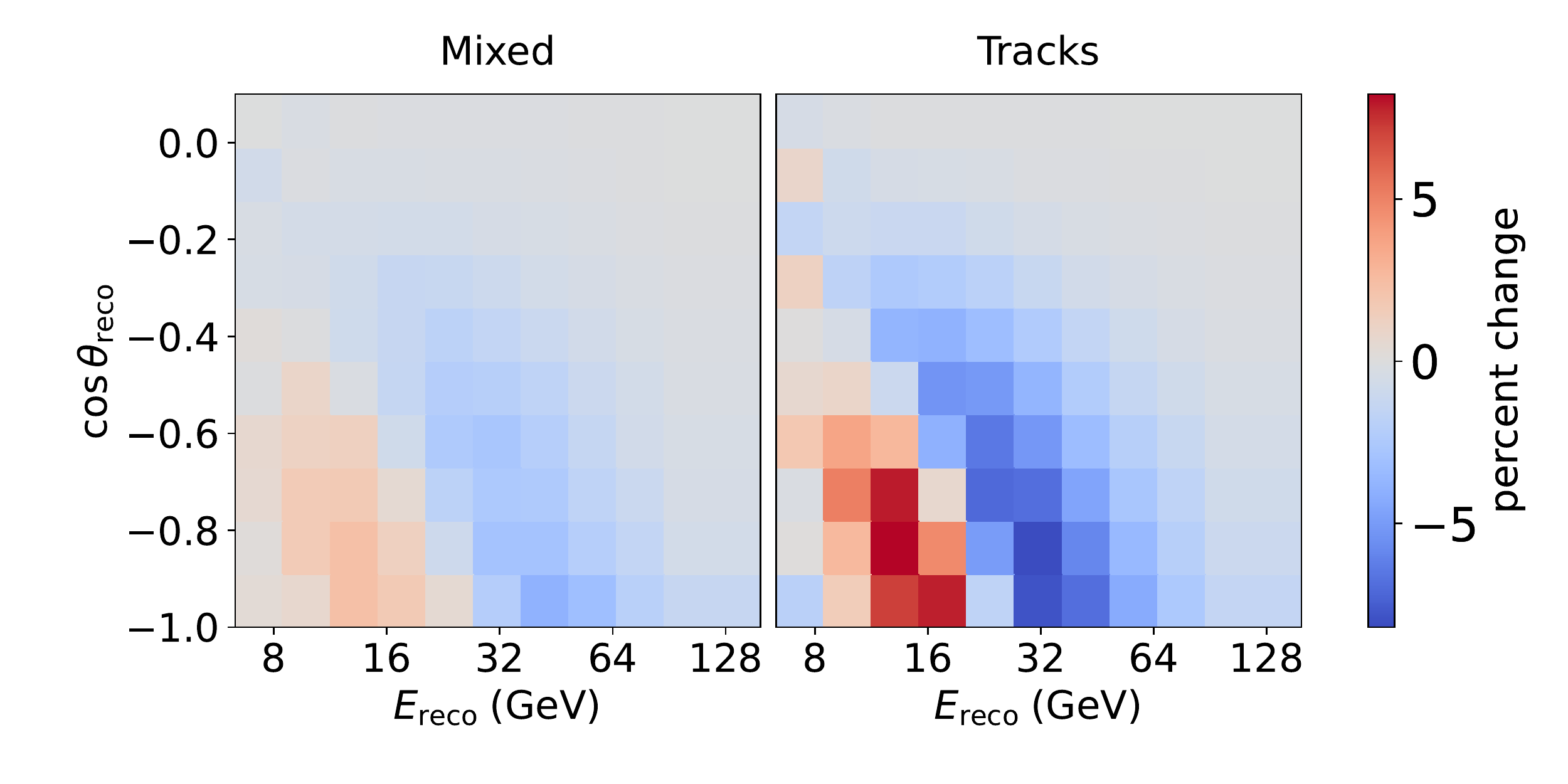}
    \includegraphics[width=0.99\linewidth]{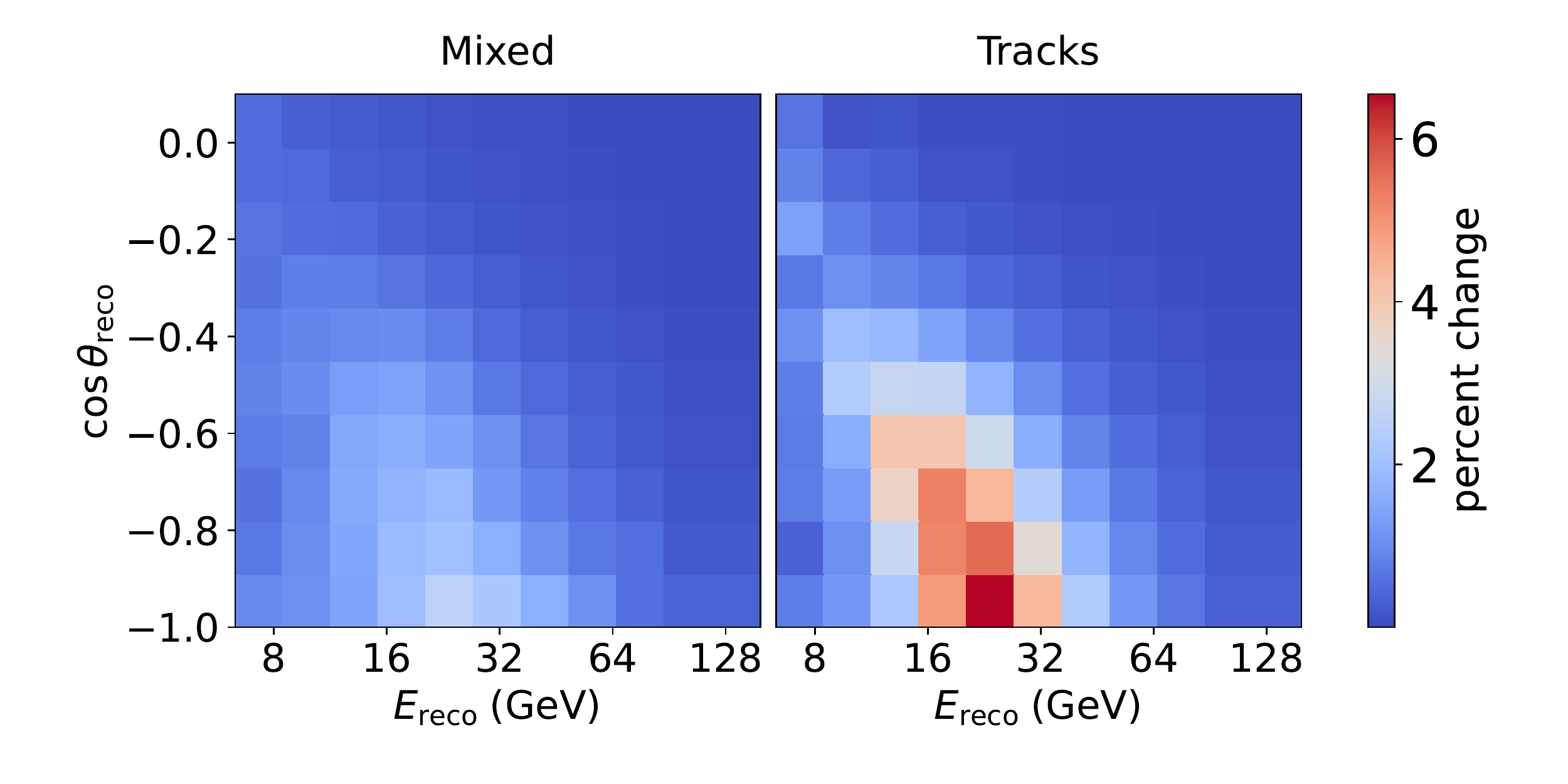}
    \caption{Change in the expected number of events in the analysis when independently varying the values of $\Delta m^2_{32}$ (top) and $\sin^2\theta_{23}$ (bottom) to the their 90\% confidence level. The mass splitting changes the position of the oscillation probability, while the mixing angle modifies the amplitude. The largest change is observed in the track channel. }
    \label{fig:modify_osc_params}
\end{figure}

Figure~\ref{fig:modify_osc_params} shows how the expected number of events from simulation changes in the analysis binning for values of oscillation parameters that differ by the expected 90\% confidence interval that this study will produce. The $\nu_{\mu}$ disappearance is most pronounced in the track PID channel, which is to be expected since it has the higher purity of $\nu_{\mu}$ CC events. 

\section{Systematic Uncertainties~\label{sec:systematics}}
There are several sources of systematic uncertainty that can introduce a modification to the expected number of events in a bin. To include these effects in the analysis, they are modeled as a function of parameters that can be varied continuously. Whenever possible, we implement a model motivated from first principles and adjust its parameters. In many instances, however, we do not have access to the code that produces our required input (e.g. atmospheric neutrino flux) and need to create effective parameters that capture the reported uncertainties.

In this section we describe the sources of uncertainty, as well as their implementation in the fit. A summary is given in Table~\ref{table:syst_param}, and Appendix~\ref{appendix:systematics} includes figures that demonstrate the impact of each one of them in the analysis binning. Since the analysis exclusively uses the relative distribution of events, sources of uncertainty that only scale the event rate are not implemented individually. Instead, a single parameter $A_{eff}$ scale is used as a global scale factor for the total neutrino rate.

\begin{table*}[!ht]
  \centering
  \caption{List of systematic uncertainties used in this analysis along with their priors, fit values and pulls. The origin of the nominal expectation (baseline) from each systematic group is given next to their name. Prior widths are symmetric around the nominal expectation, unless stated otherwise, and their origin is described in the text. Parameters without a prior are labeled as unconstrained. Relative optical efficiencies $p_0$ and $p_1$ correspond to the models in Fig.~\ref{fig:holeice_calibration}, where the best fit is also shown. The rows with corrections to meson yields refer to the regions defined in~\cite{barr_paper}.}
  \label{table:syst_param}
\begin{tabular}{p{0.25\linewidth} p{0.3\linewidth} p{0.15\linewidth} p{0.15\linewidth}}
\toprule
\hline
Parameter  & Prior & Fit value & Pull ($\sigma$)\\ \hline
\textbf{Detector}    & \multicolumn{3}{l}{Baseline from calibration data}  \\
DOM eff. correction &  10\% & +6\% & 0.63 \\
Rel. eff. $p_0$ & \footnotesize{Unconstrained} & -0.27 & ... \\
Rel. eff. $p_1$ & \footnotesize{Unconstrained} & -0.04 & ... \\
Ice absorption & \footnotesize{Unconstrained} & -3\% & ... \\
Ice scattering & \footnotesize{Unconstrained} & -1\% & ... \\ \hline
\textbf{Flux}    &    \multicolumn{3}{l}{Baseline from Honda et al.}  \\
$\Delta \gamma_\nu$ & 0.1 & +0.07 & 0.7 \\
$\Delta \pi^\pm$ yields [A-F]   & 30\% & +10\% & 0.35 \\
$\Delta \pi^\pm$ yields G   & 30\% & -6\% & -0.18 \\
$\Delta \pi^\pm$ yields H   & 15\% & -2\% & -0.12 \\
$\Delta K^+$ yields W   & 40\% & +8\% & 0.21 \\
$\Delta K^-$  yields W   & 40\% & -1\% & -0.02 \\
$\Delta K^+$ yields Y   &30\% & +11\% & 0.35 \\\hline
\textbf{Cross-section}   &  \multicolumn{3}{l}{Baseline from GENIE}  \\
$M_{A}^{CCQE}$  & $0.99$~GeV~${}_{-15\%}^{+25\%}$ & +1\% & 0.03  \\
$M_{A}^{CCRES}$  & 1.12~GeV$ \pm$20\% & +11\% & 0.57  \\
$\sigma_{NC}/\sigma_{CC}$  & 20\% & +13\% & 0.63\\
DIS CSMS   & 1.0 & 0.04 & 0.04 \\  \hline
\textbf{Atm. muons}        &  \multicolumn{3}{l}{Baseline from Gaisser et al.+Sibyll2.1}   \\
Atm. $\mu$ scale & \footnotesize{Unconstrained} &  +39\% & ... \\ \hline
\textbf{Normalisation}   & \multicolumn{3}{l}{Baseline from calibration+flux models} \\
$A_{eff}$ scale & \footnotesize{Unconstrained} & -18\% & ... \\ \hline
\bottomrule
\end{tabular}
\end{table*}

We decide which systematic uncertainties must be included in the fit by studying the potential bias they would produce in the oscillation parameters and the change on the test statistic $\chi^2_\mathrm{mod}$ if we neglected them. We create data sets with their observed quantities set equal to their expected values for a wide range of values for $\theta_{23}$ and $\Delta m^{2}_{32}$ and perform two fits: one where the oscillation parameters are fixed to their true value and one where they are left free. In both fits, the systematic parameter being tested is fixed to a value off from its nominal expectation by either 1$\sigma$ or by an educated guess, if the uncertainty is not well defined. Parameters are included in the analysis when this test creates a significant bias in the oscillation parameters, which is conservatively defined as a difference larger than $2\times10^{-2}$ between the test statistics of the two fits.

\subsection{Detector calibration uncertainties~\label{section:syst_det}}
\label{sec:sys_detector}
The uncertainties associated with the detection process of neutrinos, such as the optical efficiency of the DOMs and the properties of the ice, have the largest impact on this study. However, there is no known analytic function that maps the detector calibration uncertainties described in Section~\ref{sec:detector_calibration} onto an effect on the expected neutrino rates. Instead, we derive these relationships for each bin in the analysis histogram using MC data.

To evaluate the expected impact of detection uncertainties, data sets are produced with different variations of detector response, processed to the final level of selection, and then they are parameterized following a model of the uncertainties to evaluate how the final sample would look like for any reasonable choice of parameters. The parametrizations are done at the analysis bin level, assuming that every effect considered is independent and that they can be approximated by a linear function. Under these assumptions we can compute a reweighting factor in every bin that depends on $N$ parameters, which correspond to the number of systematic effects being considered, plus an offset $c$, as

\begin{equation}
    f(p_1,...,p_N)=c+\sum_{n=1}^N m_n \Delta p_n.
\end{equation}
Here $m_n$ are the reweighting factors obtained from simulation sets with a systematic variation and $\Delta p_n$ is the test value of a specific systematic variation.

\begin{figure}[t!]
    \centering
    \includegraphics[width=.95\linewidth]{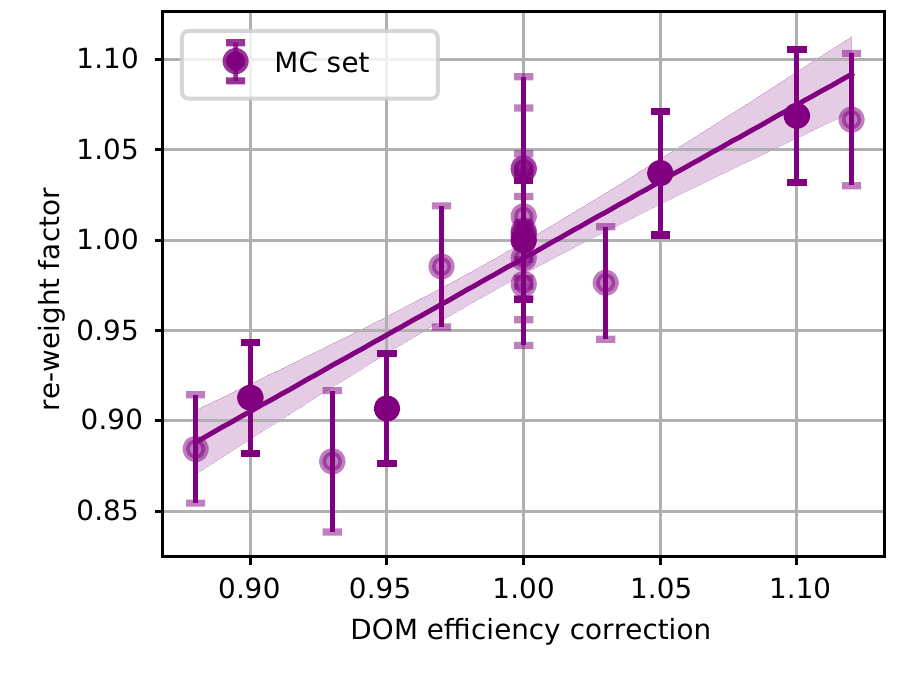}
    \caption{Example of a hypersurface function in one bin projected on the dimension of the DOM efficiency correction. Each data point corresponds to one systematic set. Translucent datapoints are from sets where one or more systematic parameter \emph{besides} the DOM efficiency correction is off-nominal. Those points are projected along the fitted plane to the nominal point. Several systematic sets have a nominal DOM efficiency of 1.0. The band shows to the standard deviation of the fitted function.}
    \label{fig:hypersurface-example}
\end{figure}

The fit of the parameters $m_n$ is done over all systematic MC sets, reducing the uncertainty on the MC prediction in each bin as a side effect since the error on the fitted function is smaller than the statistical error from the nominal MC set. The set of all fitted functions in all histogram bins are called ``hypersurfaces". An example of such a fit from a single bin, projected onto one dimension, is shown in Fig. \ref{fig:hypersurface-example}. 

The event counts coming from different flavors and interactions have a different response to varying the same detector parameter. Therefore, the hypersurfaces in each bin are fit separately for three groups of events:
\begin{itemize}
    \item ($\nu_{\mathrm{all}} + \bar{\nu}_{\mathrm{all}}$) NC + ($\nu_e + \bar{\nu}_e$) CC: These events all produce cascade signatures in the detector.
    \item ($\nu_\tau + \bar{\nu}_\tau$) CC: These interactions may differ from the previous group because they have a production threshold of $E_\nu \gtrsim 3.5\,\mathrm{GeV}$ and also produce muons with a branching ratio of 17\%.
    \item ($\nu_\mu + \bar{\nu}_\mu$) CC: These interactions produce track-like signatures.
\end{itemize}

The distribution of $\chi^{2}$/d.o.f. from the fits in all analysis bins is used as a diagnostic to ensure that the fitted, linear hypersurfaces provide a good estimate for the expected number of events for the full range of simulated detector configurations. We find that the means of these $\chi^{2}$/d.o.f. distributions are all consistent with 1.0 as expected from good fits for each of the three categories described above (NC + $\nu_{e}$ CC, $\nu_{\tau}$ CC and $\nu_{\mu}$ CC). Attempts to use higher order polynomial fits did not yield a significantly improved $\chi^{2}$/d.o.f., and in fact often rendered the fits less stable. 

\begin{figure}[t!]
    \centering
    \includegraphics[width=1.0\linewidth]{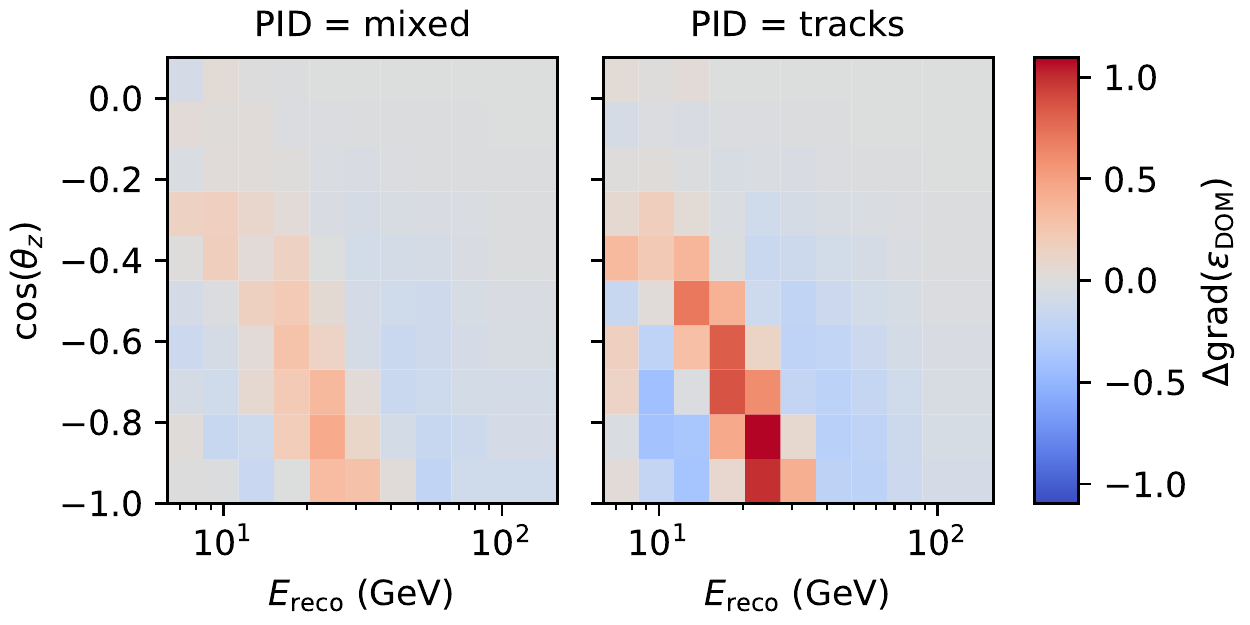}
    \caption{Difference between the fit hypersurface gradient for DOM optical efficiency in each analysis bin, comparing the gradient fit assuming $\Delta\mathrm{m}^{2}_{32}=3.0\times10^{-3}~\mathrm{eV}^{2}$ minus the gradient fit for $\Delta\mathrm{m}^{2}_{32}=2.5\times10^{-3}~\mathrm{eV}^{2}$.}
    \label{fig:interpolatedHS}
\end{figure}

To produce the histograms for fitting the hypersurfaces, a choice must be made for the values of flux, cross section and oscillation parameters. We found that the hypersurface fits are sensitive to the choice of parameters that have correlations with the effect they encode. Most notably, this effect is observed between the mass splitting and DOM optical efficiency as demonstrated in Fig.~\ref{fig:interpolatedHS}, which shows the difference between fitted hypersurface gradients for the DOM efficiency dimension for two values of $\Delta m^{2}_{32}$. 

This problem arises because we are only fitting the hypersurfaces in reconstructed phase space, without accounting for the different true energy and zenith distributions of MC in each analysis bin, which change with each detector systematic variation. To mitigate this problem, we fit the hypersurfaces for 20 different values in mass splitting between $1.5\times 10^{-3}\,\mathrm{eV}^2$ and $3.5\times 10^{-3}\,\mathrm{eV}^2$, and then apply a piece-wise linear interpolation to all slopes, intercepts and covariance matrix elements. The oscillation parameter fit can then dynamically adapt the hypersurfaces for each value of $\Delta m^{2}_{32}$ that is tested using these interpolated functions. The effects of other parameter choices were evaluated as well, but none were found to introduce a significant bias.

In this study, a 5-dimensional hypersurface is used to parameterize the most relevant detection uncertainties, namely the absolute optical efficiency of the DOMs, the relative angular acceptance of the modules (two parameters, see Fig.~\ref{fig:holeice_calibration}), and a global scaling of absorption and scattering lengths for the bulk of the medium. As motivated in Section~\ref{sec:detector_calibration}, the DOM efficiency is constrained by a Gaussian prior to the value of 1.0 $\pm$ 0.1. The ice model parameters are unconstrained in the fit, and allowed to vary within conservative ranges determined from calibration data. The hole ice model parameters are bounded within the ranges $-2.0<p_{0}<1.0$ and $-0.2<p_{1}<0.2$. The bulk ice model parameters are bounded within $-0.90 < \mathrm{Absorption} < 1.10$ and $-0.95 < \mathrm{Scattering} < 1.15$.

We also tested the impact of a new, more detailed ice model~\cite{tc-2022-174}, depth-dependent uncertainties of the ice~\cite{IceCube:2019lxi}, a simulation including the shadow cast by the cables onto the DOMs, a modification of the quantum efficiency as a function of wavelength for DeepCore DOMs and variations in the noise rate. All of these effects were found to be negligible, so they were not included in the fit.

\subsection{Atmospheric neutrino fluxes~\label{section:syst_flux}}
\label{sec:flux}

The flux of atmospheric leptons depends on the spectrum and composition of cosmic rays, the atmospheric conditions at the interaction site and the hadronic interaction model describing the development of the cosmic-ray showers. The uncertainties of each of these ingredients produces uncertainties in the lepton fluxes that need to be accounted for in the analysis.
Our studies focus on how these uncertainties impact the neutrino fluxes, since our data sample is expected to have a 98\% neutrino purity. Our starting point is the atmospheric $\nu$ flux from Honda \textit{et al.}~\cite{PhysRevD.92.023004}. We introduce parameters to encode the sources of uncertainty in the flux, effectively fitting the atmospheric neutrino spectrum as part of the study.

\subsubsection{Cosmic-ray flux uncertainties}

Neutrinos of a given energy can be produced in showers initiated by cosmic-ray primaries with energies up to 100 times higher~\cite{sibyll2.3}. This study is focused on atmospheric $\nu$ between a few and 100~GeV, so the relevant range of the cosmic-ray spectrum is between 10~GeV and 10~TeV. In this range, the spectrum is almost entirely composed of protons and helium~\cite{gsf}. The main uncertainty in this portion of the spectrum can be encoded by a power-law correction $E^{\Delta\gamma}$~\cite{barr_paper, nuflux_unc_manch}, and since the neutrinos follow closely any modification to the primary spectrum, the same form can be used for modifying their flux. The corrected flux as used in this study is then given by

\begin{equation}
    \Phi_{\nu,\mathrm{mod}} = \Phi_\nu \left( \frac{E}{E_\mathrm{pivot}}\right)^{\Delta \gamma} ,
\end{equation}
where $\Delta \gamma$ is the parameter that can be varied. Here $E_\mathrm{pivot}=24$~GeV was chosen to reduce the correlation between $\Delta \gamma$ and the overall scale of the flux. The estimate of the uncertainty on the spectral index is $\pm$0.05 \cite{nuflux_unc_manch}, but here we increase it to $\pm$0.1 to account for a similar response that is expected from independent effects, such as modifications to charged hadron multiplicity in the final state of DIS interactions~\cite{Katori_2015_pythia_had_mult}, variations to higher-twist parameters and valence quark corrections in GENIE~\cite{IceCube:2019dqi} and changes to the overall optical absorption of the ice.

\subsubsection{Hadronic interaction uncertainties}
The atmospheric $\nu$ we detect come from the decay of hadrons produced over a large region of parameter space with few measurements, interpolated by phenomenological models~\cite{barr_paper}. The authors of the atmospheric neutrino flux we use as a baseline have evaluated their impact, showing it as a function of energy~\cite{Honda:2006qj}. For this study, however, we require an uncertainty estimate as a function of energy, direction and neutrino flavor. We therefore use the MCEq\footnote{https://github.com/afedynitch/MCEq} package~\cite{Fedynitch:2018cbl}, which computes inclusive distributions of atmospheric leptons, to evaluate the impact of variations on each of the ingredients that go into the calculation. In general, we do this by introducing small variations $dB$ on a parameter $B$ of some model that gives a lepton flux $\Phi_l$, compute the change in flux $\frac{d\Phi_l}{dB}$, and introduce them scaled by a value $b$, as

\begin{equation}
\Phi_{l,\mathrm{mod}} = \Phi_l + \left( b \cdot \frac{\mathrm{d} \Phi_l}{\mathrm{d}B} \right).
\end{equation}

The modifications introduced to MCEq follow the work of Barr \textit{et al.}~\cite{barr_paper}, where the parameter space in primary energy $E_i$ and the fraction of energy taken by the secondary meson $x_{lab}=E_s/E_i$ is divided into regions with different uncertainties, chosen by the availability of fixed target experiments data used in hadronic models. Here we adopt the same scheme and compute the variations of the fluxes $\frac{d\Phi}{dB}$ after modifying the expected hadronic yields of a model by a constant factor over each region of $E_i, x_{lab}$. These modifications were computed using the Sibyll2.3c~\cite{sibyll2.3} hadronic interaction model and the GSF cosmic-ray flux~\cite{gsf} as a starting point, since the software that produces the baseline model for this analysis is not openly available~\cite{PhysRevD.92.023004}. The choice of cosmic-ray spectra and hadronic interaction model used to compute the modifications $\frac{d\Phi}{dB}$ were found to have negligible impact.

The variations of the yields of $K^\pm$ and $\pi^\pm$ were computed individually and their impact on the flux was evaluated. As the pion ratio is well-measured, the uncertainty on $ \pi^- $ is defined by the uncertainty on $ \pi^+ $ plus an uncertainty on the pion ratio, following~\cite{barr_paper}. The uncertainty on $ K^\pm$ production is kept uncorrelated. 

\begin{figure*}
    \centering
        \centering
        \includegraphics[width=0.45\textwidth]{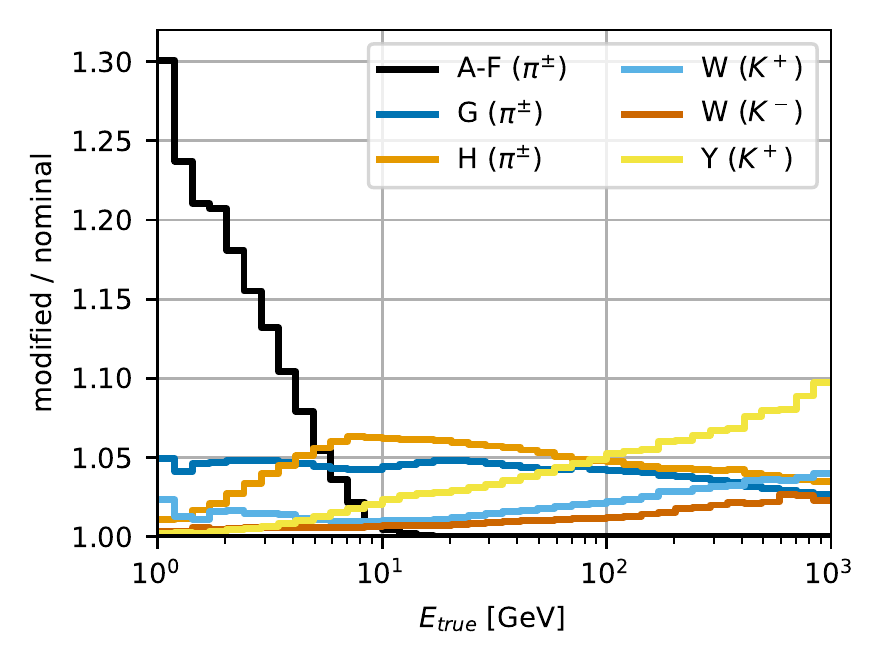}
        \centering
        \includegraphics[width=0.45\textwidth]{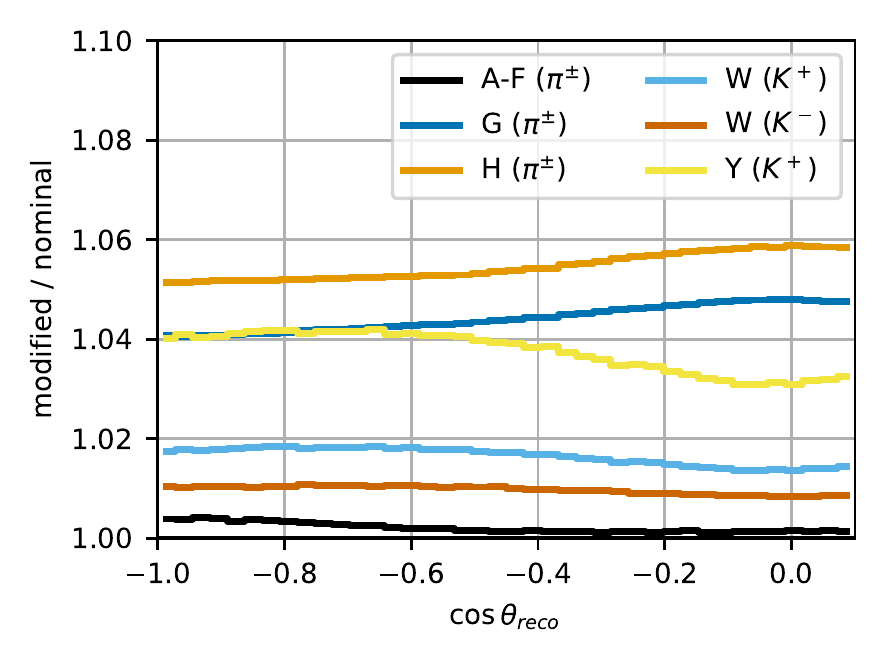}
    \caption{Impact of the variations of the flux parameters used as a function of reconstructed energy (left figure) and reconstructed zenith angle (right figure). The parameters where changed by +1$\sigma$, one at a time. By construction, the variations are symmetric around the nominal expectation. The phase space that each parameter affects follows that given in~\cite{barr_paper}.}
    \label{fig:flux_syst}
\end{figure*}

The entire suite of potential variations is encoded in 17 variables, but not all of them have a significant impact on the neutrino flux in the energy range of this analysis. Here, only the six parameters listed under the Flux section in Table~\ref{table:syst_param} were found to be relevant. The uncertainties on the pion yields resulting from incident parent particles with $E<30$~GeV (regions A to F in Fig.~\cite{barr_paper}) have been added in quadrature and grouped into one parameter, as their modifications have a degenerate effect on the atmospheric neutrino flux. The impact of these six parameters on the flux of muon neutrinos is shown in Fig.~\ref{fig:flux_syst} as function of reconstructed energy and zenith angle. 

We further investigated the impact of hadron-air inelastic cross section uncertainties that drive energy losses during shower development. Using the Glauber approach~\cite{Glauber1955,GLAUBER1970135} to propagate hadron-proton cross sections uncertainties~\cite{PhysRevD.85.074020} to hadron-air interactions as a function of the center of mass energy, we find that their impact on the atmospheric flux model in our energy range of interest ($<1$ TeV) is below 1\% and therefore negligible.

\subsubsection{Atmospheric density}

 The effect of atmospheric density uncertainty in this analysis was studied by obtaining a variation of atmospheric density profile, perturbing the Earth’s atmospheric temperature within a prior range given by the NASA Atmospheric InfraRed Sounder (AIRS) satellite \cite{AIRS} temperature data. The resulting atmospheric density profiles are injected into MCEq to calculate new fluxes. This is performed for a variety of cosmic-ray models and hadronic interaction models available in MCEq. 
 It was found that even the largest variation from density perturbation has a negligible effect in this energy range, so this uncertainty was not included in the fit.

\subsection{Neutrino-nucleon cross-section~\label{section:syst_xsec}}
The DeepCore neutrino event samples can span an energy range from GeV to TeV. For energies above 20~GeV, the interactions are dominated by deep inelastic scattering (DIS). Below this threshold, interactions with the nucleons as a whole become more relevant~\cite{RevModPhys.84.1307}. Since there is no coherent theoretical frame that explains both regimes, we keep the same approach as event generators in our study and divide the uncertainties from neutrino interactions into regions.

\subsubsection{Deep Inelastic Scattering}

The interactions of neutrinos with individual quarks can be calculated precisely. The uncertainties on the predictions of these calculations come from the data set used to describe the probability of finding a quark within the nucleons being considered. These parton distribution functions (PDFs) are mainly obtained from electron scattering data, and the predicted neutrino-nucleon cross section differs depending on which one is used.

Our interaction generator GENIE uses an outdated PDF for the calculation, GRV98~\cite{DIS:GRV98}, because it was the only PDF that incorporated the extrapolations of $Q^2$ below those available from measurements at the time that this study was performed~\cite{Bodek_2003}, but that are relevant for the energies in question\footnote{See~\cite{Candido:2023utz} for a new calculation that extends to low $Q^2$.}. At high energies it differs from other more sophisticated and newer computations. To address this point, we have studied the different predictions from multiple DIS calculations, comparing them to GENIE.

The comparison included GENIE (GRV98) and three additional cross section calculations: CSMS \cite{DIS:CSMS}, BGR \cite{DIS:BGR} and a calculation using the CTEQ6 \cite{DIS:CTEQ6} PDF set. The corresponding cross sections were obtained from the NeutrinoGenerator (based on \cite{NuGen}) and GENIE-HEDIS \cite{DIS:HEDIS} codes. 
We looked at the total and differential cross section $d\sigma/dE\,dy$ of neutrinos and antineutrinos. The cross sections seem to divide in two, with CSMS and CTEQ6 giving very similar results, and BGR and GENIE in rough agreement. The largest differences are observed in the CC cross section of antineutrinos. 

The differences were traced back to the light sea quark contributions to the cross section with the $s$-quark content being the most different among calculations. Both CSMS and CTEQ6 have a significantly larger contribution from these quarks, which results in a noticeable difference for the antineutrino cross sections.  In CSMS the overall interaction rate of antineutrinos is higher by about 10\%, with those events coming mainly from interactions with high inelasticity, defined as $y=E_\mathrm{had}/E_\nu$. Neutrino cross sections, on the other hand, are dominated by valence quarks at our energies, and thus are largely unaffected by changes to the sea quarks contribution. 

\begin{figure*}[!t]
    \centering
        \includegraphics[width=0.45\textwidth]{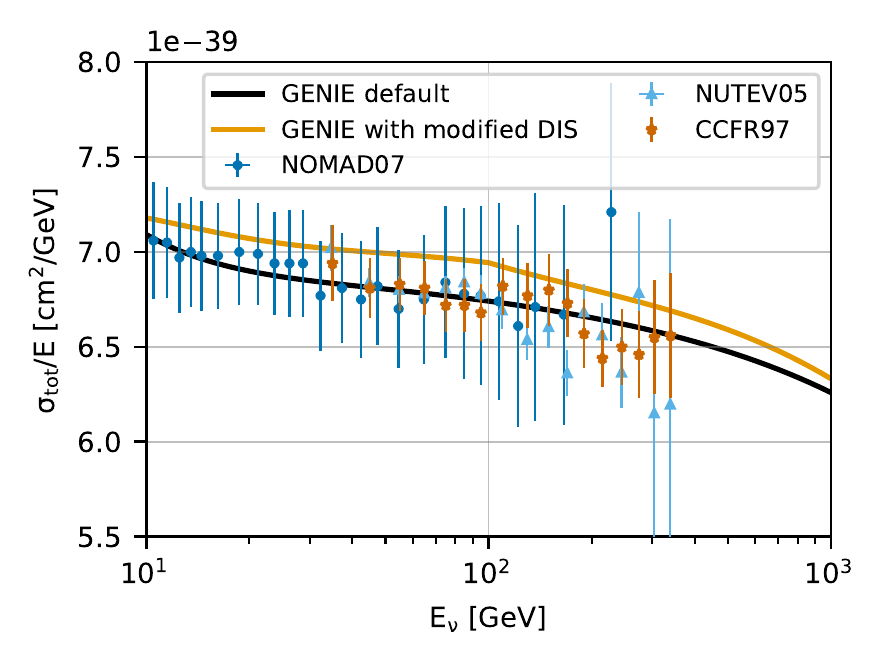}
        \includegraphics[width=0.45\textwidth]{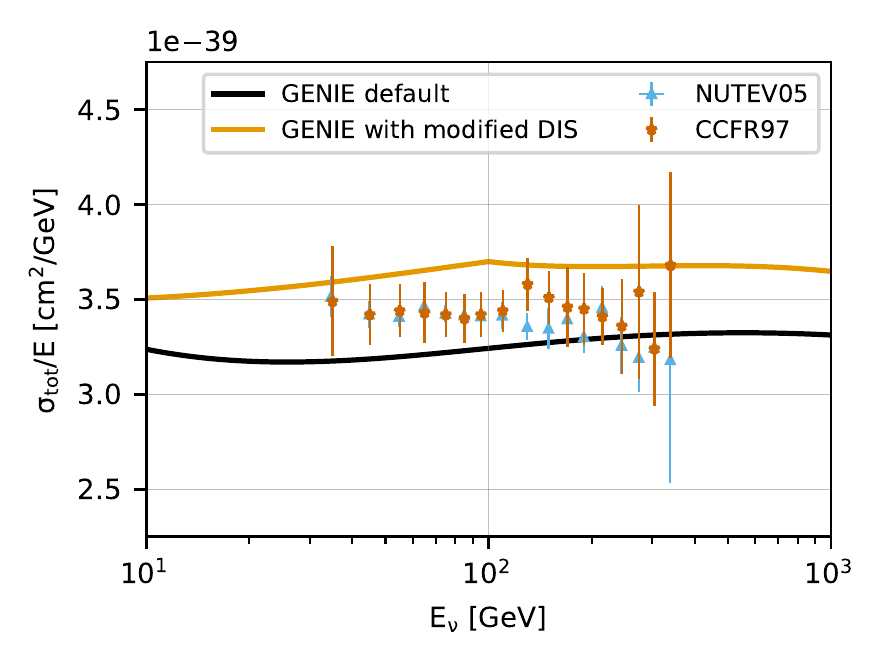}
  
  \caption{Inclusive total neutrino-nucleon cross section for neutrinos (left) and antineutrinos (right) on an isoscalar target (black line) from GENIE, compared to measurements from CCFR~\cite{DIS:tot_xsec_data:CCFR}, NUTEV~\cite{DIS:tot_xsec_data:NUTEV} and NOMAD~\cite{DIS:tot_xsec_data:NOMAD}.  The GENIE cross section with its DIS fraction fully converted to ``CSMS mode" (brown line) is also shown.
  \label{fig:DIS_uncertanty_total_xsec}}
\end{figure*}

The difference between CSMS and GENIE, the largest one observed, was parameterized as function of energy and inelasticity in a single parameter added to the analysis that can reweight every event in our simulation to mimic the predictions of the CSMS calculation. The DIS calculations discussed, however, are not valid below 100~GeV because of their limited $Q^2$ range. Without further information on how to extrapolate this effect to lower energies, we decided to avoid discontinuities in the DIS cross section by applying the same correction derived at 100~GeV for DIS events at lower energies and thus guaranteeing consistency in our approach. We tested other extrapolation methods, such as using the correction at 50~GeV or using the last few energy points to come up with a linear function to go to lower energies, and found that the analysis presented here is robust against this choice. 

In our parameterization, a value of DIS CSMS equal to zero corresponds to the cross section in GENIE, while a value of one approximates the prediction from CSMS. To evaluate the data we set a prior centered at zero and an uncertainty of 1.0 to ensure the fit is stable while allowing the parameter to choose between models.  The parameter that controls these changes can be converted to a fractional value to end up with a cross section in between the models. Comparisons of the resulting total cross section from GENIE and the \textit{CSMS-like} modification are shown on Fig.~\ref{fig:DIS_uncertanty_total_xsec}, compared to data. The correction brings the flux-averaged inelasticity from 0.40 at $E_\nu$=40~GeV in GENIE to 0.47 when corrected.

Additional studies were performed to test the impact of the parameters used in the Bodek-Yang model to extrapolate PDFs in the low $Q^2$ region and on the impact of the hadron multiplicity resulting from these interactions, as described in~\cite{IceCube:2019dqi}. We also considered the impact of nuclear effects, which have been demonstrated to have an impact on neutrino-nucleon cross sections~\cite{Klein:2020nuk}, in particular on the inelasticity of neutrino interactions with water. We found these effects to be either negligible or fully degenerate with existing parameters and therefore are not included explicitly in the fit.

\subsubsection{Non-DIS interactions}

The two main non-DIS processes that contribute to events in our sample are resonance production (RES) and charged current quasielastic scattering (CCQE). Similar to what was done in previous studies~\cite{Aartsen_2018,IceCube:2019dqi}, two systematic parameters are included to account for uncertainties in the nucleon form factors for each interaction. Both these form factors have a dependency on $Q^2$ of the form\\

\begin{equation}
    F(Q^{2}) \propto \frac{1}{1-(Q^{2}/M_{A}^{2})^{2}},
\end{equation}
where $M_{A}$ is the \textit{axial mass}, an effective parameter that can be measured experimentally. GENIE allows one to  re-weight the cross section to different values for different interaction types, so we have introduced them as free parameters in the fit: $M_{A}^{CCQE}$ and $M_{A}^{CCRES}$. Altering these parameters can change the total cross section predictions for CCQE/CCRES interactions and hence changing the expected number of CCQE/CCRES events seen in the sample.

\subsubsection{$\nu_\tau$ charged current cross section}

The charged current cross section of tau neutrinos differs from that of other flavors because the mass of the tau lepton is comparable to the target nucleon mass and the neutrino energies being studied. In particular, the terms of the form $\frac{m_{\mathrm{lepton}}^2}{2 M_{\mathrm{nucleon}} E_{\nu}}$, which are heavily suppressed for light lepton ($e$ or $\mu$) cases at all energies, become relevant for tau neutrinos up to tens of GeV~\cite{Jeong:2010nt}. A summary of various inclusive cross section calculations is discussed in~\cite{Conrad:2010mh}, where a proposed parameterization that can move between calculations is also given. We tested the impact of these changes in our study, which has a very small component of $\nu_\tau$, and found them to be well below our threshold, so this source of uncertainty was not included in the final result.

\subsection{Atmospheric muon contamination}
\label{sec:atm_mu}
 
The event selection was designed to eliminate atmospheric $\mu$ with high efficiency. This is achieved for this subset of the data, with an estimated contamination from simulation studies of $\sim2$\%. However, due to the efficient rejection and the computational requirements of simulating this background, few simulated events survive to the analysis binning. Therefore, even though the atmospheric muon fluxes should be correlated with those of neutrinos, changes to how they are modeled are mostly negligible in comparison to the statistical uncertainty of the simulation. Most of their variation is captured by a scaling factor, so a single parameter that globally changes their relative contribution to the sample is used (see Atm. $\mu$ scale in Table~\ref{table:syst_param}).

The few atmospheric $\mu$ events that are still present at the final level of the selection produce histograms that are sparsely populated, with large bin-to-bin fluctuations. To overcome this issue, we applied a variable bandwidth Kernel Density Estimator (KDE) \cite{KDE} to the binned atmospheric $\mu$ data, which results in smooth expectation histograms for the nominal data set as well as for variations produced with modified detector conditions. The error of this procedure was estimated adapting the theoretical upper bound variance computation described in \cite{KDE}, adjusting it so that their hypersurface fits would yield a reduced $\chi^2=1$. This conservative estimation of the error for the KDE procedure is included in the analysis as $(\sigma^\mu_i)^2$ in Eq.~\ref{eq:mod_chi2}.

\subsection{Oscillation parameters~\label{section:syst_osc}}
Oscillation probabilities are computed using a three-flavor scheme that includes the coherent forward scattering experienced by neutrinos as they cross Earth's matter. This is implemented in a custom python code~\cite{pisa} that follows the description in~\cite{PhysRevD.22.2718}. The computation requires all oscillation parameters to be defined (three mixing angles, two mass differences squared, and an imaginary phase), as well as the electron density profile of the Earth.

We approximate the Earth as a collection of 12 radial layers of constant matter density, following the Preliminary Earth Reference Model (PREM)~\cite{DZIEWONSKI1981297}. The electron-to-nucleon fraction can change depending on the chemical composition of the layers. Here we use the values proposed in~\cite{Rott:2015kwa}, with 0.4656 for the inner and outer core, and 0.4957 for the mantle. Varying these fractions, as well as including a finer radial description of the Earth, results in negligible changes, so they are kept fixed in the fit to the data.

We also find that this analysis is insensitive to the values of $\theta_{12}$, $\Delta m^{2}_{21}$ and $\theta_{13}$, within their current uncertainties. Therefore, we fix their values to recent global fit results from~\cite{Esteban_2019}, with $\theta_{12}=33.82^{\circ}$, $\Delta m^{2}_{21}=0.739\times10^{-5}$~eV$^{2}$ and $\theta_{13}=8.61^{\circ}$. These are the global fit results obtained under the normal neutrino mass ordering hypothesis ($m_{1} < m_{2} < m_{3}$), which is assumed throughout this analysis as well. A separate, more sensitive analysis to determine the neutrino mass ordering using the full DeepCore data set is currently underway. The analysis is also insensitive to the choice of a CP-violating phase, so we fix $\delta_{CP}=0$.

\section{Results\label{sec:results}}

This analysis was performed in a ``blind'' manner, such that all choices regarding the event selection criteria, analysis binning and implementation of systematic uncertainties are made prior to fitting the real detector data in order to avoid biasing the results. After finalizing all analysis choices and procedures, we perform a fit to real data by minimizing the modified $\chi^{2}$ test statistic defined in Eq.~\ref{eq:mod_chi2} over 200 bins. This yields the best-fit values of 
\begin{equation*}
\begin{split}
\sin^2\theta_{23} &= 0.51 \pm 0.05,\;\mathrm{and}\\
\Delta m^2_{32} &= (2.41 \pm 0.07) \times 10^{-3}\,\mathrm{eV}^2,
\end{split}
\end{equation*}
assuming normal neutrino mass ordering (NO). The 68\% C.L. for each parameter is derived following the Feldman-Cousins prescription~\cite{PhysRevD.57.3873}. The best fit nuisance parameter values are reported in Table~\ref{table:syst_param}, while Table~\ref{table:final_event_count} shows the observed and expected number of events at the best-fit point. 

\begin{figure}[!bt]
    \begin{center}
        \includegraphics[width=0.45\textwidth]{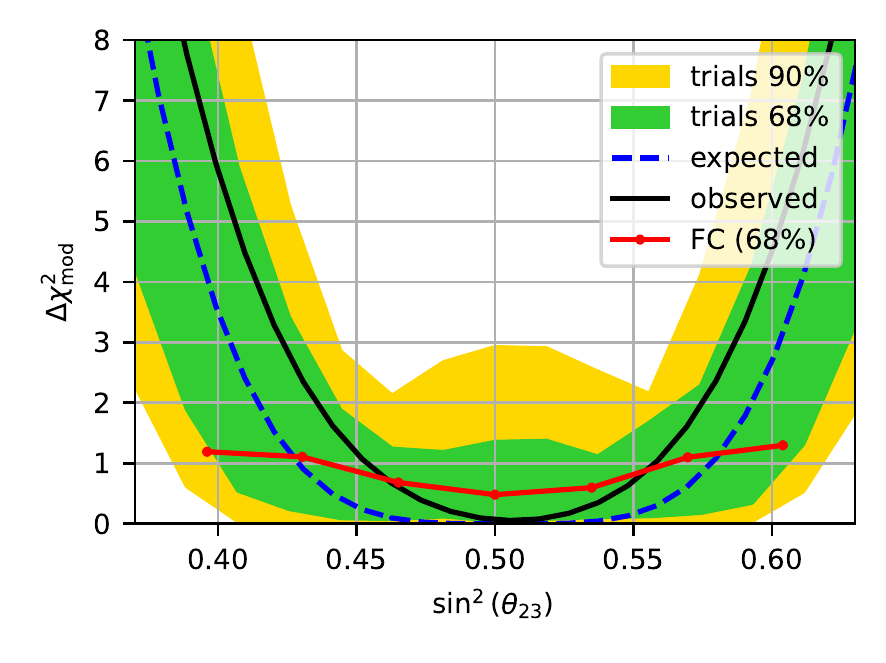}
    \includegraphics[width=0.46\textwidth]{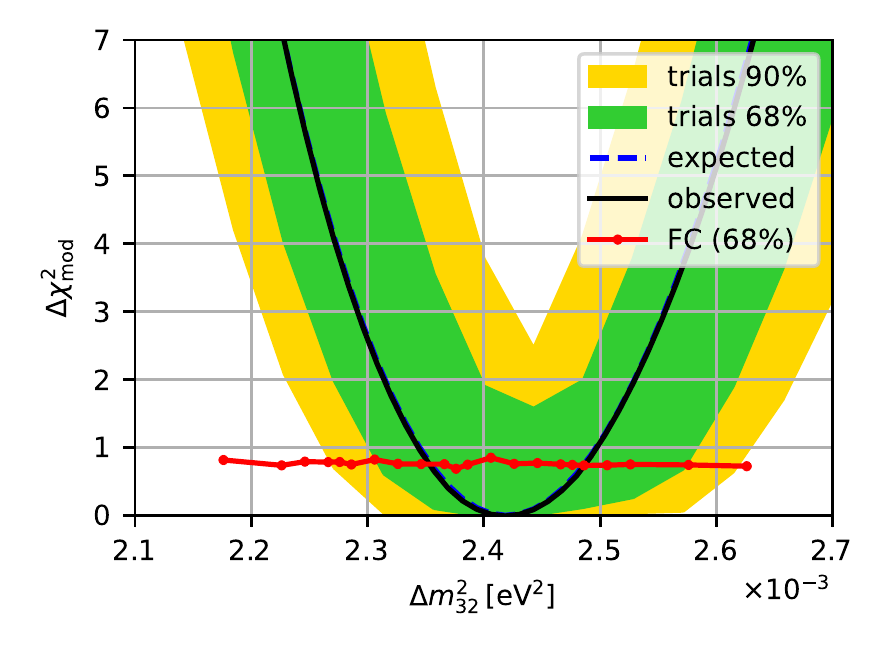}
    \caption{Observed $\Delta \chi^2_{\mathrm{mod}}$ (solid) compared to the expectation (dashed) and the distribution of 1000 pseudo-data trials (yellow and green bands) produced at the best fit point of the analysis for the atmospheric (top) mixing angle and (bottom) mass splitting. The red lines indicate the 68\% C.L. derived using the Feldman-Cousins method~\cite{PhysRevD.57.3873}.}
    \label{fig:brazil_bands}
    \end{center}
\end{figure}

\begin{table}[!b]
\centering
\caption{Best-fit number of events with 7.5 years of livetime for each neutrino flavor and interaction type, as well as atmospheric $\mu$, along with the observed counts from the data. The rate is also given for comparison to other experiments.}
\label{table:final_event_count}
\begin{tabular}{p{0.4\linewidth} r| r} \toprule
\hline
Type  &  Events & Rates [1/$10^6$s] \\ \midrule
$\nu_{\mu}+ \bar{\nu}_{\mu}$ CC &  17656 & 75.03\\
$\nu_{e}+ \bar{\nu}_{e}$ CC &  1820 & 7.74\\
$\nu_{\tau}+ \bar{\nu}_{\tau}$ CC &  603 & 2.56 \\
$\nu_{all}+ \bar{\nu}_{all}$ NC &  1222 & 5.19\\
Atmospheric $\mu$ &  711 & 3.02 \\
\hline
Total (best-fit) &  22012 & 93.54\\
Observed & 21914 & 93.08\\ \bottomrule
\end{tabular}
\end{table}

To assess the goodness-of-fit, we perform 1000 fits to pseudo-data trials that are generated by Poisson-fluctuating the expectation for neutrinos and atmospheric muons in the analysis binning, given the best fit values for all parameters. Using the resulting distribution of test statistics from these 1000 trials, we find that our observed $\chi^2_{\mathrm{mod}}$ has a p-value of 26.1\%, indicating good agreement between simulation and data. Figure~\ref{fig:brazil_bands} shows the expected and observed $\Delta \chi^2_{\mathrm{mod}}$ in $\sin^2\theta_{23}$ and in $\Delta m^2_{32}$, overlaid with the distribution of 1000 pseudo-data trials. The observed contours are fully contained within the $1\sigma$ fluctuations of the trials.   

\begin{figure}
  \centering
  \includegraphics[width=0.95\linewidth]{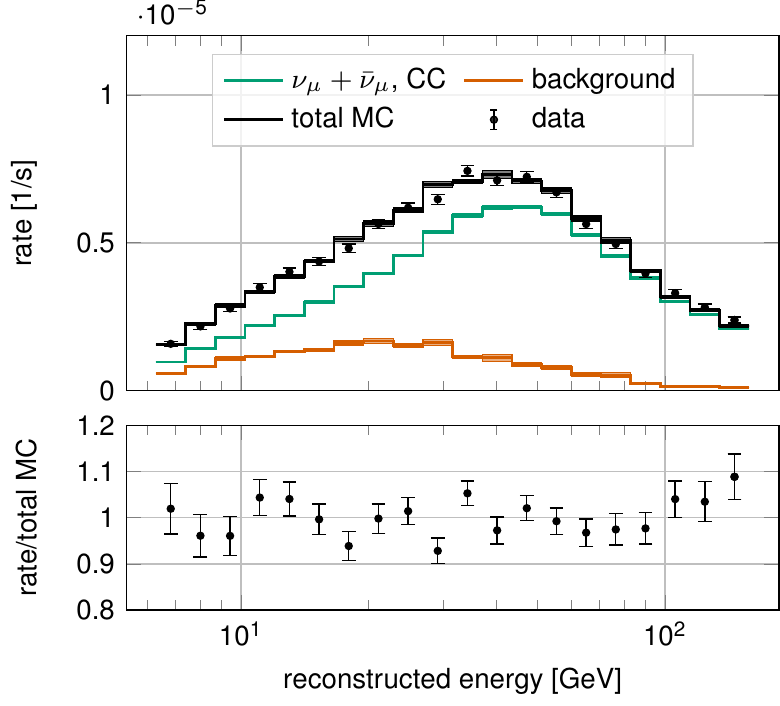}

  \includegraphics[width=0.95\linewidth]{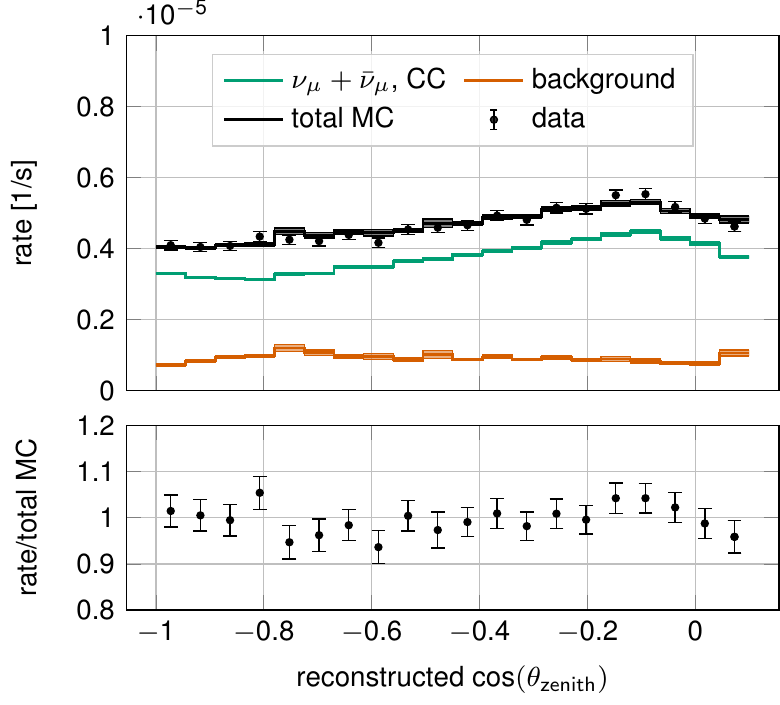}

  \includegraphics[width=0.95\linewidth]{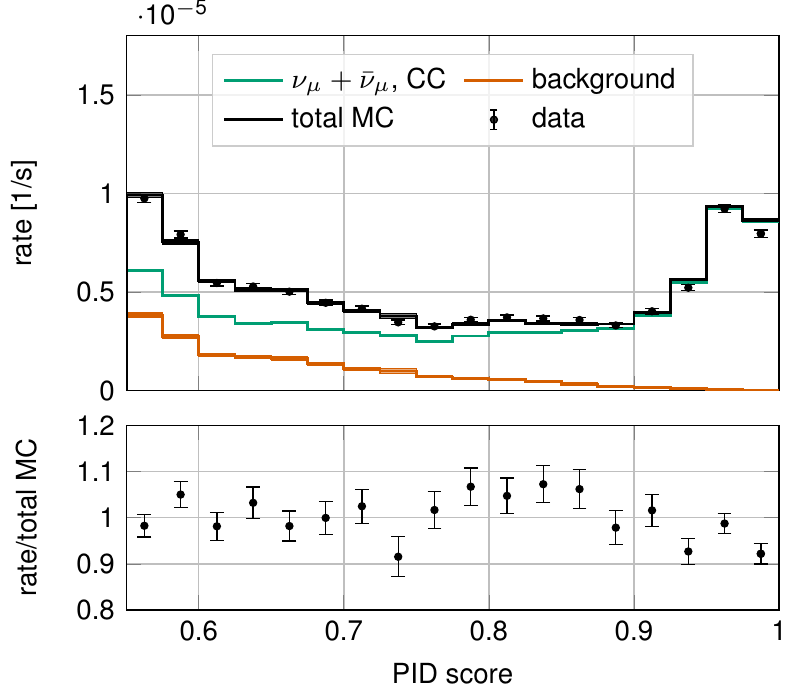}

  \caption{Distributions of reconstructed (top) energy, (middle) cosine zenith and (bottom) PID score for data compared to the best-fit simulation. Background includes atmospheric $\mu$ and all neutrino types besides $\nu_{\mu}+\bar{\nu}_{\mu}$ CC events for which this analysis was optimized.}
  \label{fig:postfit_data}
\end{figure}

\begin{figure}[!t]
  \centering
  \includegraphics[width=0.95\linewidth]{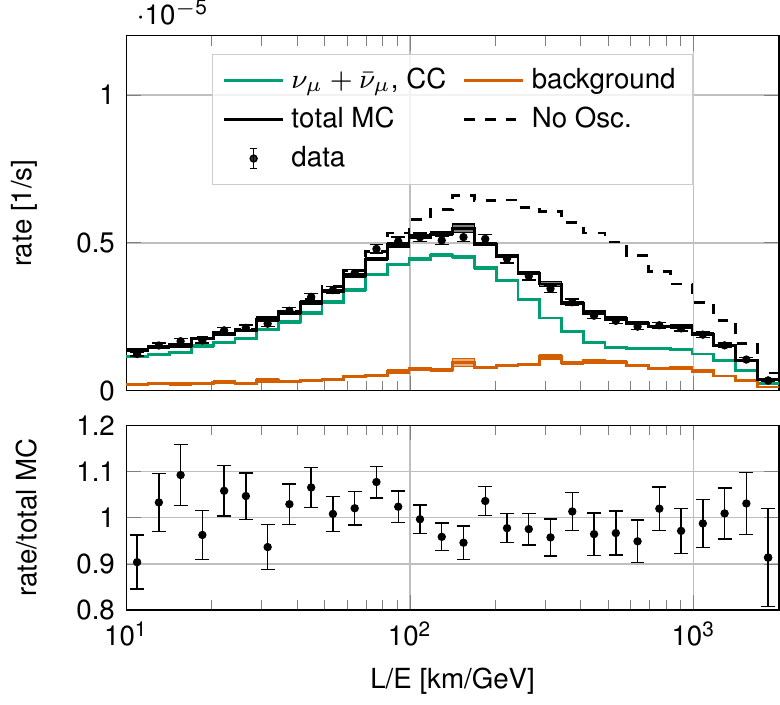}

  \caption{The L/E distribution for the best-fit expectations overlaid with the observed data. Background includes atmospheric $\mu$ and all neutrino types besides $\nu_{\mu}+\bar{\nu}_{\mu}$ CC events. The expectation at the best fit but without oscillations is shown as a dashed line.}
  \label{fig:l_over_e}
\end{figure}

The distributions of reconstructed neutrino energy, reconstructed zenith angle and the PID score for data compared with the best-fit MC simulation are shown in Fig.~\ref{fig:postfit_data}. These 1D projections show the data binned more finely than what was used in the fit by a factor of 2 for reconstructed energy and zenith, and a factor of 10 for the PID. Figure~\ref{fig:l_over_e} shows the reconstructed energy and cosine zenith projected into L/E space for all events. The best-fit expectation shows good agreement with the data for all observables. 

We find a slight excess of events compared to MC in the very highest energy bin, which is nevertheless consistent within statistical fluctuations. However, this analysis extends to higher energies than previous oscillation analyses using DeepCore data in order to better constrain systematic uncertainties in this off-signal region. Therefore as a cross-check, we fit the data again with this last, highest energy bin removed and observe a negligible shift of $\sim$1\% in $\Delta m^2_{32}$ and $\sim$0.2$^{\circ}$ in $\theta_{23}$. We further performed fits to data from each season of data-taking independently, and find that the resulting best fit atmospheric oscillation parameters and nuisance parameters are all statistically compatible across each year. 

Additional cross-checks are performed with simulation to assess the robustness of the result against perturbations to the detector model that are not parameterized by the hypersurface treatment described in Section~\ref{section:syst_det}. These include a modification to the wavelength dependence of the HQE PMTs to match alternative laboratory measurements, an implementation of the cable that shadows part of the photocathode for particular azimuthal angles~\cite{IceCube:2021lmu}, and several different bulk ice models that were developed after this analysis was finalized, which were found to fit equally-well or better to the LED calibration data. 

The simulation generated for each of these perturbations is processed through the standard event selection and used to generate pseudo-data that is weighted using the best-fit values from the original fit to data, and fit with the standard analysis procedure. No significant bias in the fitted oscillation parameters is observed for any of these perturbations. We note that the largest shift observed in the best fit point with a significance of 0.3$\sigma$ was found when using simulation produced with the birefringent ice properties incorporated in the bulk ice model~\cite{ICRC19:anisotropy}. While insignificant for this analysis, this points to the potential need for an improved treatment of systematic uncertainties related to the glacial ice optical properties for analyses with higher statistics and more scattered photons used in the reconstruction of neutrino properties.

The hypersurface treatment was also validated by generating simulation using the best-fit nuisance parameter values for detector systematics. There were no significant chnages obtained in our results after including this set in the analysis.

\begin{table}[!t]
\begin{tabular}{p{0.35\linewidth} p{0.3\linewidth} p{0.3\linewidth}}
\toprule
\hline
Systematic group & $\delta(\Delta m^{2}_{32})$ [\%]   & $\delta(\sin^{2}\theta_{23})$ [\%]  \\
\midrule
Detector         & -33.6                  & -10.6                   \\
Flux             & -5.4                   & -1.4                    \\
Cross section    & -6.8                   & -0.3    \\
$A_{eff}$ scale       & -1.0                   & -0.4                    \\
Atm. $\mu$ scale & -1.8                 & -1.1                    \\
\bottomrule
\end{tabular}
\caption{Relative change in $1\sigma$ uncertainty assuming perfect knowledge of each group of systematic uncertainties. Relative change is calculated from the width of the 68\% C.L. assuming Wilks' theorem.} 
\label{tab:error_budget}
\end{table}

\begin{figure}[b] 
    \centering
    \includegraphics[width=0.99\linewidth]{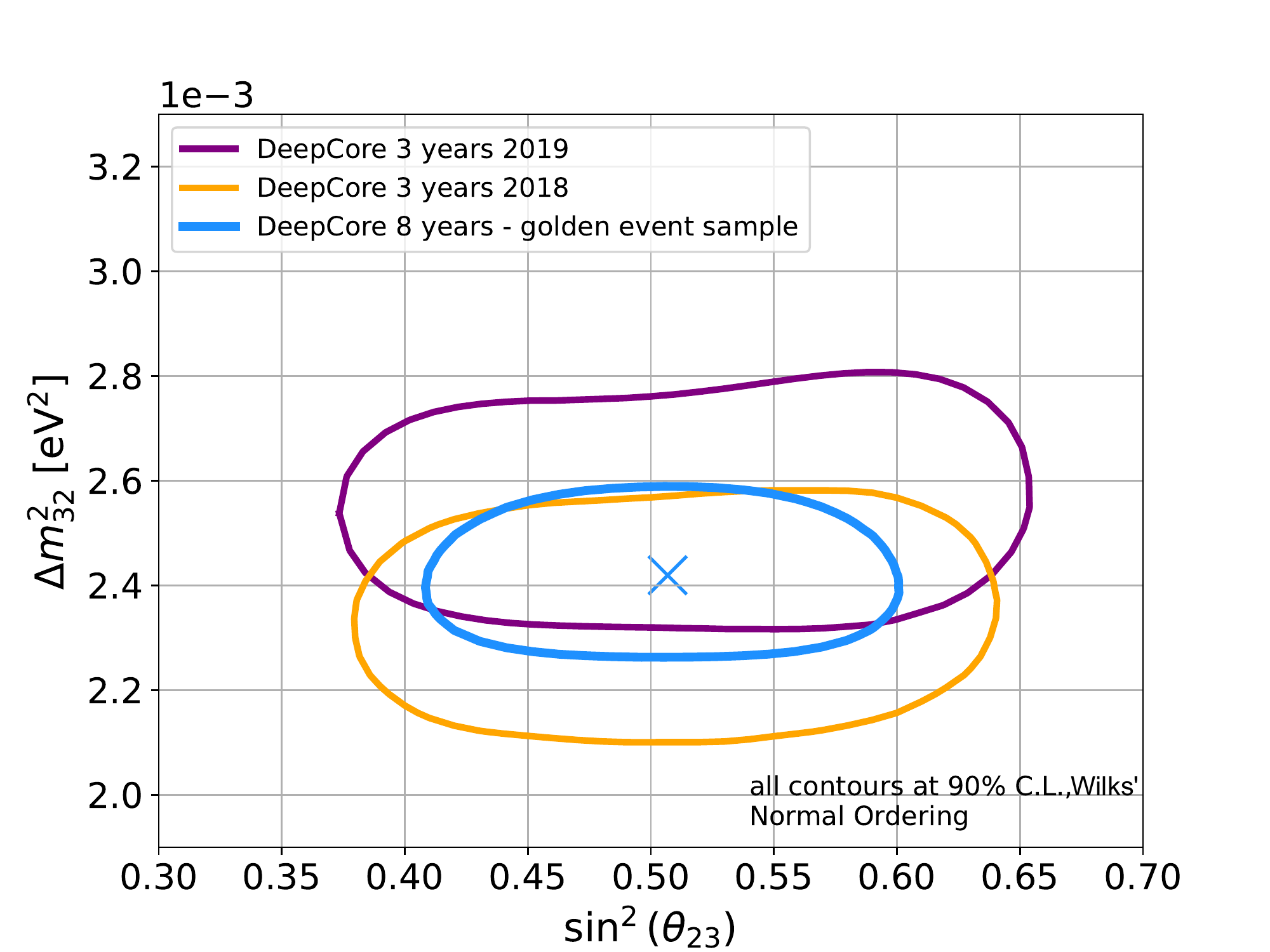}
  \caption{Contours showing the 90\% C.L. allowed region for $\Delta m^{2}_{32}$ and $\sin^{2}\theta_{23}$ from this study (blue) compared to previous IceCube DeepCore results\cite{Aartsen_2018,IceCube:2019dqi}. All confidence intervals shown are derived assuming Wilks' theorem for a consistent comparison.
  \label{fig:compare_to_DC}}
\end{figure}

\begin{figure*}[!t] 
    \centering
        \includegraphics[width=0.75\linewidth]{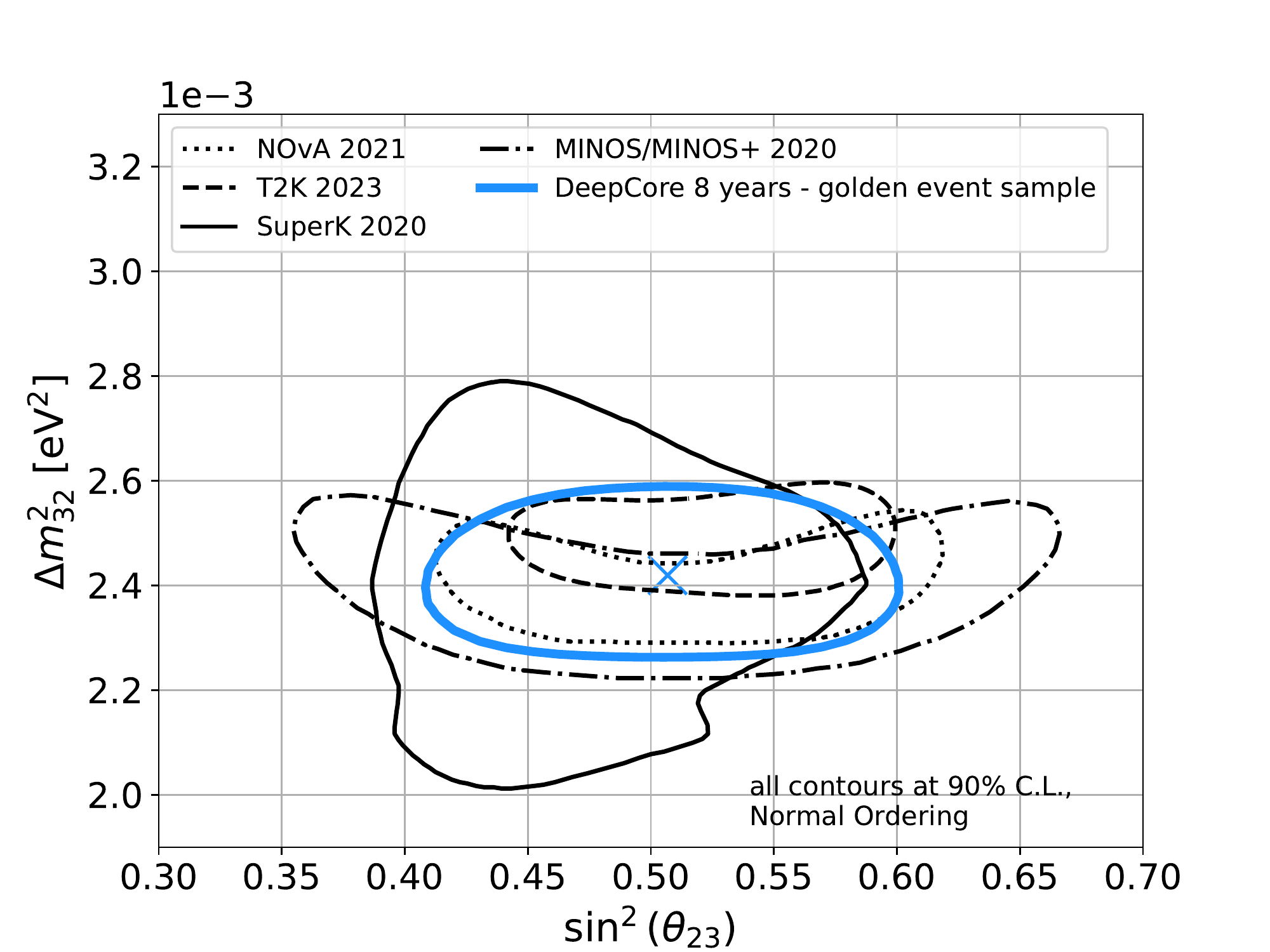}
  \caption{Contours showing the 90\% C.L. allowed region for the atmospheric neutrino oscillation parameters from this study (blue) compared to results from MINOS~\cite{MINOS:2020llm}, NOvA~\cite{NOvA:2021nfi}, Super-Kamiokande~\cite{pablo_fernandez_2021_5779075} and T2K~\cite{T2K:2023smv}. Daya Bay also measures $\Delta m^2_{32}$ in conjunction to $\theta_{13}$, but the results cannot be displayed in the format above~\cite{DayaBay:2012fng}. The DeepCore confidence interval is derived assuming Wilks' theorem.
  \label{fig:real_data_contour}}
\end{figure*}

Table~\ref{tab:error_budget} shows the relative contributions of each group of systematic uncertainties that were considered in this study to the total error. To determine the contributions, we assume perfect knowledge of each group of systematic uncertainties in the fit and calculate the relative change in the width of the 68\% CL for each oscillation parameter. The detector systematic uncertainties contribute most to the total uncertainty, with flux and cross-section systematics contributing far less to the error budget. The atmospheric $\mu$ normalization term has little effect thanks to the very small contamination in the final sample, while the overall normalization term for neutrinos has almost no effect on the error of the measurement. All sensitivity to the oscillation parameters comes from the shape of the oscillation pattern, with the total flux being relatively unimportant for this measurement. This breakdown of uncertainties also demonstrates that the measurement can still be largely improved by increased statistics.

The 90\% C.L. allowed region for the atmospheric oscillation parameters is shown in Fig.~\ref{fig:compare_to_DC} compared to previous measurements using IceCube DeepCore. All contours are derived assuming Wilks' theorem, which leads to slight over coverage due to the physical boundary of $\theta_{23}$ (see Fig.~\ref{fig:brazil_bands}). Taking into account the previously published Feldman Cousins-corrected 1$\sigma$ errors~\cite{Aartsen_2018}, we observe an improvement of 44\% and 37\% in the measurements of $\Delta m^{2}_{32}$ and $\sin^{2}\theta_{23}$, respectively. These new results therefore represent the most precise measurement of these parameters using atmospheric neutrinos to date. 

Figure~\ref{fig:real_data_contour} shows the new IceCube DeepCore result in comparison to measurements performed by other experiments, using both accelerator and atmospheric neutrinos. MINOS ~\cite{MINOS:2020llm}, T2K~\cite{T2K:2023smv} and NOvA~\cite{NOvA:2021nfi} measure these parameters using neutrinos produced in particle accelerator facilities with energies between hundreds of MeV to and a few GeV. Super-Kamiokande~\cite{pablo_fernandez_2021_5779075} uses atmospheric neutrinos, but the bulk of their statistics are around the 1~GeV region. With IceCube DeepCore we use neutrinos with energies higher than any of these experiments, interacting mainly via deep inelastic scattering, and are therefore subject to different interaction uncertainties. While our neutrino source is also the atmosphere and we use the same nominal flux calculation as Super-Kamiokande, we see a different region of the atmospheric neutrino spectrum and we include several flux-related nuisance parameters in the fit to adjust for discrepancies with data. Given the differences on how these measurements are obtained, the overlap between the results is noteworthy, but difficult to rigorously quantify using individually reported uncertainties without resimulations accounting for both correlated and uncorrelated uncertainties. This could be followed up with future studies using external data releases from each experiment.

\section{Conclusion\label{sec:conclusion}}
We have presented the most precise measurement of oscillations of atmospheric neutrinos to date, using a newly calibrated and filtered data sample from IceCube DeepCore. The measurement was made possible thanks to state-of-the-art calibrations, improved event filtering, and significant improvements to our methods for evaluating sources of uncertainty. 

The recent calibration efforts allow us to successfully describe our data with high precision, especially the response of sensors to single photons and better description of the optical properties of the ice. The data used spans a period of 8 years, over which we find stable noise rates and detector response. 

The new event selection suppresses the main sources of background by a factor greater than 10$^4$, while keeping over 20\% of all neutrinos that interact in the detector volume, resulting in a comparable rate of atmospheric $\mu$ and $\nu$ at Level 5, which is a common starting point for future DeepCore analyses. We further developed an analysis-specific selection focused on events detected with minimally scattered light, which reduced the background further to achieve a neutrino purity of 98\%, with a high fraction of $\nu_{\mu}$ CC events. 

We investigated many possible sources of uncertainty, and developed improved methods for implementing them in the analysis. The detector related uncertainties, known to have the largest impact on our results, were studied using significantly more simulation sets for known effects than in previous studies, and were parameterized in a way that accounts for correlations with other parameters. The uncertainties due to the atmospheric $\nu$ flux were significantly expanded so that possible variations to the flux were corrected for as part of the fit. Neutrino-nucleon cross sections were also evaluated with in different frameworks, where the free parameters have a more physical interpretation compared to previous analyses. The impact of atmospheric $\mu$, which has remained a challenge throughout these studies, was reduced by limiting their fraction in the final sample, making any changes in their prediction negligible.

The results presented here are consistent with measurements performed using human-made accelerator neutrino experiments~\cite{MINOS:2020llm, T2K:2023smv, NOvA:2021nfi}. At the same time, these results are obtained using much higher energy neutrinos, and are therefore insensitive to  $\delta_{CP}$, as well as many of the cross-section uncertainties that are critical to understand for accelerator neutrino experiments. Our results are of similar precision to accelerator measurements thanks to the enormous flux of neutrinos that are provided by cosmic ray interactions, and also the ability to constrain the oscillation valley across several distinct bins in L/E. Our measurements therefore provide an important, complementary probe of the parameters $\theta_{23}$ and $\Delta m^{2}_{32}$ within the context of the global neutrino oscillation landscape.

Further improvements to our oscillation measurements are currently underway on three fronts. First, future analyses of DeepCore data will start from the common event sample described herein, and employ more complex and resource-intensive reconstruction strategies in order to retain more signal events, enhance the neutrino flavor identification, and improve the directional and energy resolutions for cascade-like events in particular~\cite{lowen_reco}. Second, we continue to improve the accuracy of our simulation by testing and including higher order effects, such as direct simulation of the hole ice and the shadow that the cable casts on the modules, which are becoming more relevant as our statistical precision increases. Finally, the IceCube Upgrade detector is planned for installation in 2025/26. This detector will serve as an in-fill to the existing IceCube DeepCore array, and significantly enhance our ability to detect and reconstruct GeV-scale neutrino interactions~\cite{Ishihara:2019aao}. With these improvements we will continue to provide high precision measurements of neutrino oscillations at the highest energies and over the longest baselines.

\begin{acknowledgments}

The authors gratefully acknowledge the support from the following agencies and institutions: USA – U.S. National Science Foundation-Office of Polar Programs, U.S. National Science Foundation-Physics Division, U.S. National Science Foundation-EPSCoR, Wisconsin Alumni Research Foundation, Center for High Throughput Computing (CHTC) at the University of Wisconsin–Madison, Open Science Grid (OSG), Advanced Cyberinfrastructure Coordination Ecosystem: Services \& Support (ACCESS), Frontera computing project at the Texas Advanced Computing Center, U.S. Department of Energy-National Energy Research Scientific Computing Center, Particle astrophysics research computing center at the University of Maryland, Institute for Cyber-Enabled Research at Michigan State University, and Astroparticle physics computational facility at Marquette University; Belgium – Funds for Scientific Research (FRS-FNRS and FWO), FWO Odysseus and Big Science programmes, and Belgian Federal Science Policy Office (Belspo); Germany – Bundesministerium für Bildung und Forschung (BMBF), Deutsche Forschungsgemeinschaft (DFG), Helmholtz Alliance for Astroparticle Physics (HAP), Initiative and Networking Fund of the Helmholtz Association, Deutsches Elektronen Synchrotron (DESY), and High Performance Computing cluster of the RWTH Aachen; Sweden – Swedish Research Council, Swedish Polar Research Secretariat, Swedish National Infrastructure for Computing (SNIC), and Knut and Alice Wallenberg Foundation; European Union – EGI Advanced Computing for research; Australia – Australian Research Council; Canada – Natural Sciences and Engineering Research Council of Canada, Calcul Québec, Compute Ontario, Canada Foundation for Innovation, WestGrid, and Compute Canada; Denmark – Villum Fonden, Carlsberg Foundation, and European Commission; New Zealand – Marsden Fund; Japan – Japan Society for Promotion of Science (JSPS) and Institute for Global Prominent Research (IGPR) of Chiba University; Korea – National Research Foundation of Korea (NRF); Switzerland – Swiss National Science Foundation (SNSF); United Kingdom – Department of Physics, University of Oxford.

\end{acknowledgments}

\bibliography{MyBibFile}

\clearpage

\appendix
\section{Event selection cuts~\label{appendix:selection}}
The full list of cuts used for the common event selection, from Level~2 to Level~5, is given in Table~\ref{tab:cut_values}. The description of the algorithms is given in Sec.~\ref{sec:sample}.

\begin{table}[!h]
\begin{tabular}{p{0.6\linewidth}|l}
\hline
\hline
Cut name                                     & Keep events if                               \\ \hline
\textbf{Targeting coincident events} & \\
Uncleaned $t$ length                     & \textless{}13000 ns                     \\ 
Cleaned $t$ length                       & \textless 5000 ns                       \\ \hline
\textbf{Targeting noise} & \\
NChannel cleaned                                           & \textgreater{}= 6                       \\ 
Noise engine                                          & Pass                                    \\ 
MicroCount                                            & \textgreater{}2                         \\ 
DC Fiducial hits                                       & \textgreater{}2                         \\ 
L4 noise classifier                                      & \textgreater{}0.85                       \\ \hline
\textbf{Targeting muons} &\\
NAbove 200                                            & \textless{}10                           \\ 
VertexGuess Z                                         & \textless -120m                         \\ 
Causal veto Hits                                        & \textless{}7                            \\ 
Veto/fiducial hits                               & \textless{}1.5                          \\ 
C2HR6                                                 & \textgreater{}0.37                      \\ 
RTVeto (Nfid \textless 75)                           & \textless{}4                            \\ 
RTVeto (75\textless Nfid\textless 100)              & \textless{}5                            \\ 
L4 muon classifier score                                       & \textgreater{}0.9                      \\ 
Vertex position in $\rho$                                & \textless 150m                          \\ 
Vertex position in $z$                                  & -490m\textless{}$p_z$\textless{}-220m \\ 
$\cos(\Delta \theta)$ (corridor-fit)                      & $\leq$ 0.7 rad                    \\ 
Number of corridor hits                                         & $\leq$ 2                          \\ \hline
\hline
\end{tabular}
\caption{List of variables and cut values used for the common DeepCore events election. The description of the variables is given in Sec.~\ref{sec:sample} as well as in~\cite{IceCube:2019dqi}. \label{tab:cut_values}}
\end{table}

\FloatBarrier
\section{Impact of systematic uncertainties~\label{appendix:systematics}}
\begin{figure}
    \centering
    \includegraphics[width=0.95\linewidth]{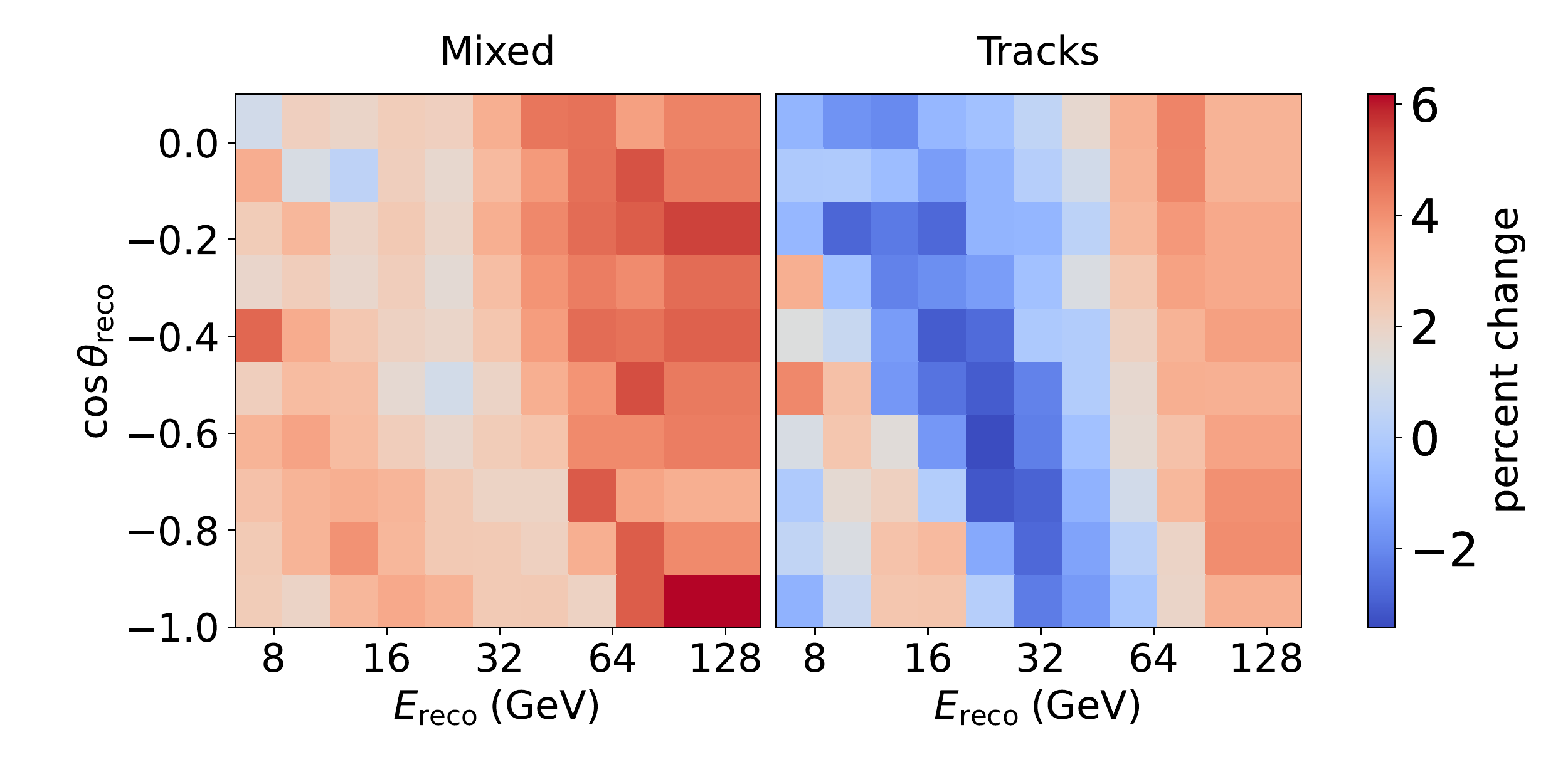}
    \caption{DOM optical efficiency.\label{fig:app_domeff}}
\end{figure}

\begin{figure}
    \centering
    \includegraphics[width=0.95\linewidth]{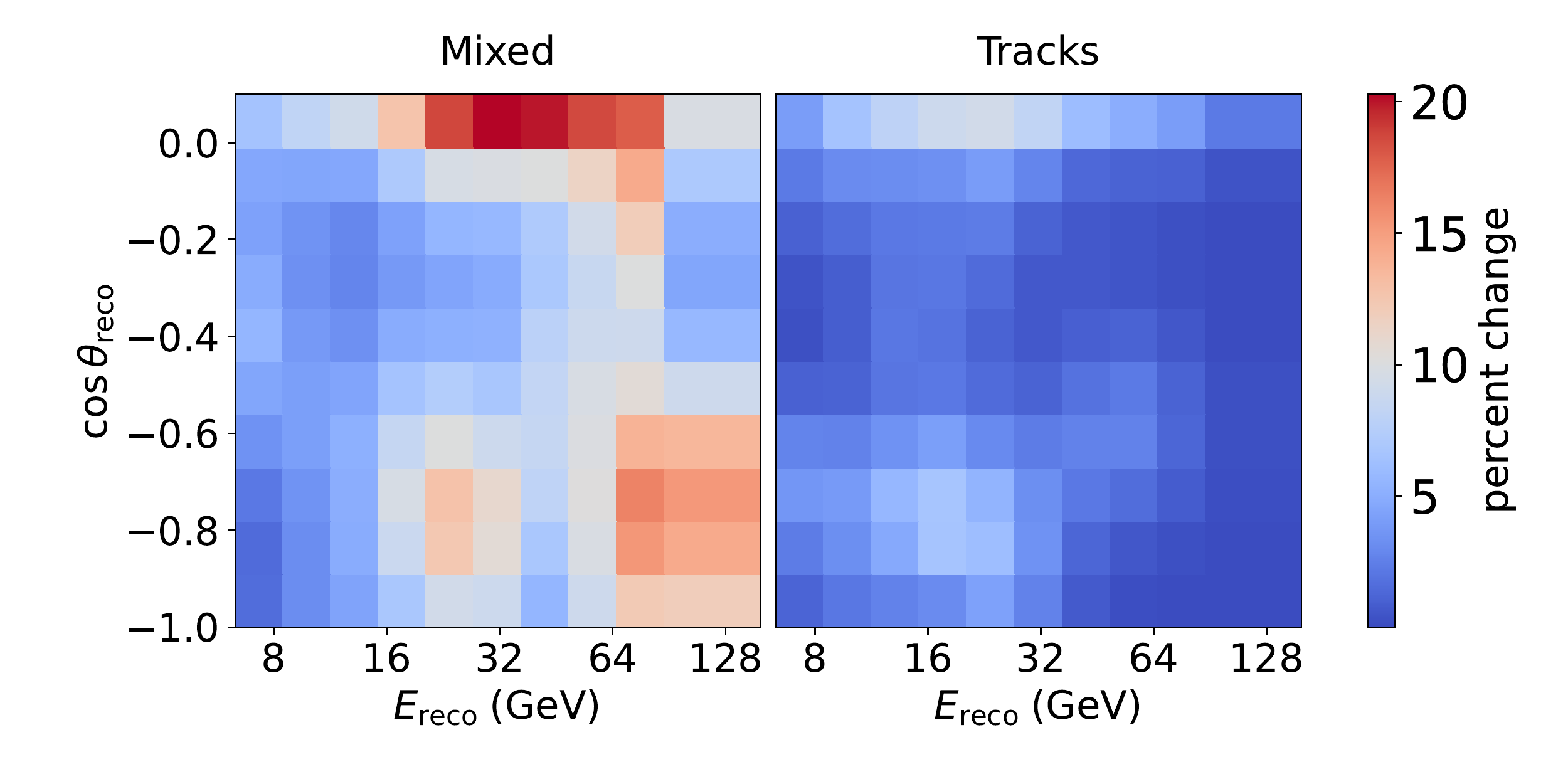}
    \caption{Atmospheric $\mu$ scale.\label{fig:app_muons}}
\end{figure}

Figures~\ref{fig:app_domeff} to~\ref{fig:app_muons} show the percent change in the expected number of events in the analysis histogram when the value of the nuisance parameters included in the study is changed. The values chosen for this depiction correspond to the +1$\sigma$ of the posterior distributions of each parameter as obtained from ensemble fits to pseudo-data and therefore are representative of the scale of the variations that the study is sensitive to. For example, for Fig.~\ref{fig:app_domeff} the DOM optical efficiency was increased by 2\%, which increases the expectation of number of events by up to 6\% in the highest energies of the mixed-channel histogram, while simultaneously reducing the number of events by about 2\% in the track histogram along the region where neutrino oscillations lead to $\nu_\mu$ disappearance.

\begin{figure}
    \centering
    \includegraphics[width=0.95\linewidth]{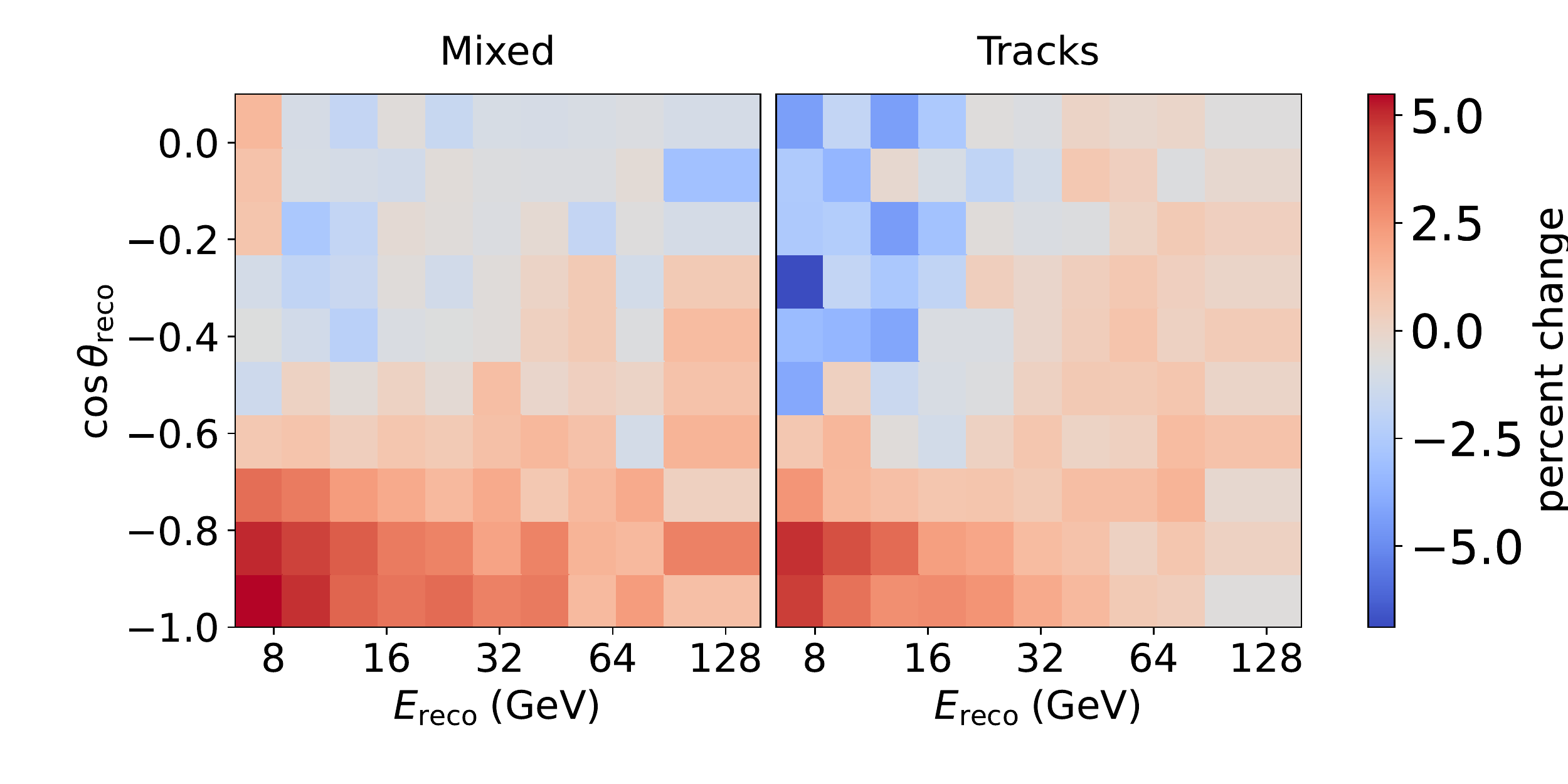}
    \caption{Hole ice p0.}
\end{figure}

\begin{figure}
    \centering
    \includegraphics[width=0.95\linewidth]{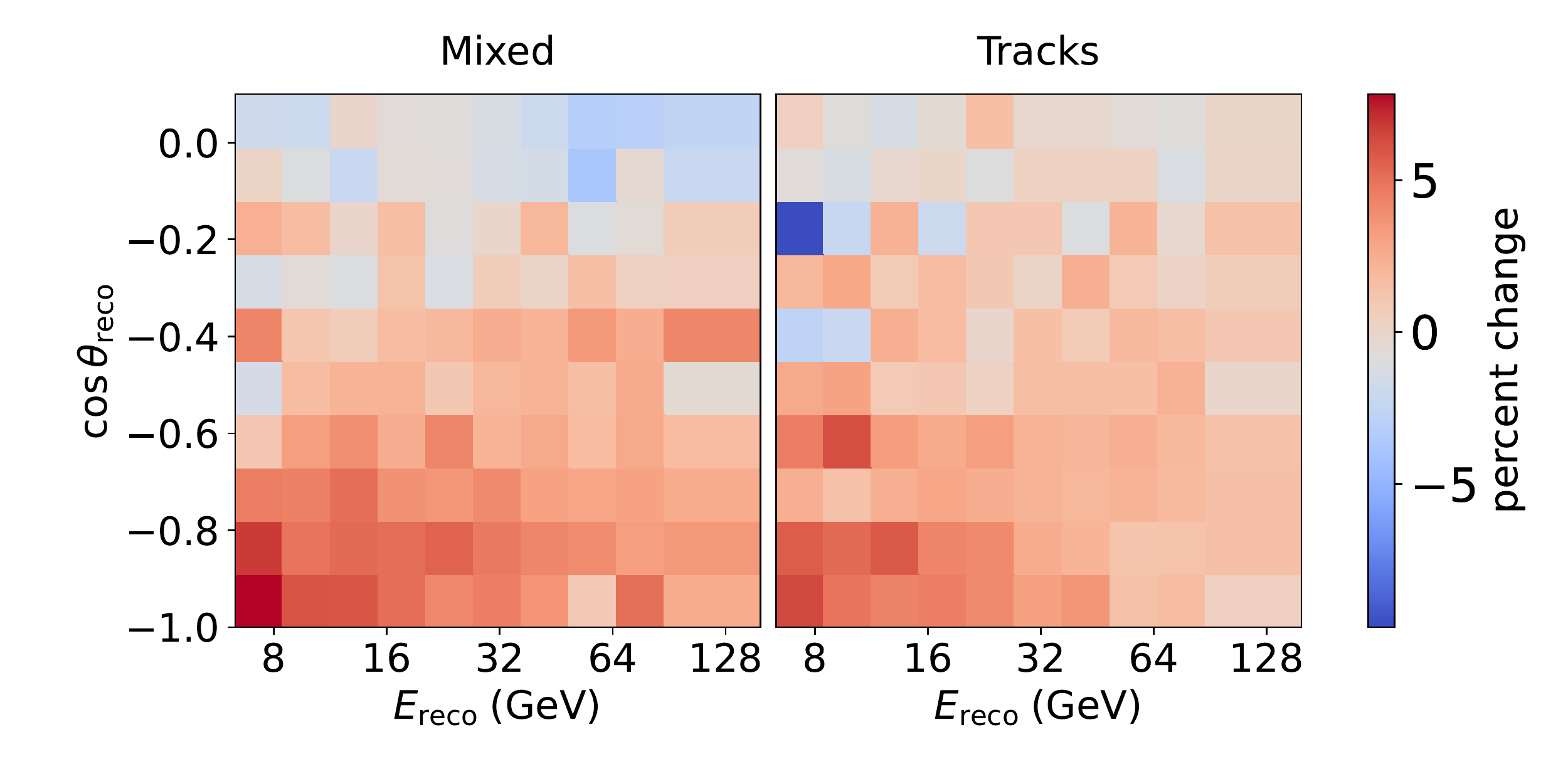}
    \caption{Hole ice p1.}
\end{figure}

\begin{figure}
    \centering
    \includegraphics[width=0.95\linewidth]{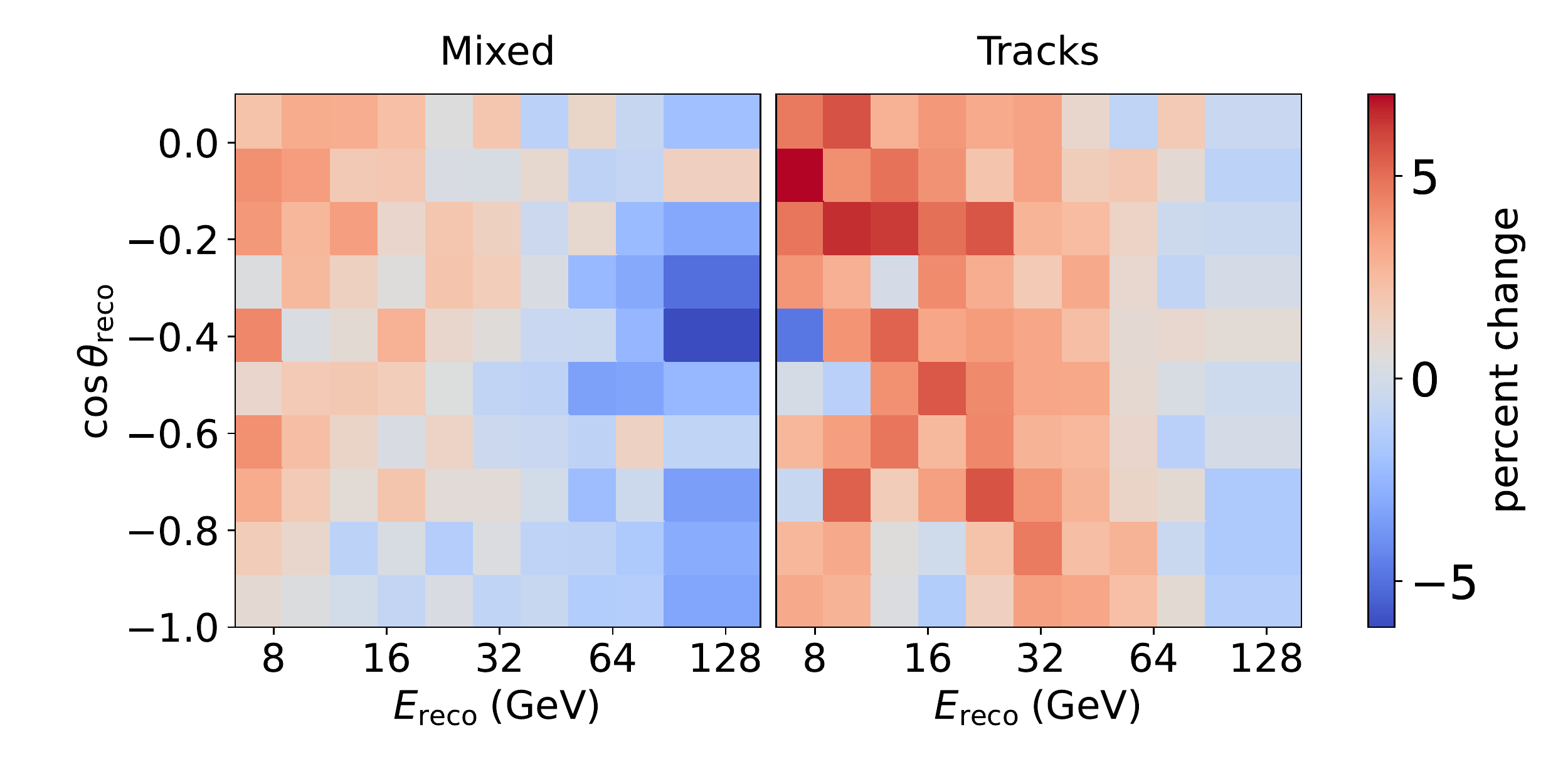}
    \caption{Ice absorption.}
\end{figure}

\begin{figure}
    \centering
    \includegraphics[width=0.95\linewidth]{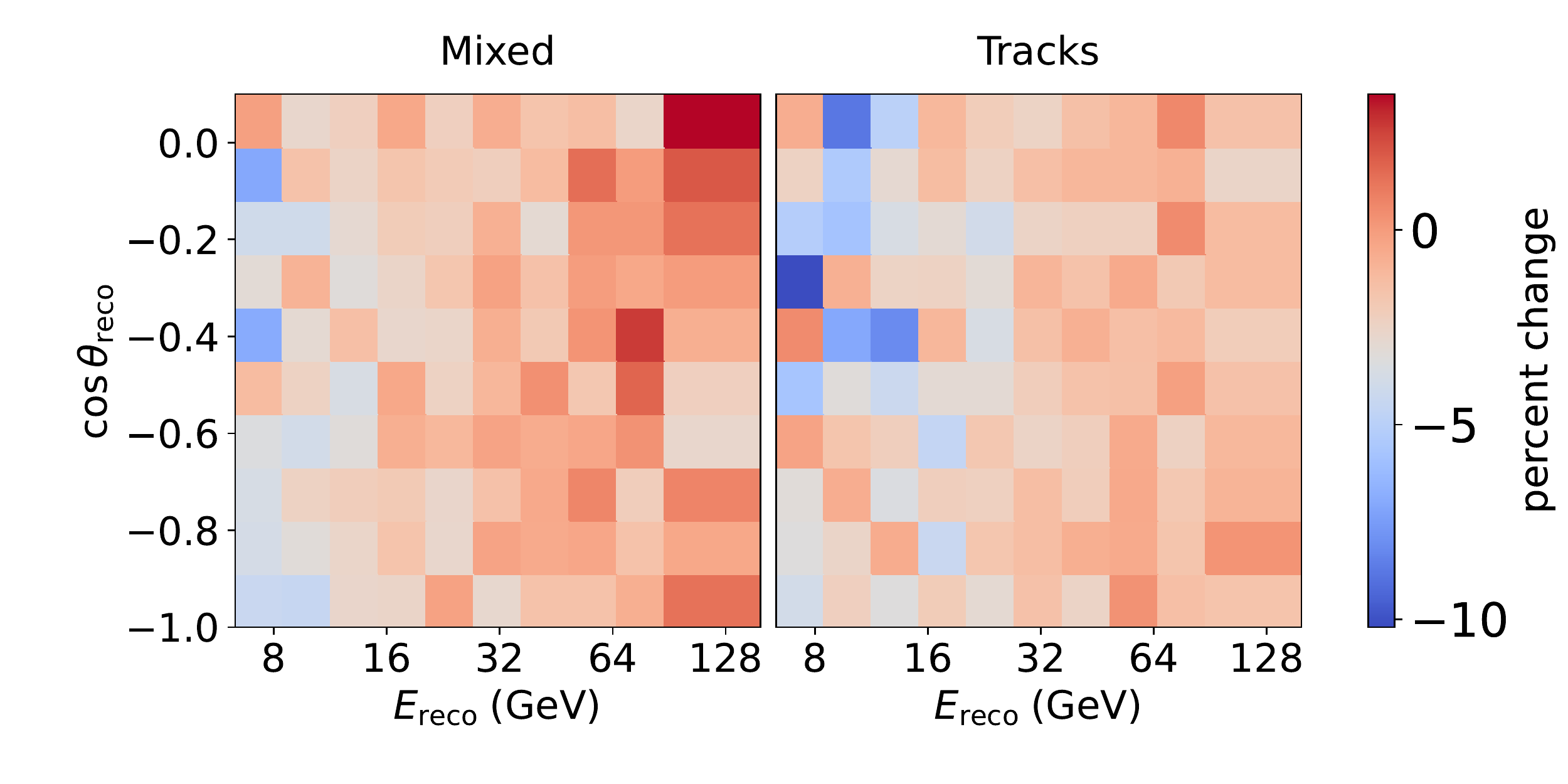}
    \caption{Ice scattering.}
\end{figure}

\begin{figure}
    \centering
    \includegraphics[width=0.95\linewidth]{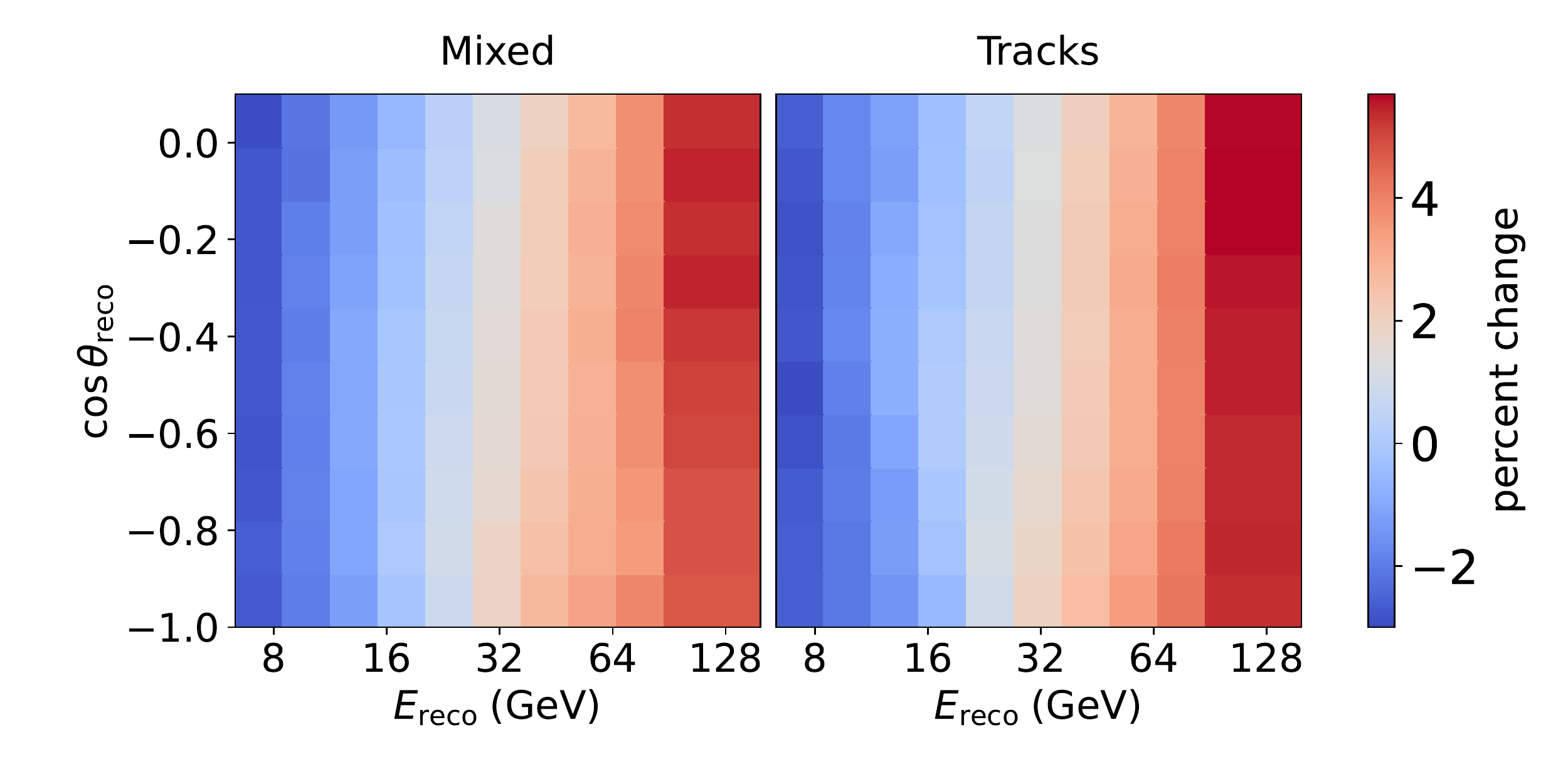}
    \caption{Atm. flux $\Delta\gamma$.}
\end{figure}

\begin{figure}
    \centering
    \includegraphics[width=0.95\linewidth]{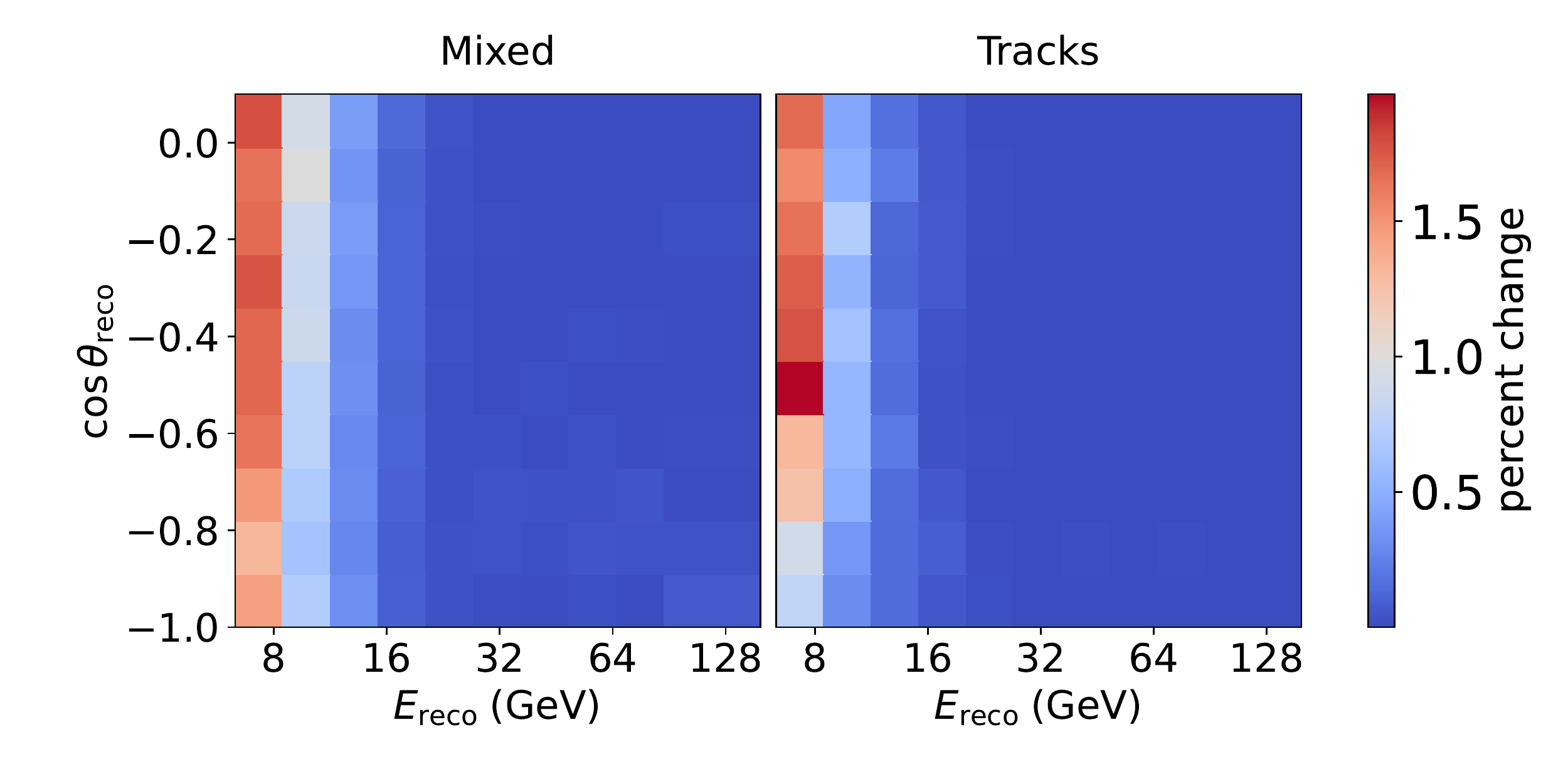}
    \caption{Atm. flux A-F parameters.}
\end{figure}

\begin{figure}
    \centering
    \includegraphics[width=0.95\linewidth]{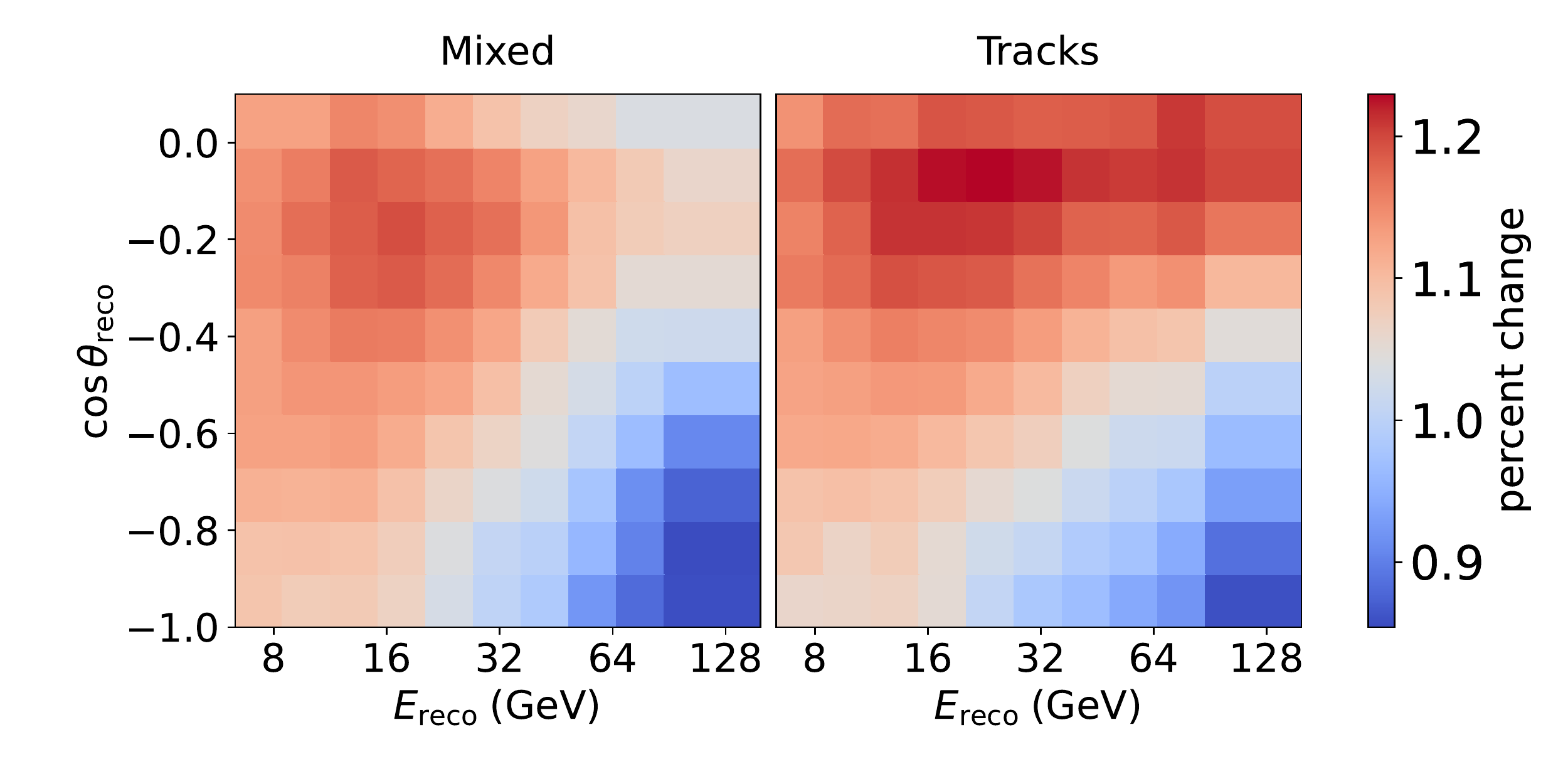}
    \caption{Atm. flux G parameter.}
\end{figure}

\begin{figure}
    \centering
    \includegraphics[width=0.95\linewidth]{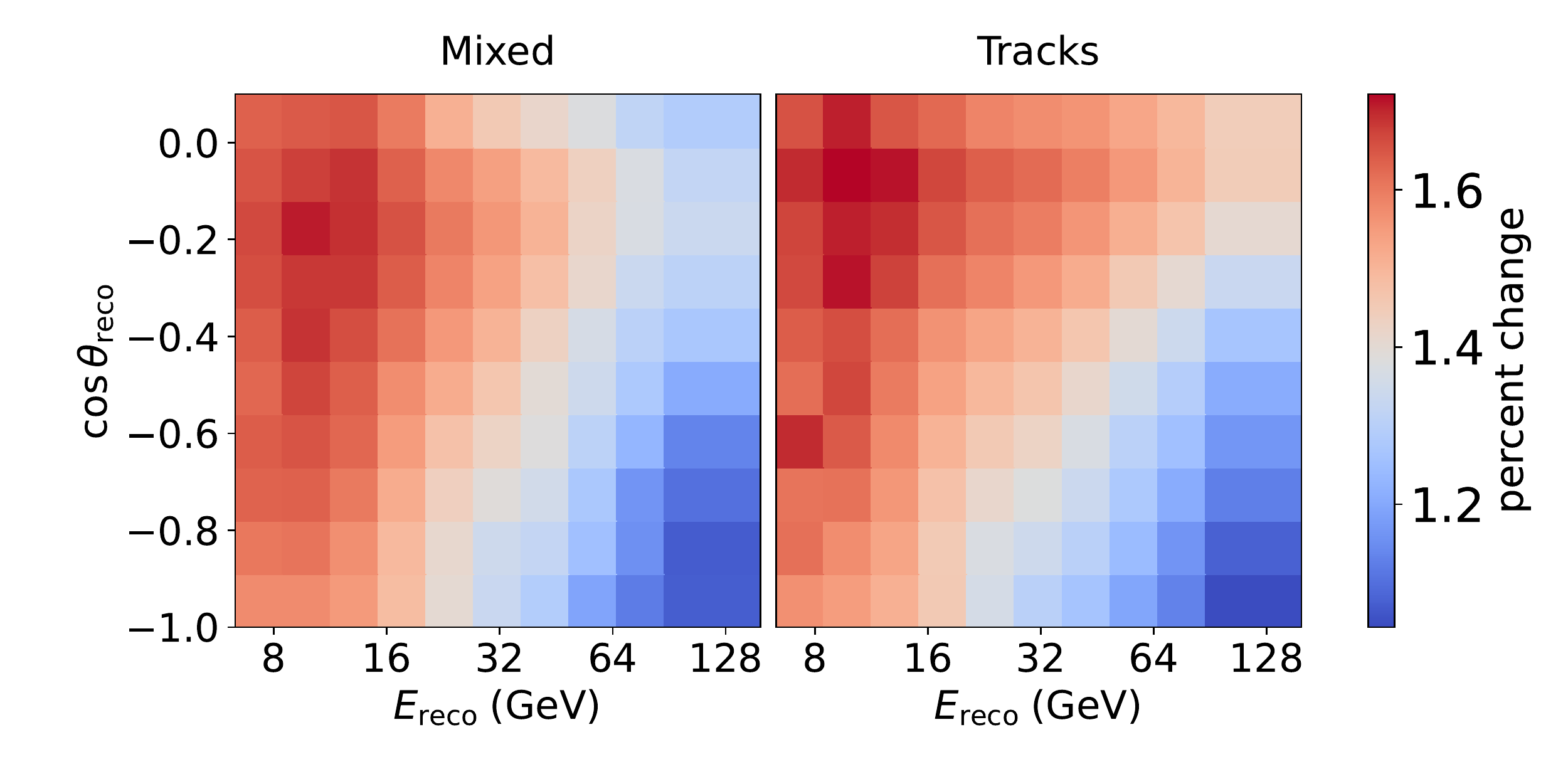}
    \caption{Atm. flux H parameter.}
\end{figure}

\begin{figure}
    \centering
    \includegraphics[width=0.95\linewidth]{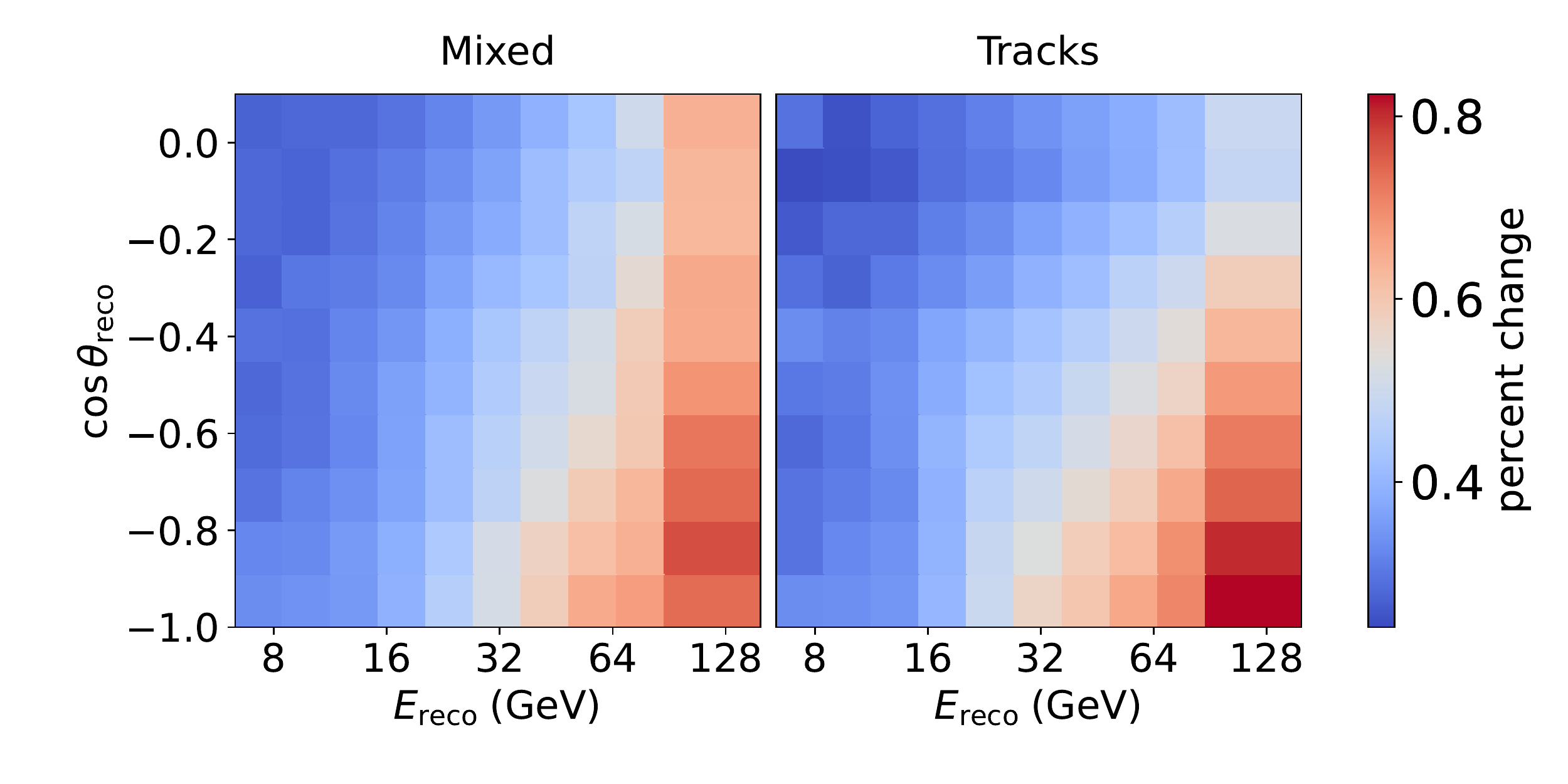}
    \caption{Atm. flux W parameter ($K^+$).}
\end{figure}

\begin{figure}
    \centering
    \includegraphics[width=0.95\linewidth]{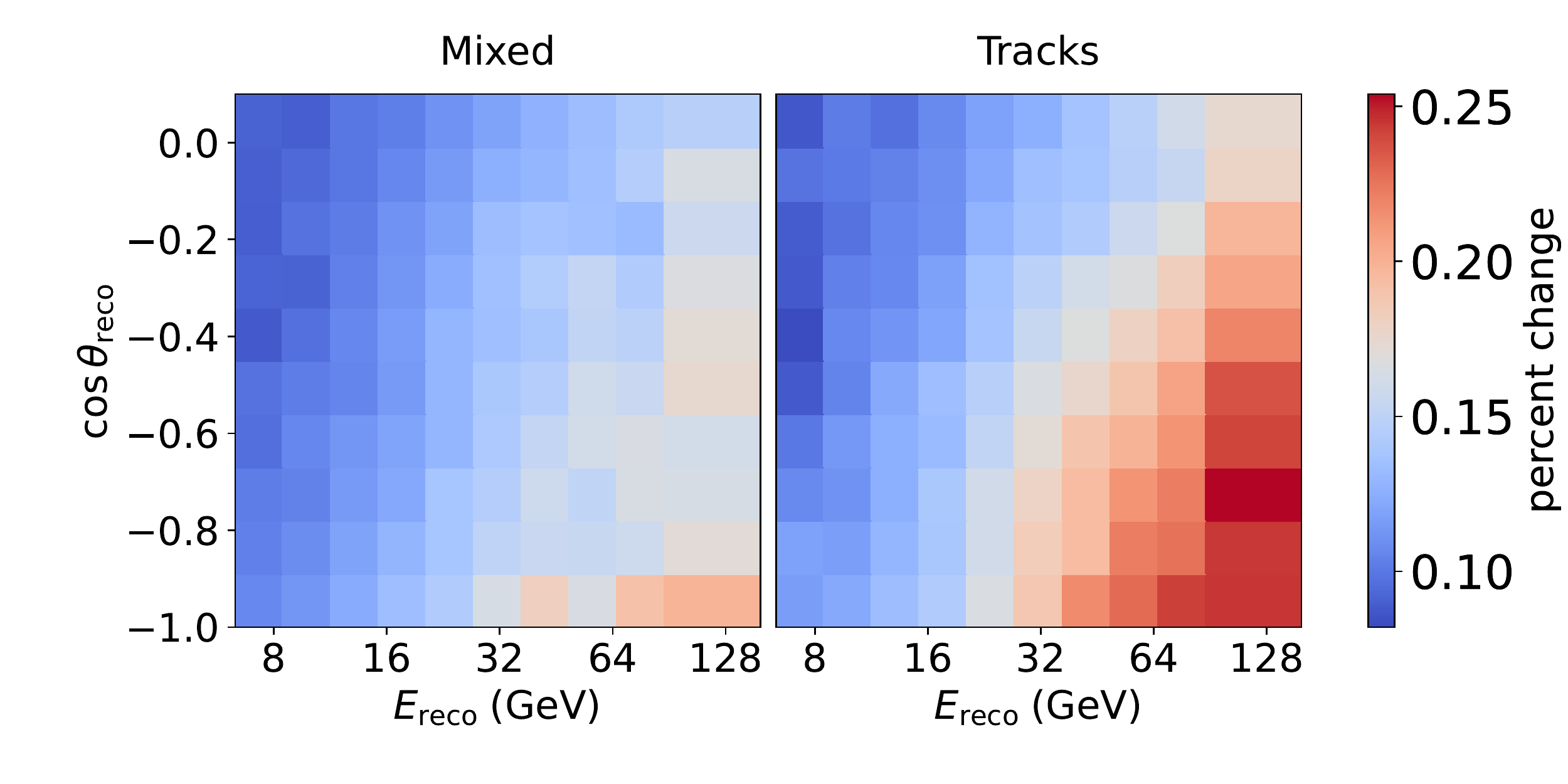}
    \caption{Atm. flux W parameter ($K^-$).}
\end{figure}

\begin{figure}
    \centering
    \includegraphics[width=0.95\linewidth]{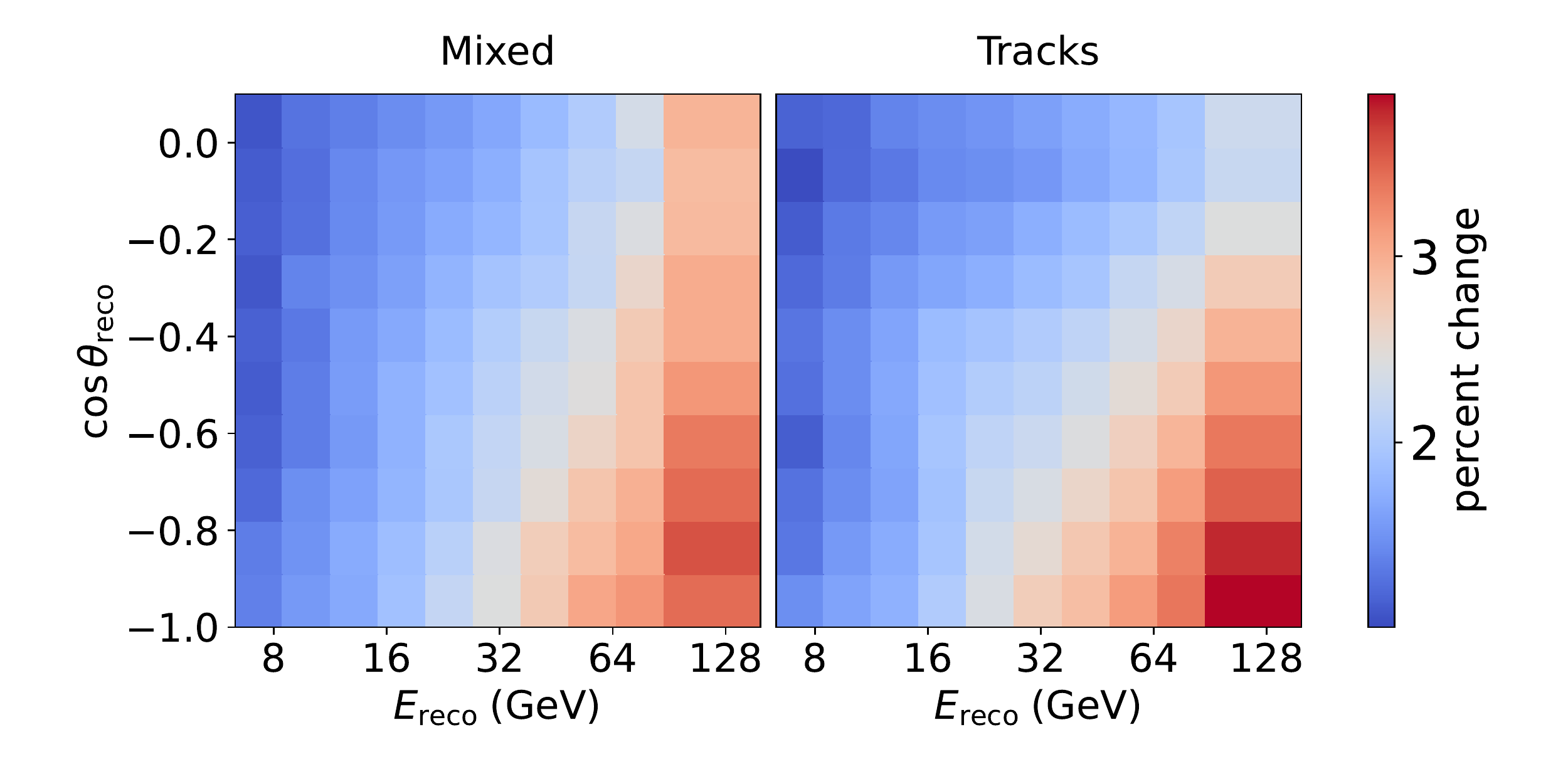}
    \caption{Atm. flux Y parameter.}
\end{figure}

\begin{figure}
    \centering
    \includegraphics[width=0.95\linewidth]{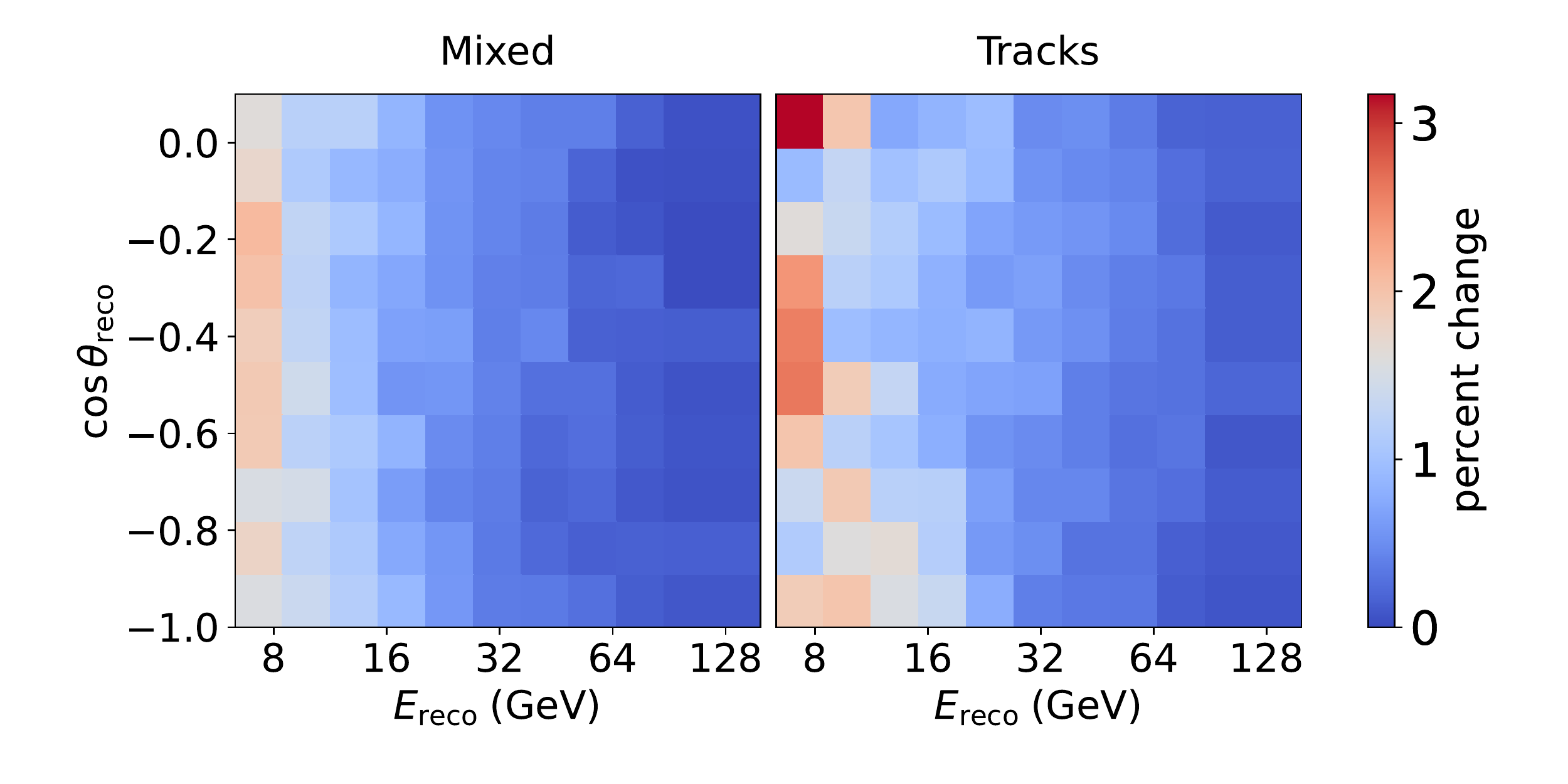}
    \caption{Axial mass CCQE.}
\end{figure}

\begin{figure}
    \centering
    \includegraphics[width=0.95\linewidth]{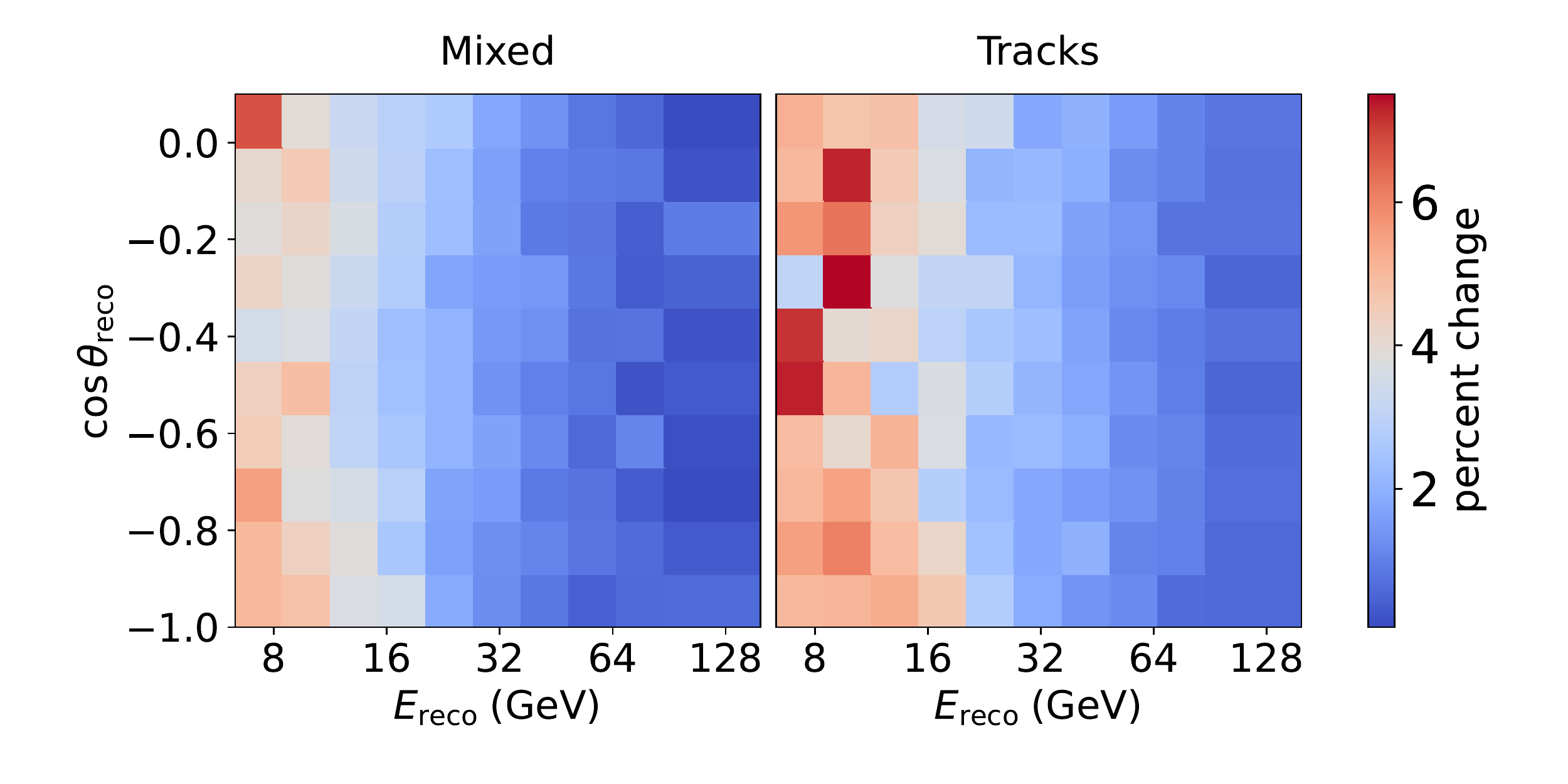}
    \caption{Axial mass RES.}
\end{figure}

\begin{figure}
    \centering
    \includegraphics[width=0.95\linewidth]{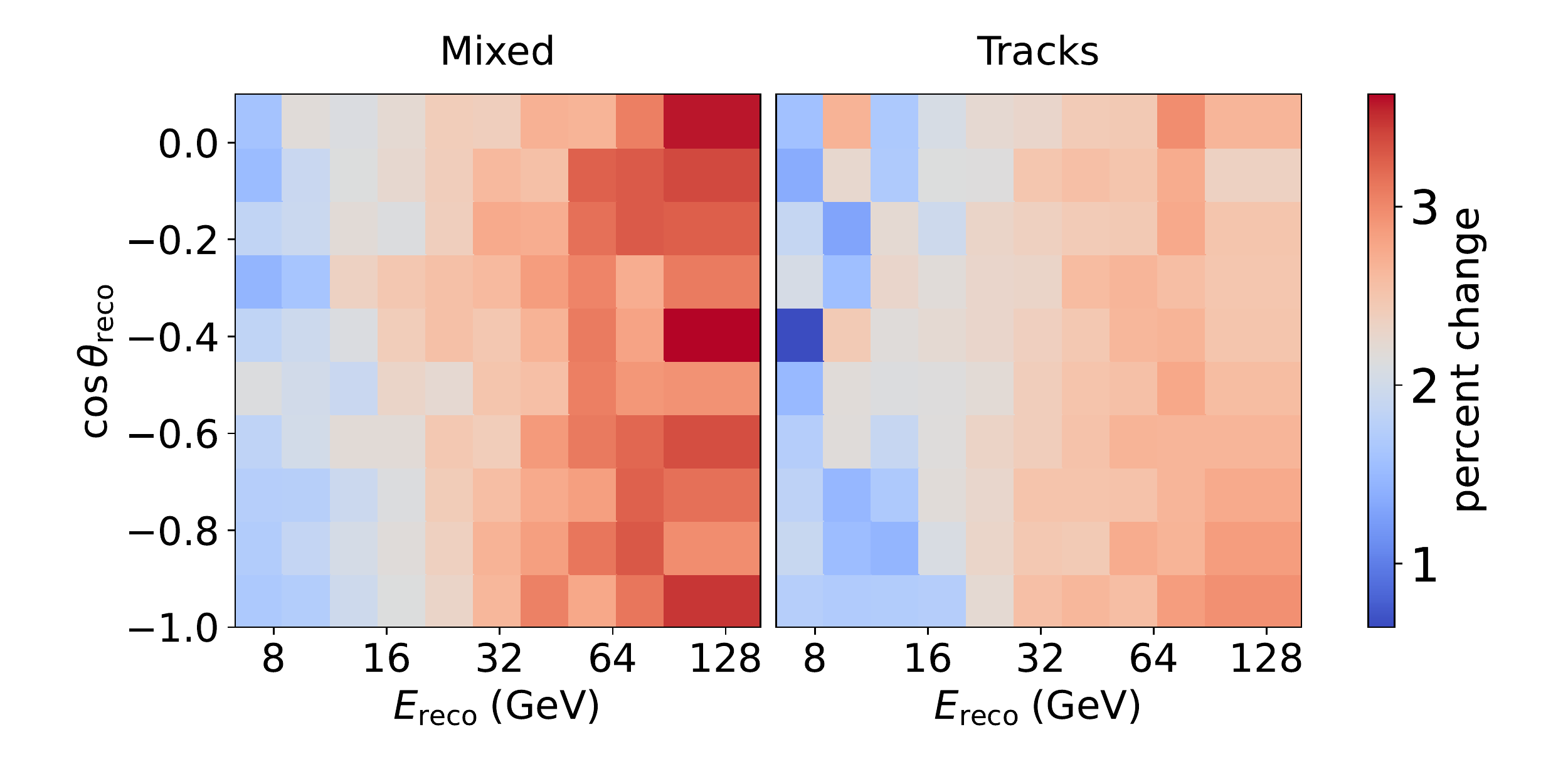}
    \caption{Deep inelastic scattering correction to CSMS.}
\end{figure}

\begin{figure}
    \centering
    \includegraphics[width=0.95\linewidth]{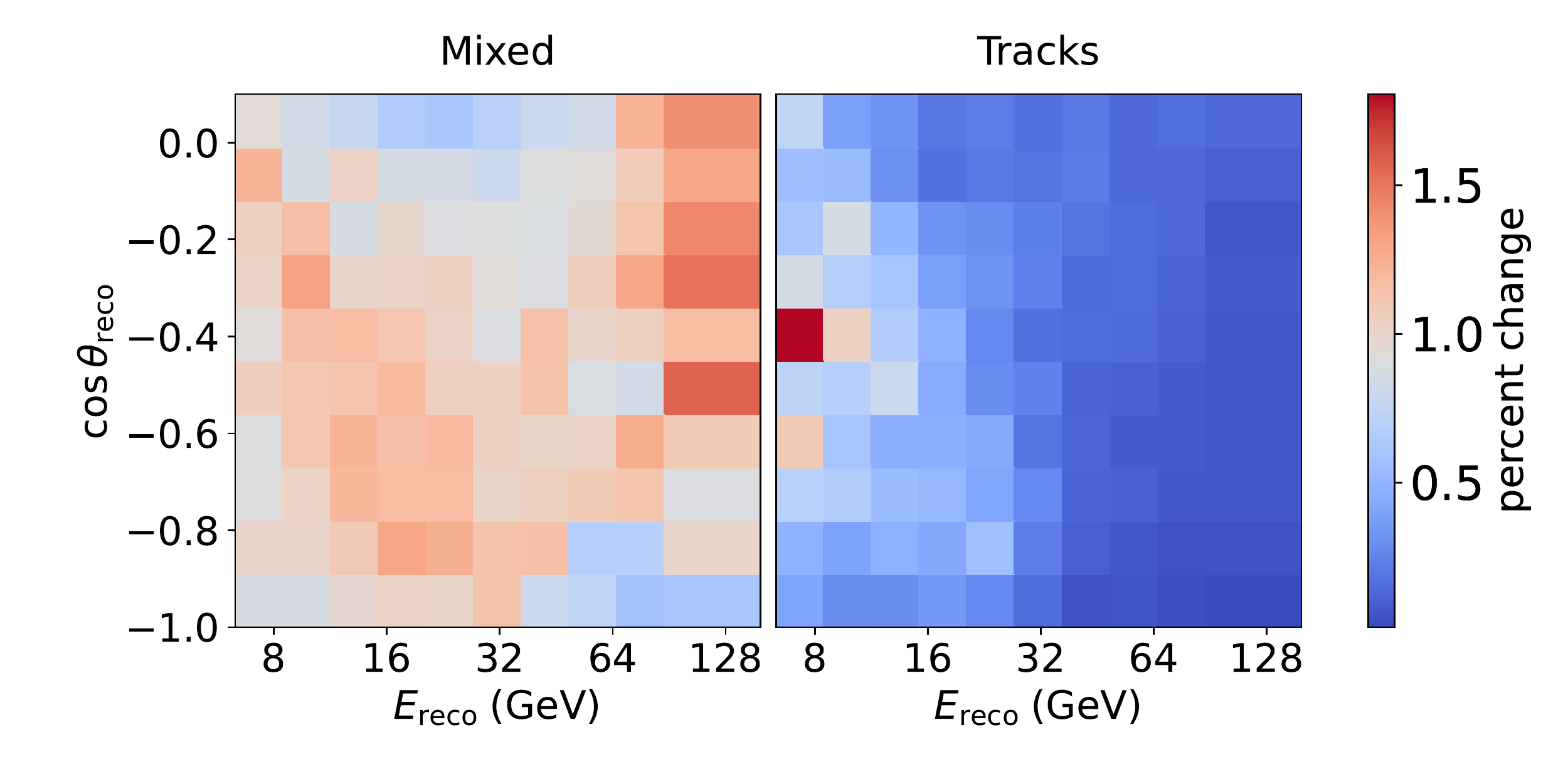}
    \caption{NC normalization.}
\end{figure}

\FloatBarrier
\onecolumngrid

\section{Data long-term stability~\label{appendix:data_quality}}
\begin{figure*}[!h]
  \begin{center}      
      \includegraphics[width=0.3 \linewidth]{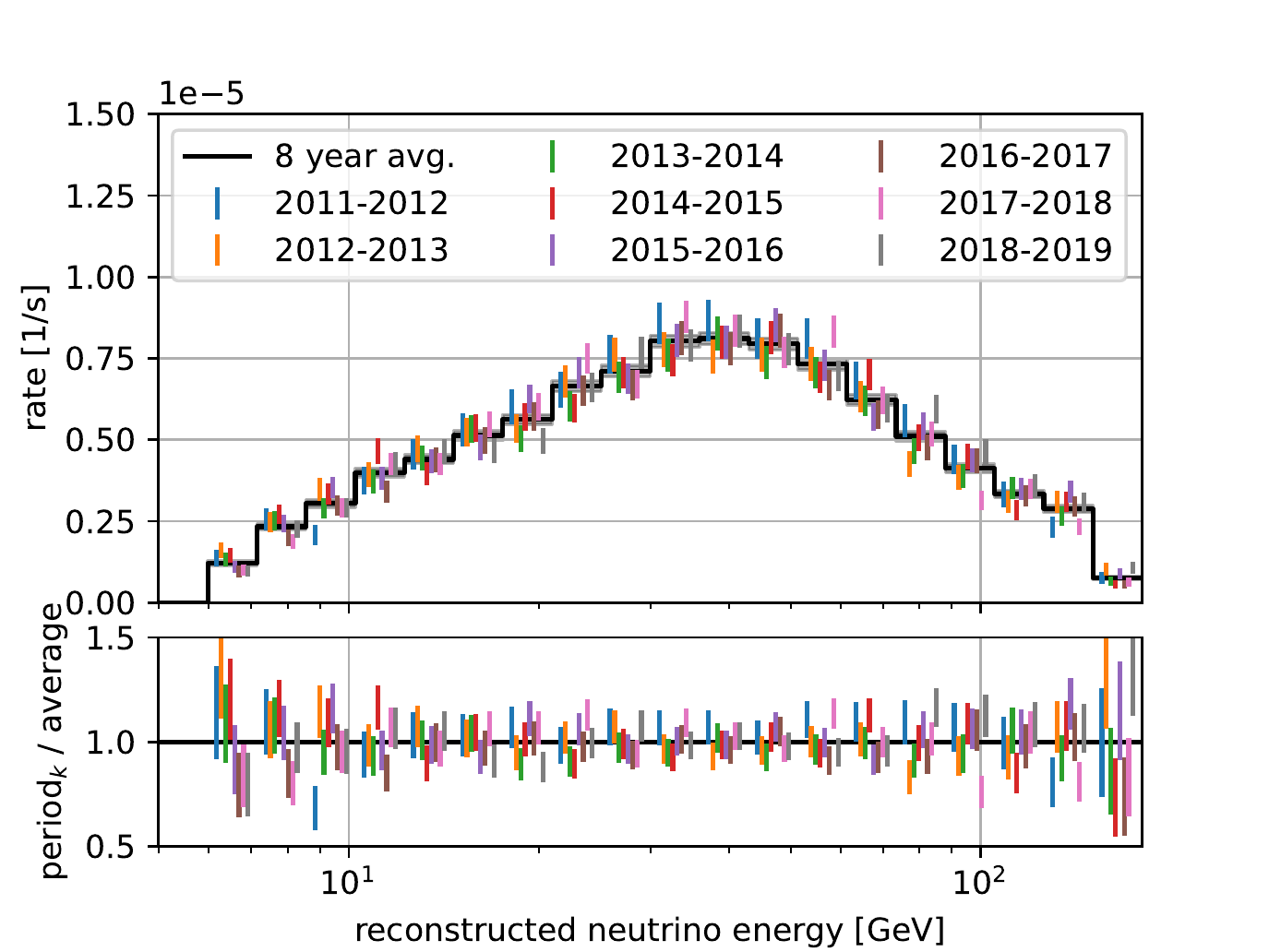}
      \includegraphics[width=0.3\linewidth]{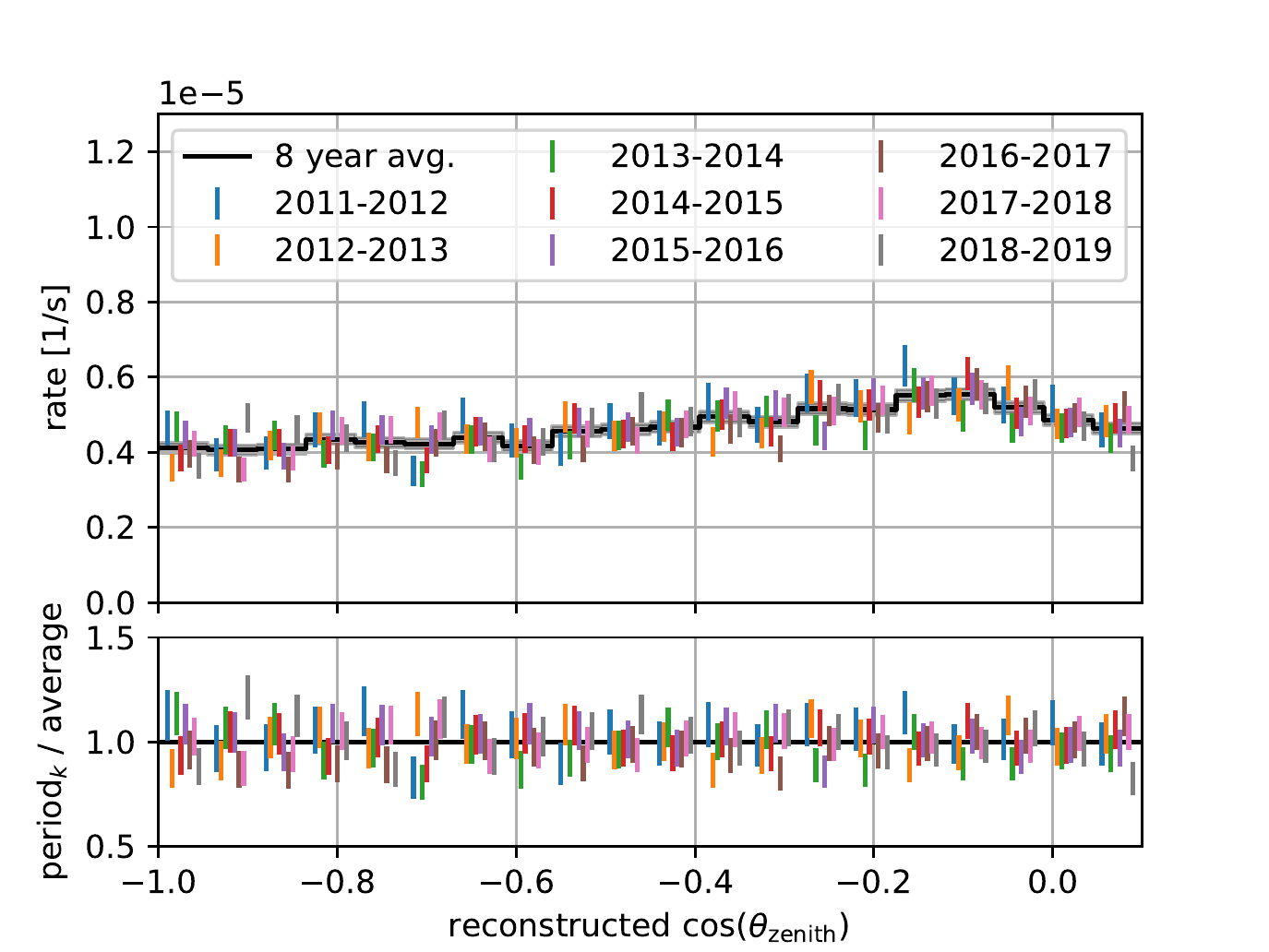}
      \includegraphics[width=0.3 \linewidth]{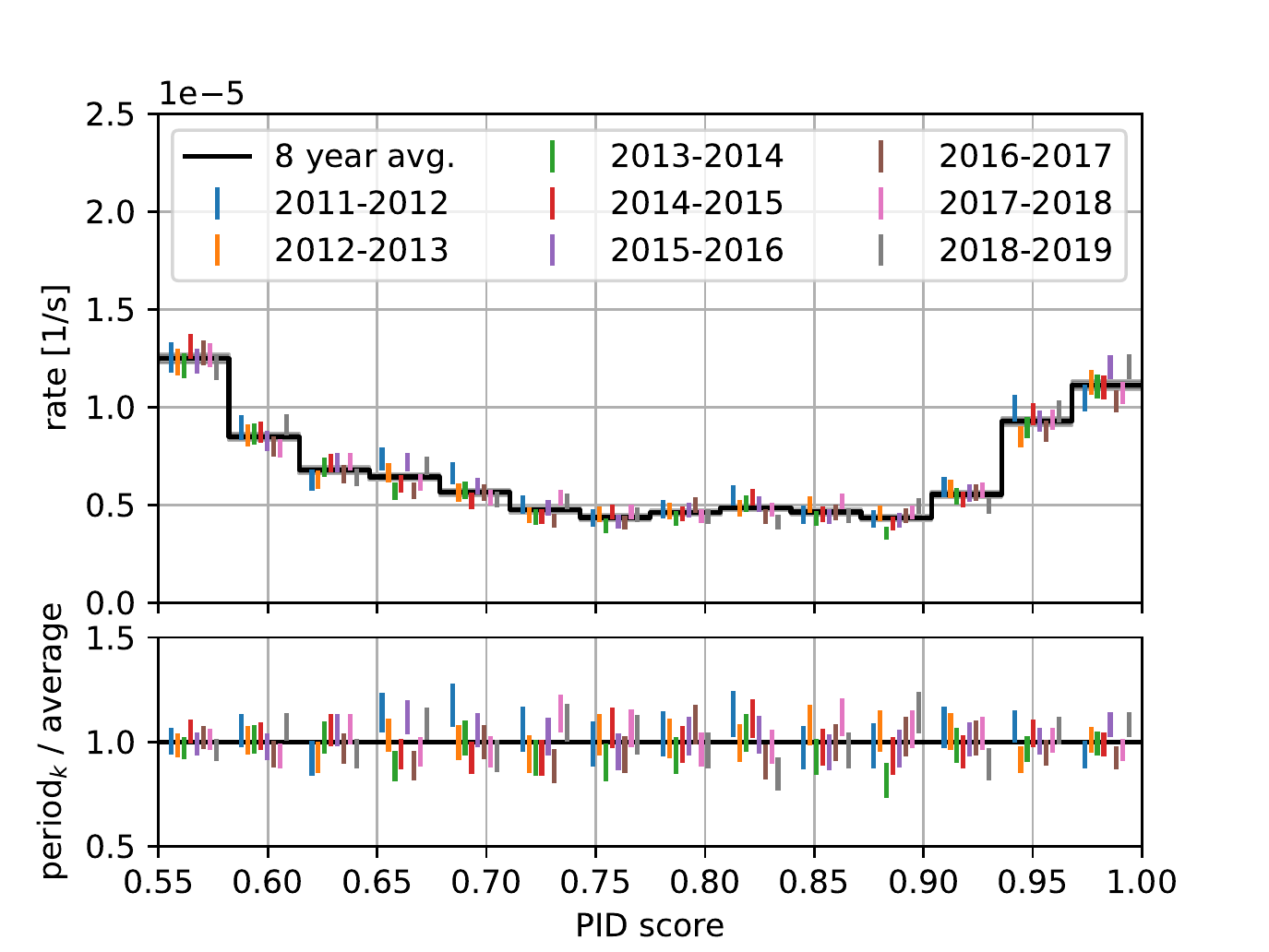}
  \end{center}
  \begin{center}      
      \includegraphics[width=0.3\linewidth]{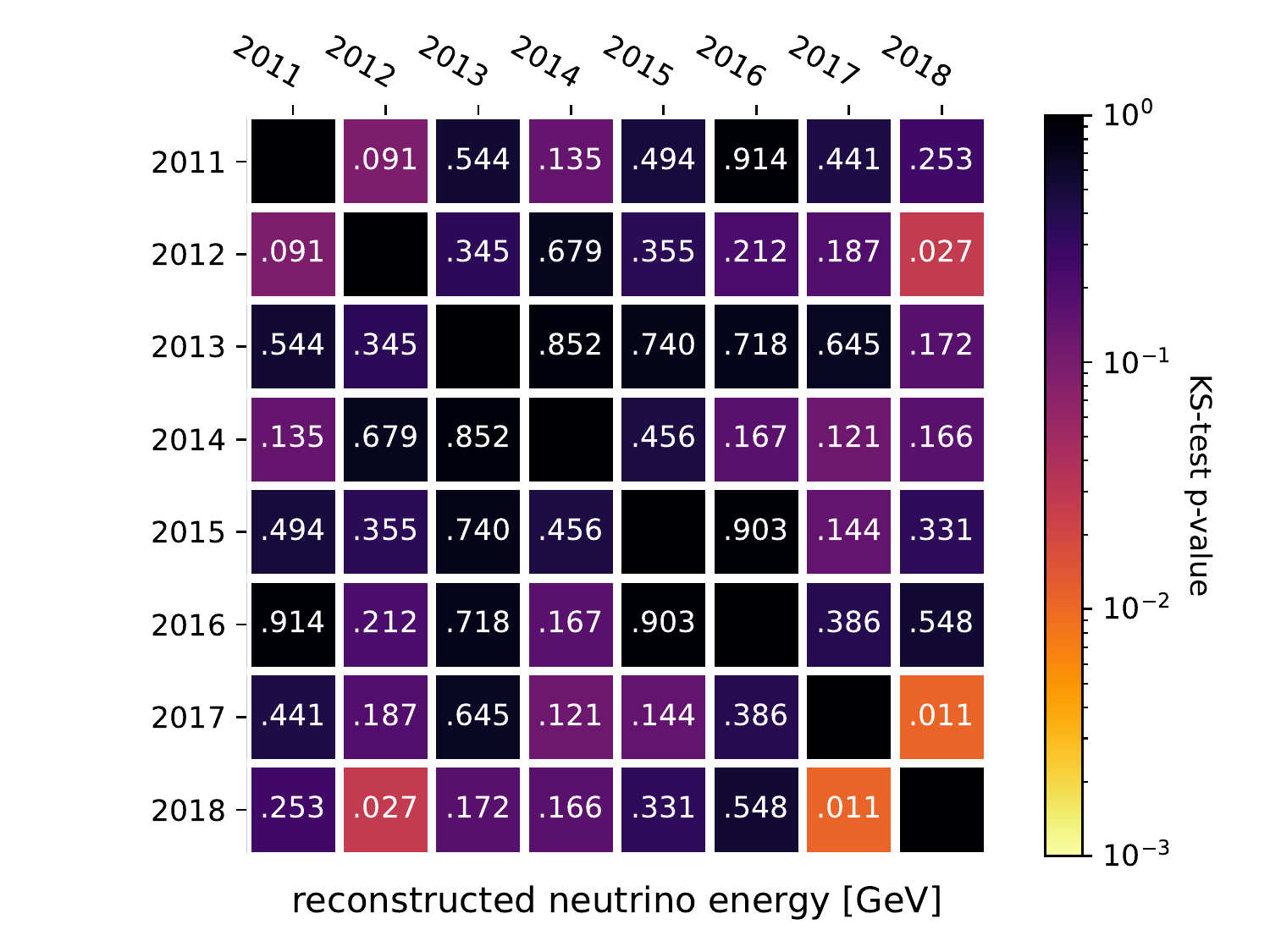}
      \includegraphics[width=0.3 \linewidth]{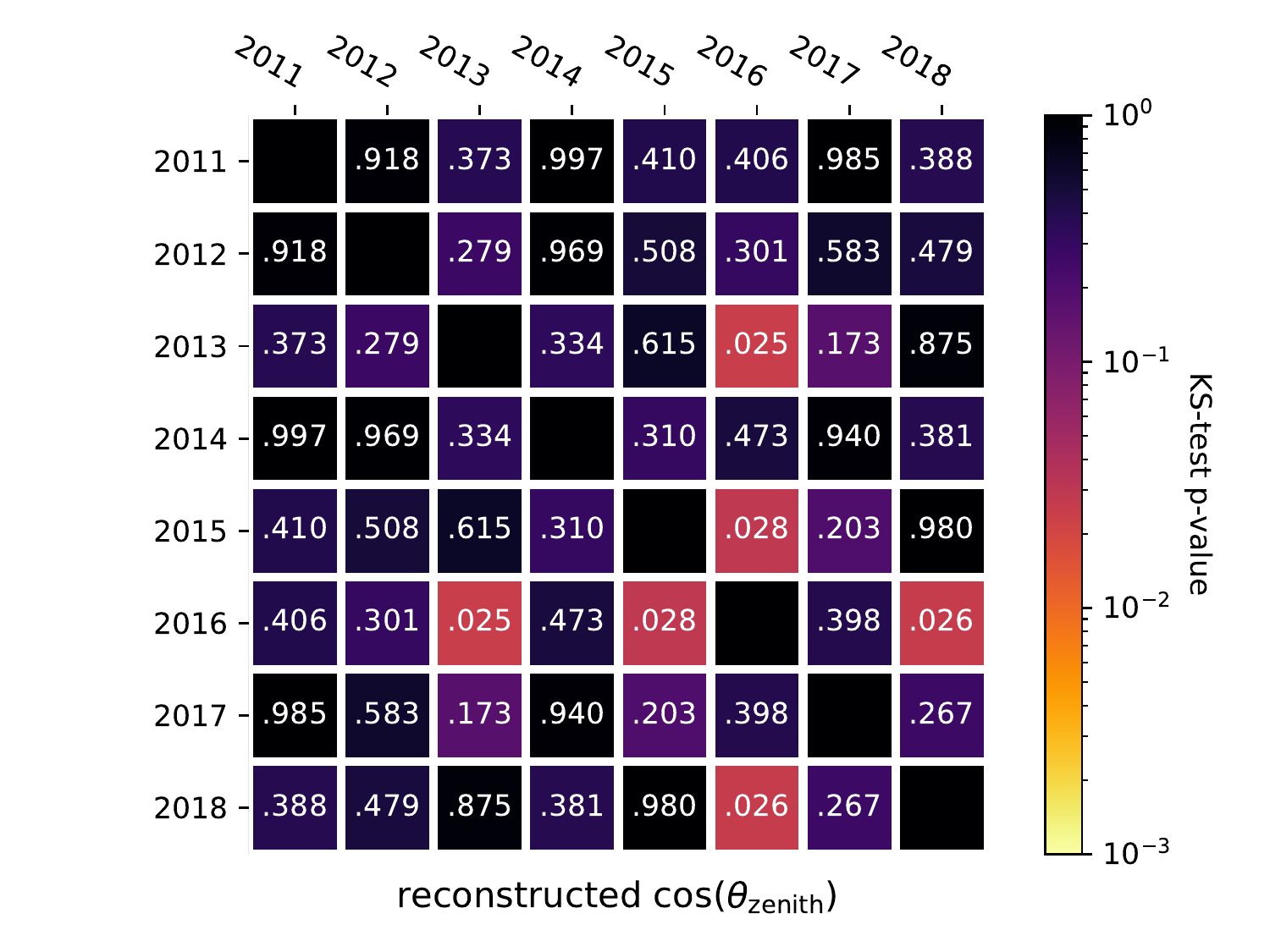}
      \includegraphics[width=0.3 \linewidth]{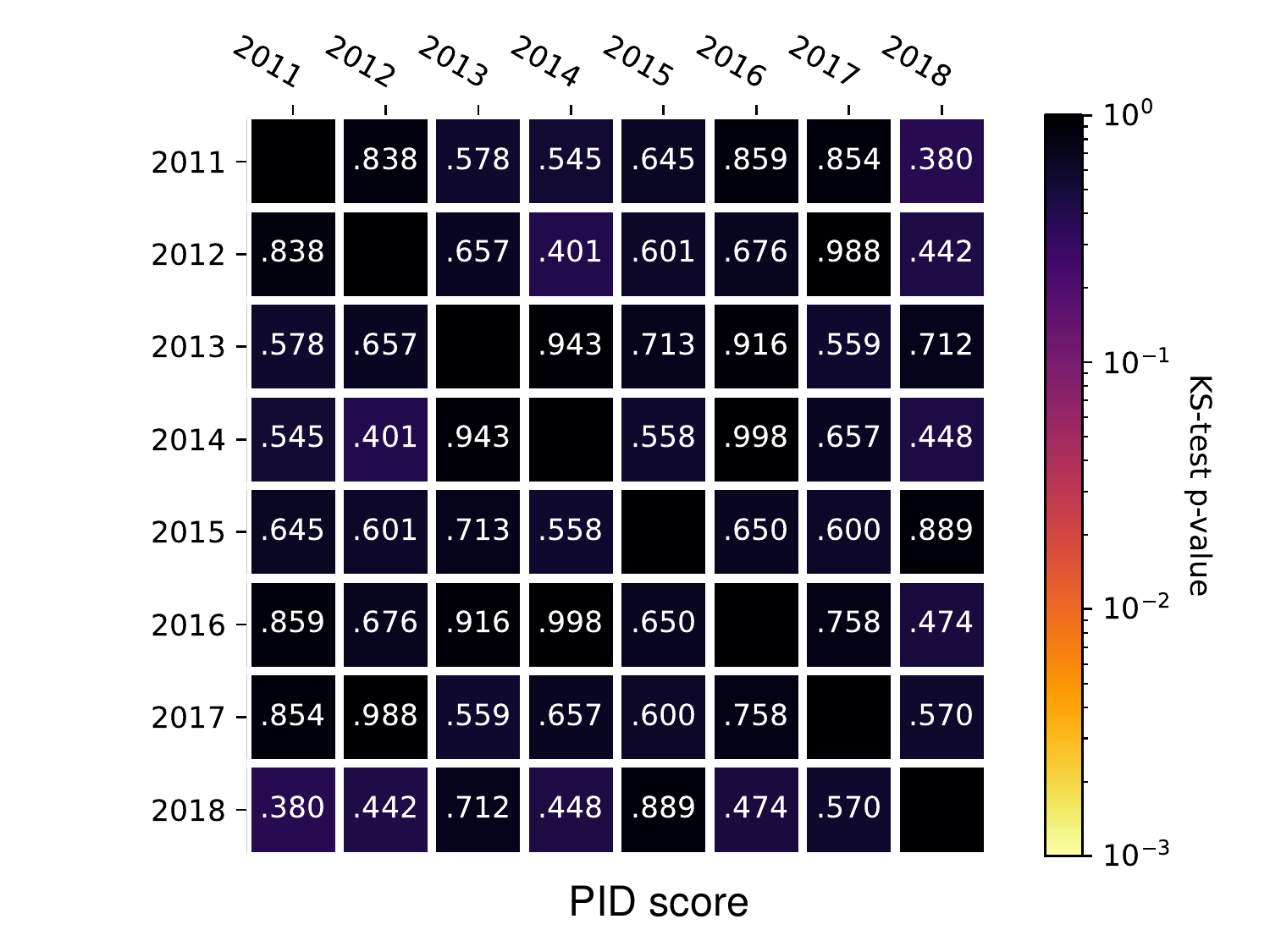}
  \end{center}
  \caption{Agreement between years for the reconstructed quantities used in the fit. Top panels show the distribution for every year, compared with the average. Bottom panels display the Kolmogorov-Smirnov p-values calculated between each season of data.}
  \label{fig:data_stability_2D_control_KS}
\end{figure*}

\begin{figure*}[!h]
  \begin{center}      
      \includegraphics[width=0.3 \linewidth]{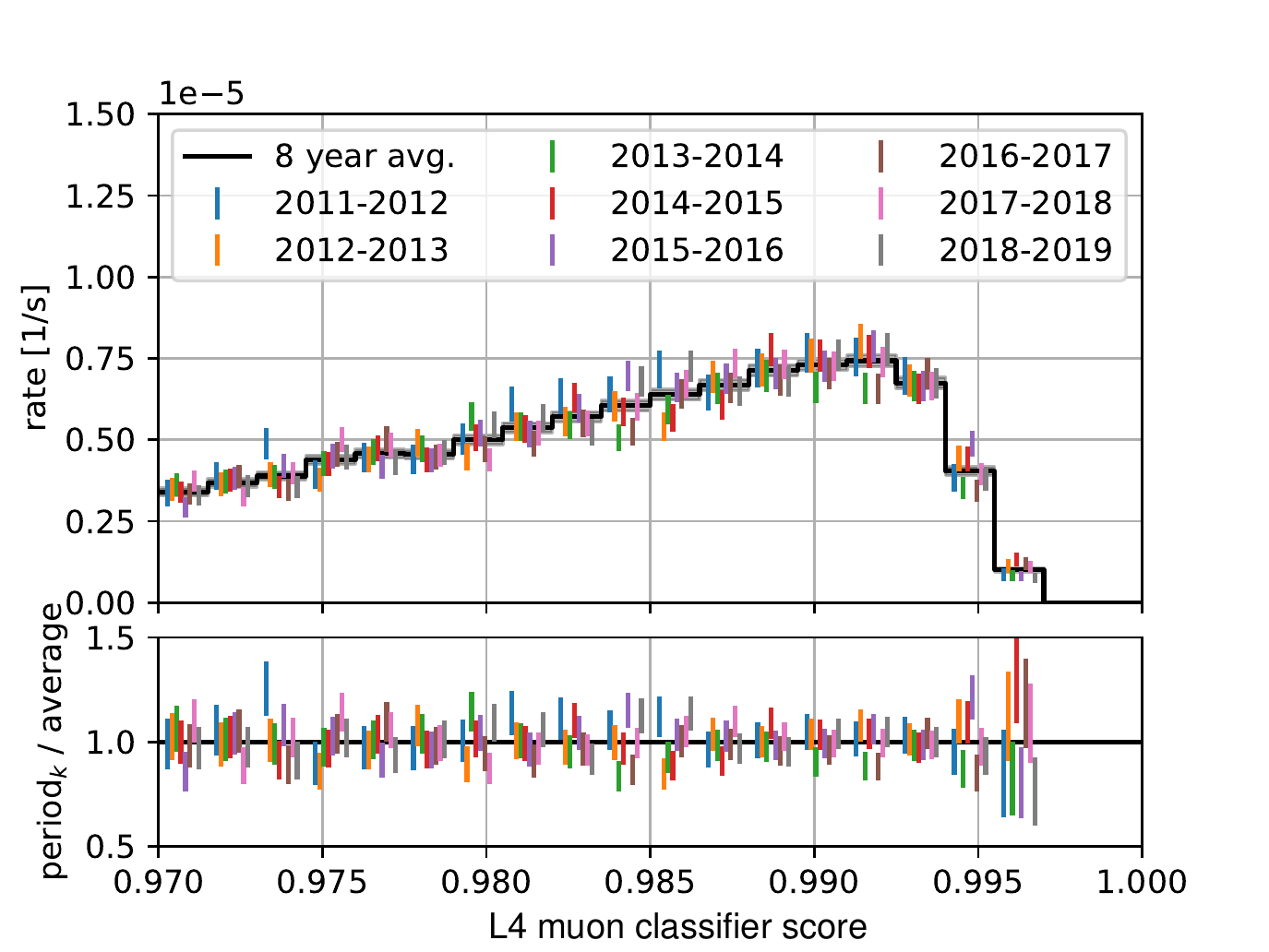}
    \includegraphics[width=0.3 \linewidth]{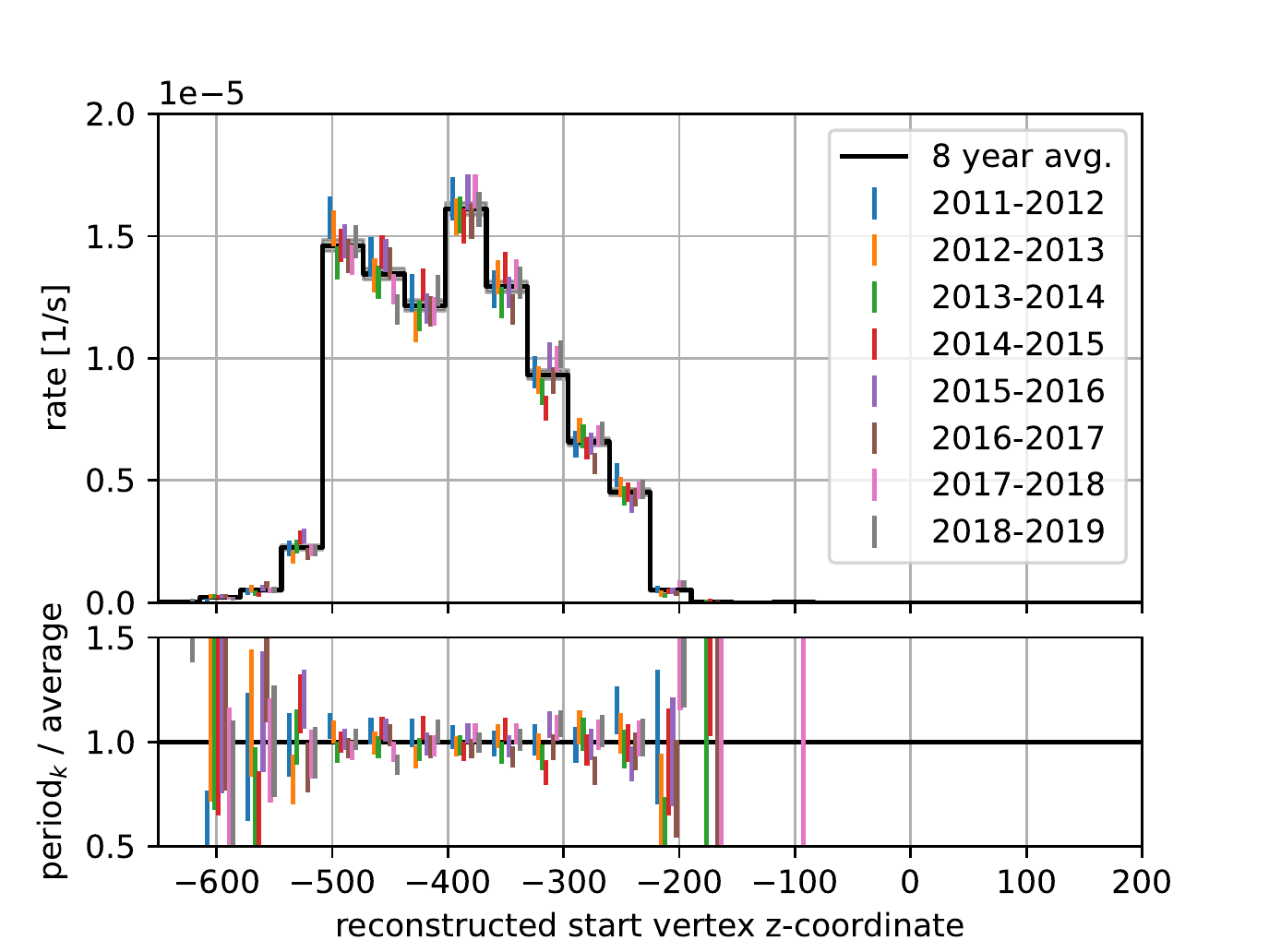}
      \includegraphics[width=0.3\linewidth]{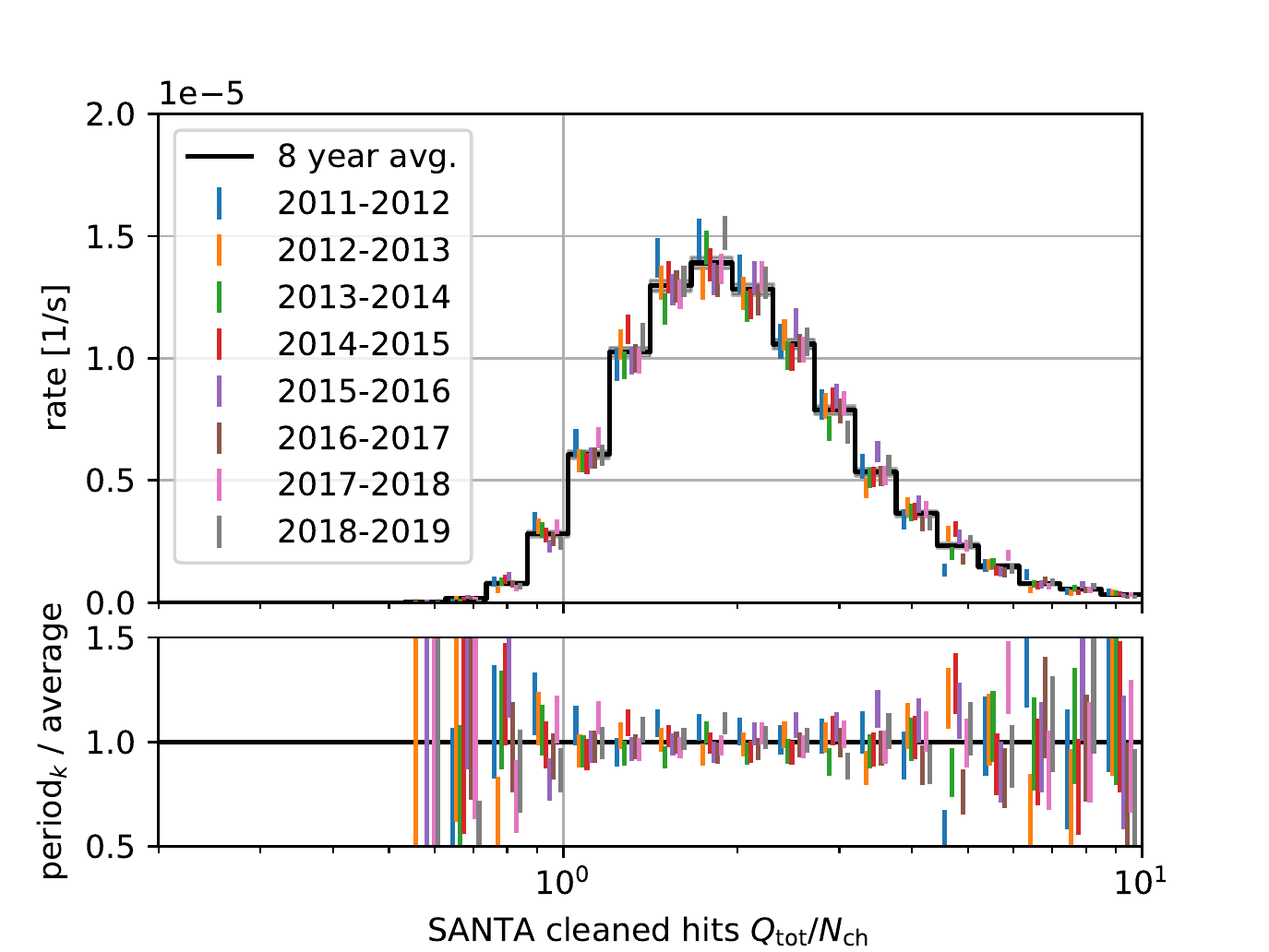}
  \end{center}
  \begin{center}      
      \includegraphics[width=0.3\linewidth]{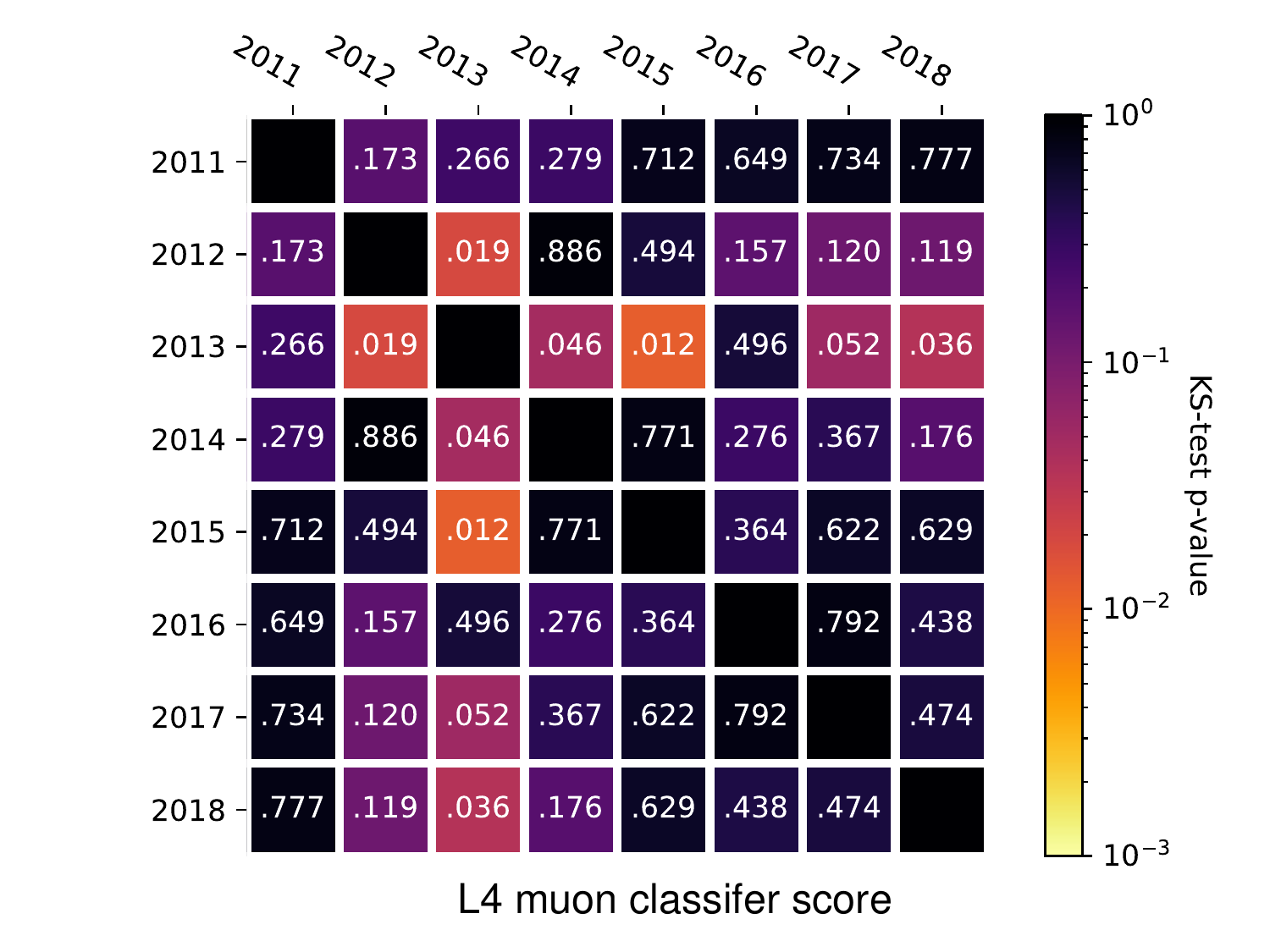}
      \includegraphics[width=0.3 \linewidth]{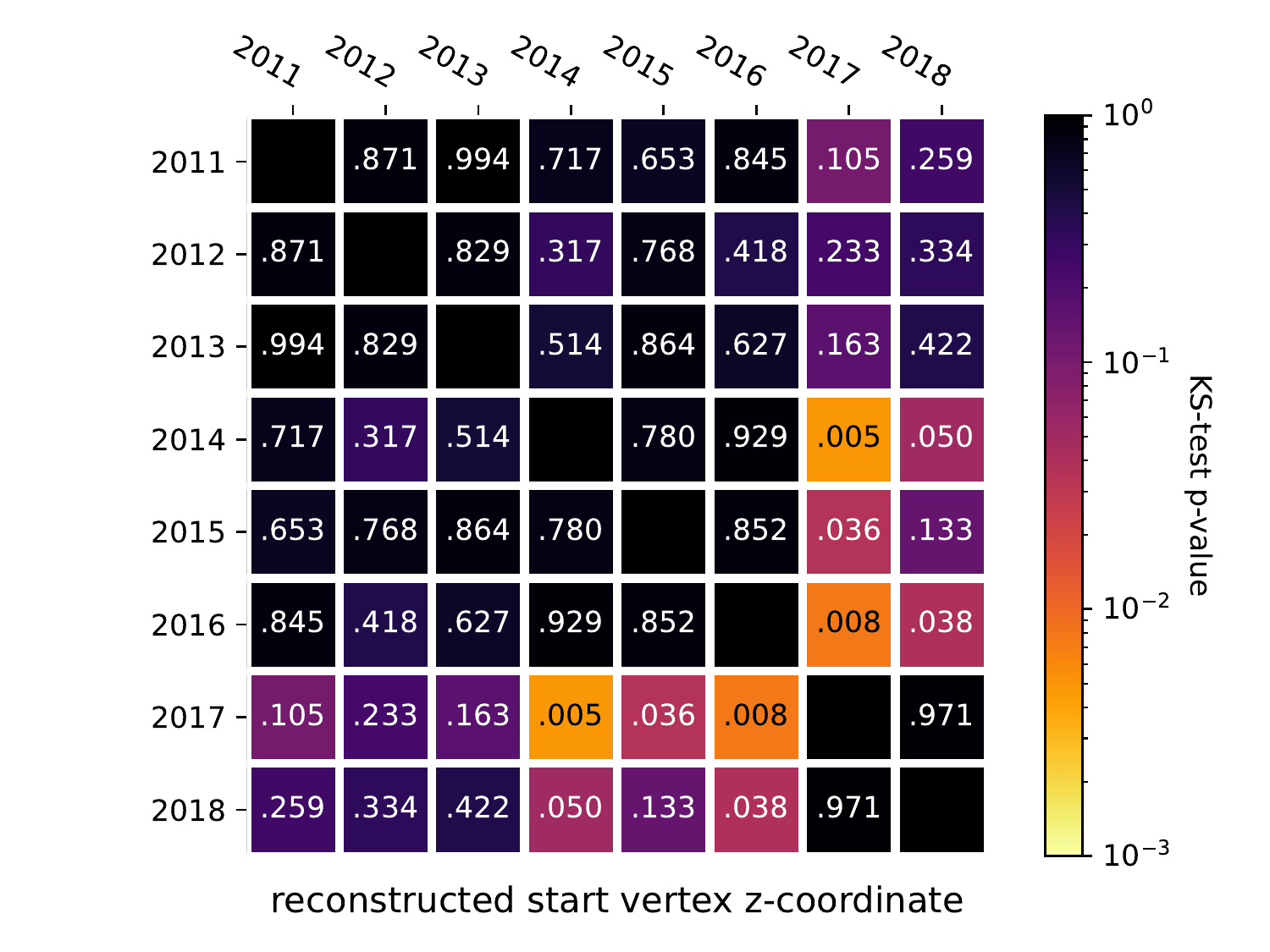}
      \includegraphics[width=0.3 \linewidth]{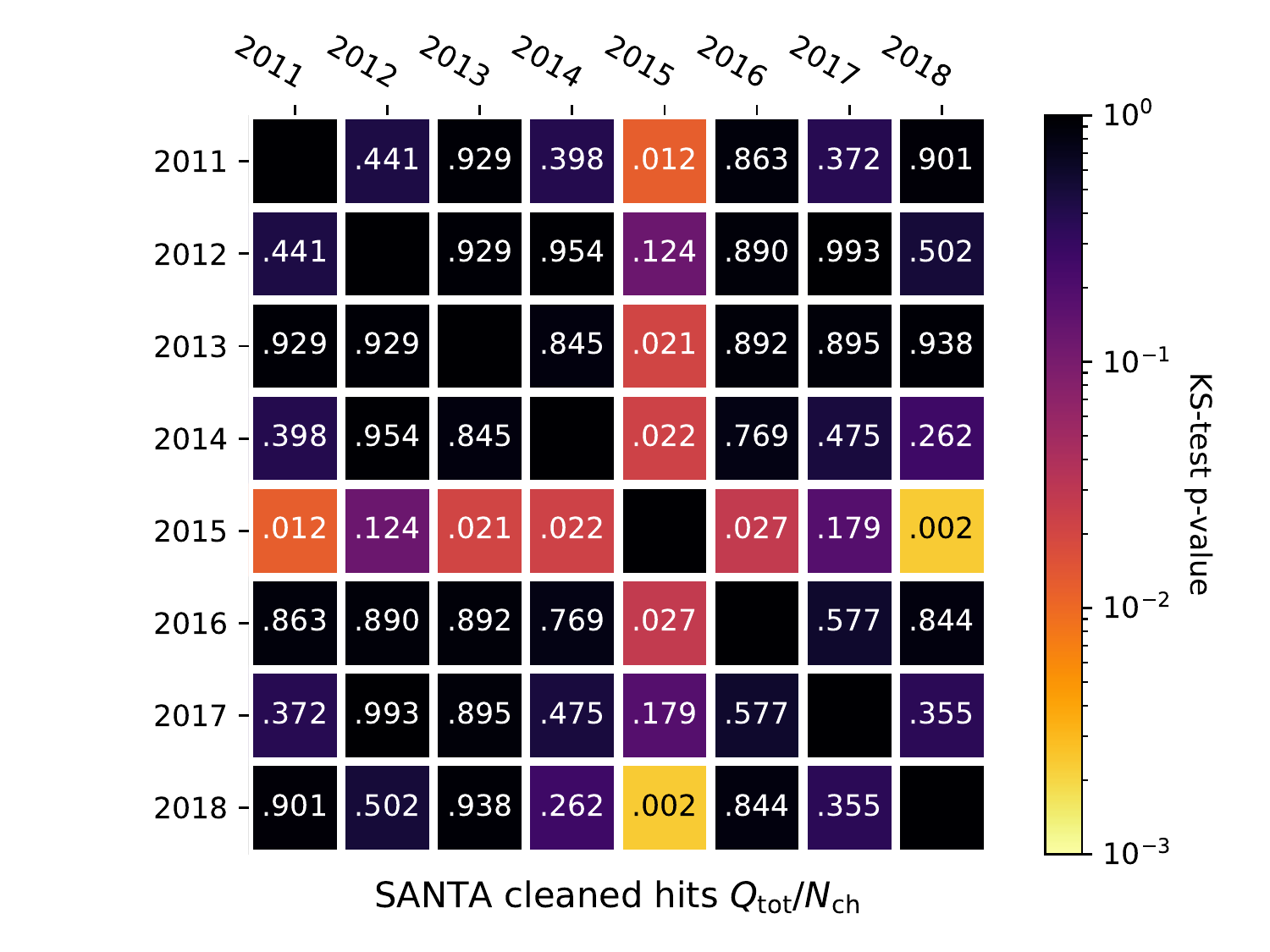}
  \end{center}
  \caption{Agreement between years for two of the variables used in the event selection (L4 muon classifier score and depth of the event vertex) and one low-level control variable that shows the typical charge per hit observed. For a description of the content of the panels, see Fig.~\ref{fig:data_stability_2D_control_KS}.}
  \label{fig:data_stability_2D_control_KS_2}
\end{figure*}

We verified the stability of the data by studying the compatibility of several variables across all years in the study using a Kolmogorov-Smirnov test. The observables used to bin the data are shown in Fig.~\ref{fig:data_stability_2D_control_KS}. Two selection variables and one control variable are also shown in Fig.~\ref{fig:data_stability_2D_control_KS_2}. The L4 muon classifier score is of special interest, since it was trained using a subset of data from 2014. As shown in Fig.~\ref{fig:data_stability_2D_control_KS_2}, all of the years have a similar behavior.

\end{document}